%% file: main.tex
\documentclass[authoryear,reqno,11pt,a4paper]{elsarticle}
\usepackage{etoolbox}
\usepackage{amsmath,amssymb,graphicx,hyperref,setspace,ifthen,soul,amsthm,float,color}
\usepackage{booktabs}
\usepackage{algorithm}
\usepackage{subfigure}
\usepackage{multirow}
\usepackage{makecell}
\usepackage{cleveref}
\crefformat{appendix}{#2#1#3}
\Crefformat{appendix}{#2#1#3}
\usepackage{microtype}
\usepackage{array}
\usepackage{tabularx}
\usepackage{enumitem}
\usepackage{threeparttable}
\usepackage{graphicx}
\usepackage{tikz}
\usetikzlibrary{arrows.meta,calc,positioning,fit,patterns}

\usepackage[authoryear]{natbib}
\robustify\bfseries
\journal{Mathematical Finance}

\hypersetup{
	pdftitle={},
	pdfauthor={},
	pdfkeywords={},
	colorlinks=true,
	linkcolor=blue,
	citecolor=blue,
	urlcolor=blue,
	bookmarksnumbered=true
}

\date{\today}

\newtheorem{definition}{Definition}
\newtheorem{lemma}{Lemma}
\newtheorem{theorem}{Theorem}
\newtheorem{proposition}{Proposition}
\newtheorem{remark}{Remark}
\newtheorem{corollary}{Corollary}

\newcommand{\E}{\mathbb{E}}
\newcommand{\Q}{\mathbb{Q}}
\newcommand{\Pp}{\mathbb{P}}
\newcommand{\dd}{\mathrm{d}}
\newcommand{\llevy}{L\'evy}

\newcommand{\safeincludegraphics}[2][]{%

  \IfFileExists{#2}{\includegraphics[#1]{#2}}{%

    \fbox{\parbox{0.85\textwidth}{Missing figure: \texttt{#2}}}%

  }%

}

\begin{document}

\setstretch{1.05}

\begin{frontmatter}

\title{Stochastic Mortality Model with Fractional L\'evy Dynamics}

\author[1]{Congxin He}

\author[2]{Lilian Hu}

\author[1]{Yue Kuen Kwok}

\author[3]{Yifan Ye\corref{cor1}}
\ead{yifanye@bnbu.edu.cn}

\cortext[cor1]{Corresponding author}

\address[1]{
Financial Technology Thrust, Hong Kong University of Science and Technology (Guangzhou), China
}

\address[2]{
Department of Statistics and Actuarial Science, University of Waterloo, Canada
}

\address[3]{Faculty of Business and Management, Beijing Normal-Hong Kong Baptist University, China}

\begin{abstract}
{A substantial body of empirical evidence suggests} that stochastic mortality models ignoring long range dependence tend to underestimate life expectancy, which may lead to profound implications for pension schemes and funding arrangements. This paper addresses the modelling of stochastic mortality via a mixture of {a} fractional \llevy{} process and standard Brownian motion. Our stochastic mortality model exhibits nice {analytical tractability} in actuarial valuations and flexibility in the choice of underlying \llevy{} specifications. 
The long range dependence feature embedded in our stochastic {mortality model} is well reflected in our empirical studies on the mortality shocks during World War II and COVID-19.  For efficient numerical pricing of longevity derivatives, we construct an effective singular value decomposition approximation scheme to overcome the computational challenges arising from {the non-Markovian nature} of the fractional \llevy{} process.  
Truncation errors in singular value decomposition approximation can be reduced by an effective residual correction scheme.    
\end{abstract}

\begin{keyword}
Stochastic mortality model \sep fractional L\'evy process \sep long-range dependence \sep singular value decomposition \sep longevity derivatives
\end{keyword}

\end{frontmatter}

\section{Introduction}\label{sec:intro}
\noindent
Mortality models are a cornerstone of actuarial science, playing a fundamental role in pricing, valuation, and risk management of life insurance, annuities, and pension plans. Their importance stems from the fact that accurate mortality assumptions directly determine the financial stability and solvency of insurance companies and pension funds. Mortality models are essential for calculating the probability of death at various ages, which directly determines insurance premiums. Insurers rely on these models to estimate the likelihood of policyholders dying within a given timeframe, ensuring that premiums are sufficient to cover future claims while remaining competitive.

Mortality modelling is a fundamentally interdisciplinary field that sits at the intersection of demography, actuarial science, biology, and statistics. Its core purpose is to mathematically describe and predict how death rates vary across age, time, and populations.
The most enduring observation in mortality modelling is the Gompertz Law (\citealp{Gompertz1825}). This simple deterministic law states that for adult humans, the force of mortality increases approximately exponentially with age. With the recent development of more complex life insurance products and more rigorous regulations on controls of demographic risks, researchers have chosen to adopt stochastic modelling of mortality. 

{The widely used Lee--Carter model} (\citealp{LeeCarter1992}) is a foundational discrete-time stochastic mortality model that decomposes log-mortality rates into an age-specific component and a time-varying mortality index. It is widely used for forecasting mortality trends and understanding longevity risk.  Extensions to the Lee-Carter model are numerous. \citet{renshaw2006cohort} generalized the Lee-Carter model by incorporating a cohort effect to capture the mortality experience of specific birth cohorts. This model has been shown to improve mortality prediction, particularly for populations with significant cohort patterns.  Extensions that add multiple time-varying factors allow for capturing different aspects of mortality evolution. For example, the Cairns-Blake-Dowd model (\citealp{CairnsBlakeDowd2006}) uses two factors: a level factor and a slope factor, making it particularly suitable for older-age mortality. \cite{ToczydlowskaPetersFungShevchenko2017} developed a multi-factor extension of the Lee-Carter family that incorporates exogenous observable demographic features as additional factors. This approach employs robust dimensionality reduction techniques and a Bayesian state-space estimation framework, demonstrating improved in-sample fit and out-of-sample forecast performance on European Human Mortality Database data.	

Continuous-time stochastic mortality models have gained increasing popularity since the early 2000s, offering a tractable framework for pricing and hedging longevity risk. The key developments began with \citet{MilevskyPromislow2001}, who first applied the Cox modeling approach to individual survivorship. This seminal work introduced the mean-reverting Brownian Gompertz model, where a mean-reverting process captures the uncertainty of the future hazard rate. Subsequent papers developed this idea further. \citet{Dahl2004} modelled mortality intensity as a stochastic process and studied market reserves and mortality-linked insurance contracts. \citet{Biffis2005} focused on the affine processes, which are analytically tractable and provide closed-form expressions for survival probabilities. The affine framework has become a cornerstone of continuous-time mortality modelling. For example, \citet{LucianoVigna2005} proposed non-mean-reverting affine processes for stochastic mortality. \citet{BlackburnSherris2013} developed and calibrated consistent multi-factor affine mortality models for longevity risk applications. The affine setting allows for closed-form survival probabilities, making it particularly suitable for valuation and hedging under the martingale pricing approach.

For more accurate modelling of mortality risk, long range dependence (LRD) feature has emerged as a critical concept that fundamentally challenges traditional actuarial modelling approaches. The LRD feature, also known as long memory or fractional persistence, is a statistical property where correlations between observations decay more slowly than in standard random processes. Essentially, LRD is characterized by a power-law decay of the autocovariance function. It is opposed to the exponential decay observed in short-memory processes. This means that observations far apart in time remain correlated, a property that traditional Markovian models cannot capture. A substantial body of empirical research has established that mortality rates exhibit LRD across multiple dimensions. \citet{YanPetersChan2020} developed multivariate long-memory cohort mortality models. Using mortality data from 16 countries, \citet{YanPetersChan2021LongMemory} documented long-memory behaviour across genders and age groups. Also, \citet{PetersYanChan2021} provided further empirical evidence of persistence and long memory in mortality data. Additional empirical support comes from multiple sources. \citet{DelgadoVencesOrnelas2019} provided strong evidence of LRD using Italian mortality data from 1950 to 2004, employing a fractional Ornstein–Uhlenbeck process. \citet{YayaGilAlanaAmoateng2019} identified long-memory patterns in under-five mortality rates of G7 countries. Recent studies using publicly available life tables have consistently identified LRD in national mortality data.
Ignoring LRD tends to underestimate life expectancy, which may 
have profound implication for longevity risk management.

Incorporating LRD into mortality models affects the pricing of various longevity derivatives and insurance products. For example, annuities and death benefits show different valuations when LRD is considered. Longevity bonds require adjustments for LRD in their pricing formulas. The Volterra mortality model, proposed by \citet{WangChiuWong2021}, provides a class of dynamic stochastic mortality models that simultaneously renders actuarial valuation tractable and captures the LRD property. 
\citet{WangWong2021} derived the first time-consistent mean-variance longevity hedging strategy using a stochastic mortality model with LRD. They found that the optimal hedging strategy depends on whether mortality rates exhibit LRD. 
Also, the survival probability formulas derived from the Volterra models account for historical health records, enabling more accurate pricing. Besides, \citet{ZhouZhouLi2022} proposed a mortality model using a combination of independent Brownian motion and fractional Brownian motion to capture LRD. This approach uses directional derivatives to obtain mortality sensitivity measures in the presence of LRD. On the other hand, \citet{JiangZhangZhu2024} proposed a stochastic mortality process driven by a mixture of Brownian motion and a modified fractional Poisson process to capture LRD. This non-Gaussian fractional model offers flexibility while maintaining consistency with existing affine-form mortality models. \citet{ChiuWangWong2026} examined LRD mortality modelling with cointegration, providing explicit and unique equilibrium longevity hedging strategies for cointegrated forces of mortality with LRD.

In this paper, we propose a stochastic mortality process driven by a combination of Brownian motion and fractional \llevy{} process. {The fractional L\'evy component accommodates several specifications, including Poisson, Gamma, and tempered stable subordinators.} Following the technical tools developed by \citet{Marquardt2006}, our stochastic mortality model shows nice {analytical tractability} in actuarial valuations of survival probability and annuity products.  Our mortality model allows a wide choice of \llevy{} processes, providing better flexibility in incorporating LRD feature in empirical fitting of mortality data. For pricing of exotic longevity derivatives, we develop an efficient singular value decomposition approximation scheme that is designed to overcome the computational challenges associated with the non-Markovian nature of the fractional \llevy{} kernel. Truncation errors in singular value decomposition approximation can be reduced by an effective residual correction scheme.
	
The later sections of the paper are organized as follows. In Section 2, we present the mathematical preliminaries of the fractional \llevy{} process. We illustrate the LRD feature of the mixed Brownian motion and fractional \llevy{} driver. Various conditional expectations of exponentials of fractional \llevy{} integrals are derived. In particular, we derive the singular value decomposition of the fractional \llevy{} kernel. In Section 3, we present the Gompertz-type mortality model with mixed fractional \llevy{} shocks. We derive the conditional survival probability under the physical measure. In Section 4, we consider the mortality dynamics under the risk-neutral measure. We derive the integral representation formula for the log-survival index under the risk-neutral measure. We also 
illustrate the convergence of the log-survival index under the singular value decomposition. In Section 5, we discuss valuation of various longevity derivatives, including the continuous life annuity, longevity swap and longevity corridor contract. 
Section 6 presents the numerical tests on the singular value decomposition approximation schemes for pricing longevity derivatives. We illustrate how an effective residual correction scheme can reduce the truncation error in the low-rank approximation of the singular value decomposition.
In Section 7, we present the empirical calibration of our stochastic mortality model to fit the mortality shocks during World War II and COVID-19. 
The paper is ended with conclusive remarks in the last section.


\section{Model formulation of mixed fractional {L\'evy} process with Brownian motion}
\label{sec:preliminaries}
\noindent
We present the {model formulation} of our proposed stochastic mortality model, which is a mixture of Brownian motion 
and fractional \llevy{} process. We start with a summary of mathematical 
preliminaries of the fractional \llevy{} process. We then construct 
the mixed driver of combination of independent Brownian motion and 
fractional \llevy{} process on modelling stochastic mortality. 
The LRD feature of the mixed driver is illustrated. 
We then derive some conditional exponential formulas, which are useful 
for deriving survival probability and characteristic function of the survival 
index. Lastly, we present the low-rank approximation of the fractional 
\llevy{} process using the singular value decomposition. This serves 
to facilitate actuarial valuations by overcoming the challenge inherited 
in the non-Markovian structure of the fractional \llevy{} process.

\subsection{Mathematical preliminaries of fractional \llevy{} process}
\noindent
Following \citet{Marquardt2006}, the class of fractional \llevy{} process is defined 
by replacing the Brownian motion component in fractional Brownian motion by a 
general \llevy{} process with zero mean and finite variance. We present the definition and properties of 
fractional \llevy{} process as follows.


Let $(\Omega,\mathcal F,\mathbb F,\Pp)$ be a complete filtered probability space satisfying the usual conditions, where $\mathbb F=(\mathcal F_t)_{t\geq0}$. Let $X^{\Pp}=(X_t^{\Pp})_{t\geq0}$ be a non-decreasing L\'evy subordinator under measure $\Pp$. 
The subordinator may be chosen as either Poisson, Gamma, or Tempered Stable (TS) process. We suppose that its exponential moments are finite on a non-empty interval containing zero. Its cumulant under measure $\Pp$ is defined by
\begin{equation*}
 \Psi_X^{\Pp}(z)=\log \E^{\Pp}\left[e^{zX_1^{\Pp}}\right].
\end{equation*}
The compensated L\'evy driver under $\Pp$ is obtained by setting
\begin{equation*}
 L_t^{\Pp}=X_t^{\Pp}-t(\Psi_X^{\Pp})'(0),
\end{equation*}
so that $\E^{\Pp}[L_t^{\Pp}]=0$. The corresponding compensated cumulant under measure $\Pp$ is given by
\begin{equation}
 \Psi_L^{\Pp}(z)
 =\log \E^{\Pp}\left[e^{zL_1^{\Pp}}\right]
 =\Psi_X^{\Pp}(z)-z(\Psi_X^{\Pp})'(0).
\end{equation}
Assuming $\mathrm{var}^{\Pp}(L_1^{\Pp})<\infty$, we write
\begin{equation}
 s_L^2=\mathrm{var}^{\Pp}(L_1^{\Pp})
 =(\Psi_L^{\Pp})''(0).
\end{equation}
Later, we use $s_L$ to normalize the fractional \llevy{} process.



Following \cite{MolchanGolosov1969}, for $H\in(1/2,1)$, we define \(K_H(t,v)\) to be the fractional kernel used to generate LRD, where
\begin{equation}
 K_H(t,v)=c_H v^{1/2-H}\int_v^t u^{H-1/2}(u-v)^{H-3/2}\,\dd u,
 \qquad 0<v<t.
 \label{eq:MG_kernel}
\end{equation}
The positive constant \(c_H\) fixes the scale of the kernel. It is chosen such that
\[
\int_0^t K_H(t,v)^2\,\dd v=t^{2H}.
\] 
Imposing this normalization condition yields
\begin{equation*}
    c_H
    =
    \frac{1}{\Gamma\left(H-\frac12\right)}
    \left[
        \frac{
            2H
            \Gamma\left(H+\frac12\right)
            \Gamma\left(\frac32-H\right)
        }{
            \Gamma(2-2H)
        }
    \right]^{1/2}.
\end{equation*}

\begin{definition}
For $H\in(1/2,1)$, the standardized fractional L\'evy
process driven by the compensated square-integrable L\'evy process
$L^{\Pp}$ is defined by
\begin{equation}
    \Upsilon_t^{H,\Pp}
    =
    \frac{1}{{s_L}}
    \int_0^t
    K_H(t,v)
    \, {\dd L_v^{\Pp}},
    \qquad
    t\geq0,
    \label{eq:fractional}
\end{equation}
provided that the above stochastic integral exists in $L^2$.
\end{definition}
The process \(\Upsilon^{H,\mathbb P}\) is generally not a semimartingale when the driving Lévy process is non-degenerate. Therefore, stochastic integrals involving the fractional component are not interpreted as classical Itô integrals. They are defined through the deterministic Volterra representation induced by \(K_H\).


\begin{definition}
Let \(H\in(1/2,1)\), and let \(L^{\Pp}\) be a compensated
square-integrable L\'evy process under \(\Pp\). For a deterministic
function \(f\) on \([0,T]\), we define
\[
(K_T^H f)(v)
=
c_H v^{\frac12-H}
\int_v^T u^{H-\frac12} (u-v)^{H-\frac32}f(u) \,\dd u,
\qquad 0<v<T,
\]
where \(c_H\) is the normalizing constant appearing in Eq.~\eqref{eq:MG_kernel}. 
The stochastic integral with respect to the standardized fractional
L\'evy process is defined by
\begin{equation}
    \int_0^T
    f(u)\,\dd\Upsilon_u^{H,\Pp}
    =
    \frac{1}{{s_L}}
    \int_0^T
    \left(
         K_T^Hf
    \right)(v)
    {\, \dd L_v^{\Pp}},
\end{equation}
provided that the right-hand side stochastic integral is well defined.
\end{definition}
For \(0\leq t<T\), we use \(K_{t,T}^H\) as a shorthand notation for the
operator \(K_T^H\) applied to \(f\mathbf 1_{[t,T]}\):   
\[
    (K_{t,T}^Hf)(v)
    :=
    \left(
        K_T^H
        \left(
            f\mathbf 1_{[t,T]}
        \right)
    \right)(v),
    \qquad 0\leq v\leq T.
\]
Let \(\mathcal U_H^2([0,T])\) denote the space of deterministic
functions \(f\) for which \(K_T^Hf\in L^2([0,T])\), equipped with the following {seminorm}
\[
\lVert f\rVert_{\mathcal U_H^2}
:=
\left(
\int_0^T \left|(K_T^Hf)(v)\right|^2\,\dd v
\right)^{1/2}.
\]
{
For \(0\le a<b\le T\), we write
\[
\mathcal U_H^2([a,b];T)
:=
\bigl\{
f:
K_T^H(f\mathbf 1_{[a,b]})\in L^2([0,T])
\bigr\},
\]
so that \(\mathcal U_H^2([0,T];T)=\mathcal U_H^2([0,T])\). Note that \(f \in \mathcal U_H^2([0,T])\) does not automatically imply \(f\mathbf 1_{[a,b]}\in\mathcal U_H^2([0,T])\).
}

\begin{definition}
{Let \(0\leq a<b\leq T\) and \(f\mathbf 1_{[a,b]}\in\mathcal U_H^2([0,T])\), equivalently \(f\in\mathcal U_H^2([a,b];T)\).} The stochastic integral of $f$ with respect to $\Upsilon^{H,\Pp}$ over $[a,b]$ is defined by
\begin{equation}
\int_a^b f(u)\,\dd\Upsilon_u^{H,\Pp}
:=
\frac{1}{{s_L}}
\int_0^T
\left(
K_T^H(f\mathbf 1_{[a,b]})
\right)(v)
{\, \dd L_v^{\Pp}}.
 \label{eq:fractional_integral_def_full}
\end{equation}
This definition represents the fractional integral as a stochastic integral
of the deterministic function {\(K_T^H(f\mathbf 1_{[a,b]})\)} with respect to the compensated
\llevy{} process.
\end{definition}

\begin{proposition}
\label{prop:fractional_integral_existence}
{Let \(0\leq a<b\leq T\) and \(f\mathbf 1_{[a,b]}\in\mathcal U_H^2([0,T])\),} and assume
$\operatorname{var}^{\Pp}(L_1^{\Pp})<\infty$. {Then the stochastic integral in
Eq.~\eqref{eq:fractional_integral_def_full} is well-defined as an element of \(L^2(\Pp)\).} Moreover, we have
\begin{equation}
 \E^{\mathbb P}\left[\int_a^b f(u)\,\dd\Upsilon_u^{H,\Pp}\right]=0,
 \label{eq:fractional_integral_mean}
\end{equation}
and
\begin{equation}
 \operatorname{var}\left(\int_a^b f(u)\,\dd\Upsilon_u^{H,\Pp}\right)
 =\int_0^T \left(K_T^H(f1_{[a,b]})(v)\right)^2\,\dd v.
 \label{eq:fractional_integral_variance}
\end{equation}
\end{proposition}
\noindent
Readers may refer to \ref{app:fractional_integral_proof} for the proof of Proposition~\ref{prop:fractional_integral_existence}. 
In addition, we observe the following covariance structures 
as stated in \Cref{lem:fractional_levy_second_moment}.


\begin{lemma}
\label{lem:fractional_levy_second_moment}
Let \(f,g\in \mathcal U_H^2([0,T])\) and assume
\(\operatorname{var}^{\Pp}(L_1^{\Pp})<\infty\). Then we have
\begin{equation}
\E^{\mathbb P}\left[
\left(\int_0^T f(u)\,\dd\Upsilon_u^{H,\Pp}\right)
\left(\int_0^T g(u)\,\dd\Upsilon_u^{H,\Pp}\right)
\right]
=
\int_0^T
(K_T^Hf)(v)(K_T^Hg)(v)\,\dd v .
\label{eq:kernel_cov_isometry}
\end{equation}
Moreover, under the integrability condition
\(\int_0^T\int_0^T |f(y)g(z)||y-z|^{2H-2}\,\dd y\,\dd z<\infty\),
the covariance in Eq.~\eqref{eq:kernel_cov_isometry} admits the
equivalent representation
\begin{equation}
\begin{aligned}
&\E^{\mathbb P}\left[
\left(\int_0^T f(u)\,\dd\Upsilon_u^{H,\Pp}\right)
\left(\int_0^T g(u)\,\dd\Upsilon_u^{H,\Pp}\right)
\right] \\
&\quad \quad =
H(2H-1)
\int_0^T\int_0^T
f(y)g(z)|y-z|^{2H-2}\,\dd y\,\dd z .
\end{aligned}
\label{eq:fractional_cov_integral}
\end{equation}
\end{lemma}
\noindent
Readers may refer to \ref{app:proof-fractional-levy-second-moment} for the proof of Lemma~\ref{lem:fractional_levy_second_moment}.

For \(s,t\geq0\), we define the covariance function associated with $K_H$ by
\[
R_H(t,s)
:=
\int_0^{t\wedge s}K_H(t,v)K_H(s,v)\,\dd v
=
\frac{1}{2}\left(t^{2H}+s^{2H}-|t-s|^{2H}\right).
\]
For \(s\neq t\), the fractional covariance density in
Eq.~\eqref{eq:fractional_cov_integral} is the mixed derivative of
\(R_H\):
\[
\frac{\partial^2 R_H(t,s)}{\partial t\,\partial s}
=
H(2H-1)|t-s|^{2H-2}.
\]

\begin{corollary}
For \(s,t\geq0\), the mean and covariance of $\Upsilon^{H,\Pp}$ are given by
\begin{equation}
\label{eq:fraclevy_properties}
\begin{aligned}
 \E^{\mathbb P}[\Upsilon_t^{H,\Pp}]&=0,\\
 \operatorname{cov}(\Upsilon_t^{H,\Pp},\Upsilon_s^{H,\Pp})
 &=R_H(t,s),\\
 \operatorname{cov}(\Upsilon_{at}^{H,\Pp},\Upsilon_{as}^{H,\Pp})
 &=a^{2H}R_H(t,s), \qquad a>0.
\end{aligned}
\end{equation}
The process $\Upsilon^{H,\Pp}$ therefore has stationary increments in the second-order sense, and the
variance of an increment over an increment length \(m\) is \(R_H(m,m) = m^{2H}\). 
\end{corollary}

\subsection{Mixed Brownian--fractional \llevy{} driver}
\noindent
Mortality dynamics may be affected by two different types of uncertainty.
Regular fluctuations can be represented by Brownian motion, whereas sudden
non-Gaussian mortality shocks cannot be adequately described by a Brownian
driver alone. A fractional \llevy{} process is introduced to capture both the
jump behaviour of these shocks and the persistence of their effects with LRD through
the fractional kernel. In our proposed stochastic mortality model, 
we  therefore combine Brownian motion with fractional
\llevy{} process, {coined as} the mixed Brownian--fractional \llevy{} driver.

Let $W^{\Pp}=(W_t^{\Pp})_{t\geq0}$ be a standard Brownian motion under the physical measure $\Pp$.
We define the mixed Brownian--fractional \llevy{} driver by
\begin{equation}
M_t
=
p_1W_t^{\mathbb P}
+
p_2\Upsilon_t^{H,\Pp},
\label{eq:mixed_driver}
\end{equation}
where \(W^{\mathbb P}\) and \(\Upsilon^{H,\Pp}\) are independent. The 
process combines a Markovian Brownian component $W_t^{\mathbb P}$ with a non-Markovian \llevy{} component $\Upsilon^{H,\Pp}_t$. 
Here, $p_1$ and $p_2$ are their respective nonnegative weights, where $p_1+p_2=1$.

Since the Brownian motion and fractional \llevy{} component correspond to different types of
 stochastic integrals and therefore impose different admissibility conditions on their deterministic integrands. The Brownian integral requires square integrability in \(L^2([0,T])\), whereas the fractional \llevy{} integral requires \(K_T^Hf\in L^2([0,T])\), or equivalently \(f\in\mathcal U_H^2([0,T])\). Hence, for stochastic integrals involving the mixed driver \(M\), we consider \(f \in \mathcal U_H^2([0,T])\cap L^2([0,T])\).

\begin{proposition}
\label{prop:mixed_driver_second_moments}
For $s,t\geq0$, the mean and covariance of the mixed process $M_t$ are given by
\begin{align}
 \E^{\mathbb P}[M_t]&=0,\notag\\
 \operatorname{cov}(M_t,M_s)&=p_1^2(t\wedge s)+\frac{p_2^2}{2}
 \left(t^{2H}+s^{2H}-|t-s|^{2H}\right).
 \label{eq:mixed_cov}
\end{align}
For deterministic $f,g\in\mathcal U_H^2([0,T])\cap L^2([0,T])$ {with \(\int_0^T\int_0^T |f(y)g(z)||y-z|^{2H-2}\,\dd y\,\dd z<\infty\)}, we have
\begin{align}
&\E^{\mathbb P}\left[\left(\int_0^T f(u)\,\dd M_u\right)
\left(\int_0^T g(u)\,\dd M_u\right)\right]\notag\\
&\qquad \quad =p_1^2\int_0^T f(u)g(u)\,\dd u
+p_2^2H(2H-1)\int_0^T\int_0^T f(y)g(z)|y-z|^{2H-2}\,\dd y\,\dd z.
\label{eq:mixed_integral_cov}
\end{align}
\end{proposition}
\noindent
Readers may refer to \ref{app:proof-mixed-driver-second-moments} for the proof of Proposition~\ref{prop:mixed_driver_second_moments}.

\subsection{Long-range dependence}
\noindent
Although \(M\) itself is non-stationary, for any fixed \(m>0\), its
increments over intervals of length \(m\) have a common variance.
The correlation between two such increments depends only on the
distance between their endpoints. It is therefore natural to
study persistence through lagged increment correlations. The
convergence of an individual correlation to zero does not by itself
rule out LRD, since the rate of decay is decisive.
Summing the absolute correlations over all lags measures their
cumulative persistence. If this sum diverges, then the correlations decay
too slowly to be absolutely summable, thus indicating LRD.

More precisely, for a process \(Y\) whose fixed-length increment
correlations depend only on the lag, LRD is
identified when the following condition is satisfied. For some fixed \(m>0\),
this condition is stated as
\[
\sum_{n=1}^{\infty}
\left|
\operatorname{corr}
\left(
Y_{n+m}-Y_n,
Y_m-Y_0
\right)
\right|
=
\infty.
\]
We now apply this criterion to examine LRD of the mixed driver \(M\).

\begin{proposition}
\label{prop:long_range_dependence}
For the mixed driver $M_t$ defined in \eqref{eq:mixed_driver}, we fix $m>0$ and define
\begin{equation*}
 \Delta_mM_n=M_{n+m}-M_n,
 \qquad \Delta_mM_0=M_m-M_0.
\end{equation*}
Suppose $p_2\ne0$ and $H\in(1/2,1)$, we observe
\begin{equation}
 \sum_{n=1}^{\infty}|\mathrm{corr}(\Delta_mM_n,\Delta_mM_0)|=\infty.
 \label{eq:lrd_result_full}
\end{equation}
As a result, the mixed Brownian--fractional \llevy{} driver $M_t$ exhibits LRD.
\end{proposition}
\noindent
Readers may refer to \ref{app:proof-long-range-dependence} for the proof of Proposition~\ref{prop:long_range_dependence}.

The result shows the precise role of $H$ and $p_2$ in characterizing LRD in the mixed 
driver $M$. The Brownian part affects the denominator of the correlation but does not generate persistent dependence across disjoint increments. The fractional L\'evy component makes the increment correlation decay at the rate
\(n^{2H-2}\). Since \(2H-2\in(-1,0)\), this decay is slow enough to make the
sum of correlations divergent. A larger \(H\) means \(2H-2\) closer to zero, and
therefore implies stronger persistence. 
When $p_2=0$, the mixed driver reduces to a Brownian driver and LRD disappears.

\begin{figure}[H]
\centering
\includegraphics[width=0.6\textwidth]{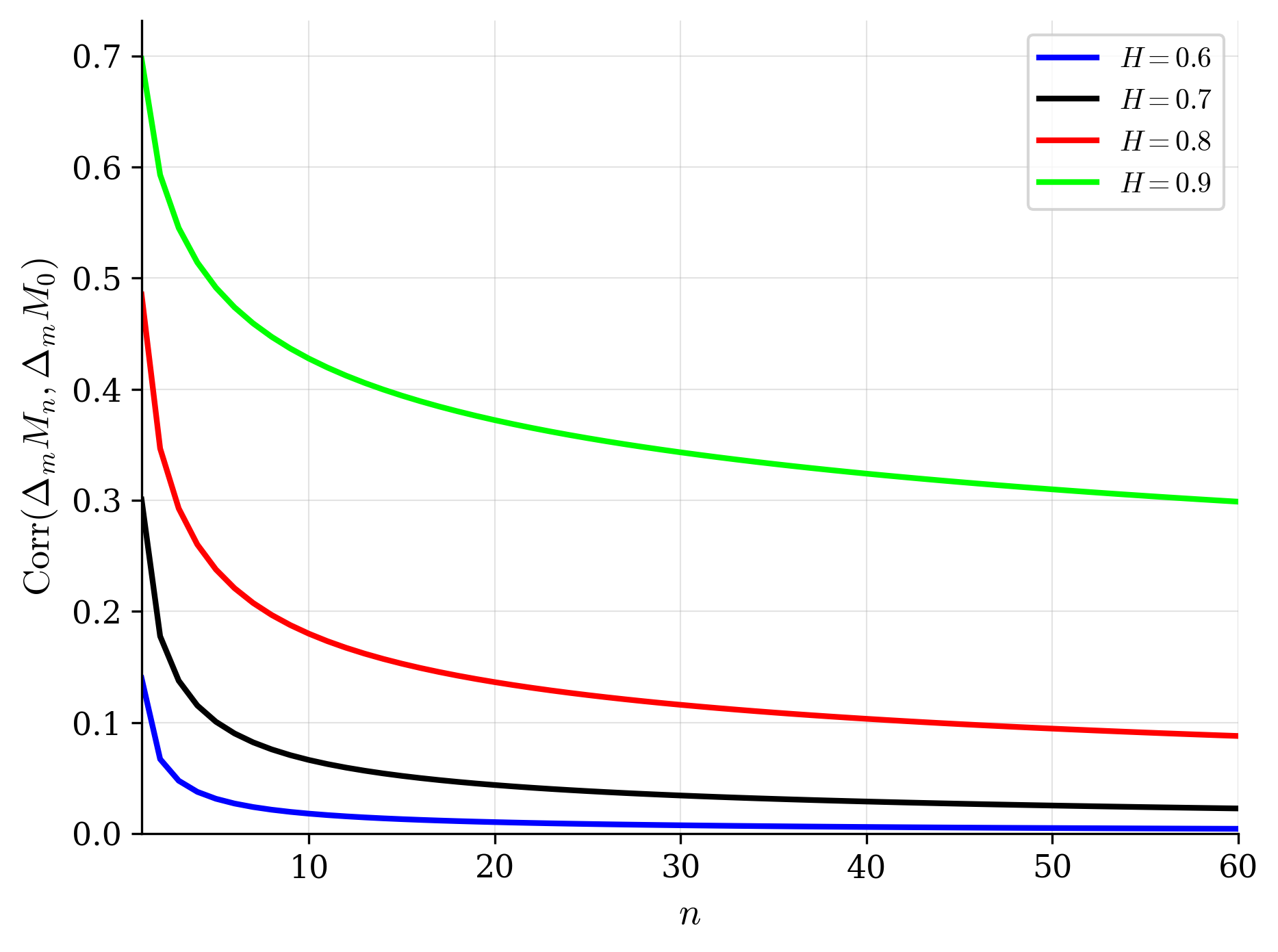}
\caption{Increment correlation $\operatorname{corr}(\Delta_m M_n,\Delta_m M_0)$ against $n$ for varying values of $H$, with $m=1$, $p_1=0.20$, and $p_2=0.80$.}
\label{fig:lrd-correlation}
\end{figure}

In Figure~\ref{fig:lrd-correlation}, we
show the plots of the increment correlation
    $\operatorname{corr}(\Delta_m M_n,\Delta_m M_0)$ against $n$ for varying
     values of $H$, with $m=1$ and
$p_1=0.20$ and $p_2=0.80$. The correlation decreases with the lag $n$ for all
values of $H$, while a larger value of $H$ produces a more visibly slower decay.
The Brownian component reduces the overall level of the correlation
through the increment variance, but it does not
alter the power-law decay rate generated by the fractional L\'evy
component. In particular, the correlation decays more rapidly for $H=0.6$,
whereas it remains substantial over long lags for $H=0.9$. The numerical
patterns illustrate that the mixed driver $M$ does have LRD. 
Also, we observe that a larger value of $H$ corresponds to stronger
LRD.

\subsection{Conditional expectation of exponential of fractional L\'evy integral}

\noindent
The determination of survival probability and characteristic function of the survival index requires
conditional expectation of exponential of fractional L\'evy integral. The following
results show how to simplify the conditional expectation into the sum
of stochastic integral with respect to the corresponding \llevy{}
process $L^{\Pp}$ and deterministic integral.

\begin{theorem}
\label{thm:conditional_exponential_formula}
Let $0\leq t<s\leq T$, let {\(f\mathbf 1_{[t,s]}\in\mathcal U_H^2([0,T])\)}, and choose $\eta\in\mathbb C$ to ensure that the cumulant below is finite. We then have
\begin{align}
&\E^{\mathbb P}\left[\exp\left(\eta\int_t^s f(u)\,\dd\Upsilon_u^{H,\Pp}\right)\middle|\mathcal F_t\right] \notag\\
&\qquad \quad \quad =\exp\left(
\frac{\eta}{{s_L}} \int_0^t  K_T^H(f1_{[t,s]})(v)\,{\dd L_v^{\Pp}}
 +\int_t^s \Psi_L^{\mathbb P}\left(\frac{\eta}{s_L} K_T^H(f1_{[t,s]})(v)\right)\,\dd v
 \right).
 \label{eq:conditional_exp_full_history}
\end{align}
For the characteristic function, we set $\eta=iu$ and obtain
\begin{align*}
&\E^{\mathbb P}\left[\exp\left(iu\int_t^s f(u)\,\dd\Upsilon_u^{H,\Pp}\right)\middle|\mathcal F_t\right] \\
&\qquad \quad \quad =\exp\left(
 \frac{iu}{{s_L}}\int_0^t  K_T^H(f1_{[t,s]})(v)\,{\dd L_v^{\Pp}}
 +\int_t^s \Psi_L^{\mathbb P}\left(\frac{iu}{s_L} K_T^H(f1_{[t,s]})(v)\right)\,\dd v
 \right).
\end{align*}
\end{theorem}
\noindent
Readers may refer to \ref{app:proof-conditional-exponential-formula} for the proof of Theorem~\ref{thm:conditional_exponential_formula}.

\begin{remark}
Since the fractional \llevy{} process is defined through a Volterra kernel,
the conditional exponential formula contains an
\(\mathcal F_t\)-measurable historical term. This term records the
effect of the L\'evy increments prior to time \(t\) on the future
 shock under the fractional L\'evy process.

At \(t=0\), the historical term vanishes. Therefore, all time-zero
survival probabilities and derivative pricing formulas derived in the later sections
retain the simpler cumulant representation involving integration
over \([0,T]\) only.
\end{remark}

\subsection{Singular value decomposition approximation of the fractional L\'evy kernel}
\label{sec:svd-lifting}
\noindent
The fractional L\'evy process depends on the past trajectory of the
L\'evy driver through the Volterra kernel \(K_H(t,v)\). This path
dependence does not lead directly to a finite-dimensional Markov
representation. For a convolution kernel of the form \(K(t-v)\), an
exponential-sum approximation can be used to obtain a finite-dimensional
Ornstein--Uhlenbeck Markovian representation. However, the kernel in our stochastic mortality model depends separately
on \(t\) and \(v\) [see Eq.~\eqref{eq:MG_kernel}], so instead we use a separable low-rank
approximation based on a truncated singular value decomposition (SVD).

Fix a finite horizon \(T>0\), we define the Volterra triangle
\[
D_T
=
\{(t,v):0\leq v\leq t\leq T\}.
\]
To formulate the SVD on the product domain \([0,T]^2\), we extend
\(K_H\) which is originally defined on \(D_T\) by setting it equal to zero
outside the Volterra triangle:
\[
\widetilde K_H(t,v)
=
\begin{cases}
K_H(t,v), & 0\leq v\leq t\leq T,\\
0, & 0\leq t<v\leq T.
\end{cases}
\]
For notational simplicity, we continue to denote the above extended kernel by \(K_H\).
Since the new extended \(K_H\in L^2([0,T]^2)\), the associated integral
operator belongs to the Hilbert--Schmidt type. For each singular value
\(\lambda_i>0\), let \(\phi_i\) and \(\psi_i\) denote the corresponding
left and right singular functions, respectively, satisfying
\begin{align*}
\int_0^T K_H(t,v)\psi_i(v)\,\dd v
&=
\lambda_i\phi_i(t),
\int_0^T K_H(t,v)\phi_i(t)\,\dd t =
\lambda_i\psi_i(v).
\end{align*}
Both $(\phi_i)_{i\geq1}$ and $(\psi_i)_{i\geq1}$ are orthonormal
systems in $L^2([0,T])$. {They are \(L^2\) equivalence classes: \(\phi_i(t)\) and \(\psi_i(v)\) are interpreted as point values after a measurable version is chosen.
The Hilbert--Schmidt theory does not yield \(\phi_i\in C^1([0,T])\).} The kernel then admits the singular value
decomposition
\begin{equation}
    K_H(t,v)
    =
    \sum_{i=1}^{\infty}
    \lambda_i\phi_i(t)\psi_i(v)
    \quad\text{in }L^2([0,T]^2),
\end{equation}
where $(\lambda_i)_{i\geq1}$ are arranged in non-increasing order.
Restricting the above decomposition to the Volterra triangle
$D_T$ gives convergence in
$L^2(D_T)$.

For positive integer $r\geq1$, we define the rank-$r$ truncated SVD kernel by
\begin{equation}
    K_H^{(r)}(t,v)
    :=
    \sum_{i=1}^{r}
    \lambda_i\phi_i(t)\psi_i(v).
    \label{eq:svd-truncated-kernel}
\end{equation}
On a finite grid, \(\phi_i\) and \(\psi_i\) are approximated by the
\(i\)-th left and right singular vectors of the discretized kernel
matrix, respectively.
Since the zero-extended \(K_H\) belongs to
\(L^2([0,T]^2)\), it defines a Hilbert--Schmidt integral operator on
\(L^2([0,T])\). The Hilbert--Schmidt operators are compact, and compact
operators between Hilbert spaces admit a singular value decomposition
[see Theorems~4.4.3 and 4.3.1 of \citet{HsingEubank2015}].
Moreover, Eq.~(4.28) and Theorem~4.4.7 in \citet{HsingEubank2015} imply that truncating
the singular value decomposition gives the best rank-\(r\) approximation
in the Hilbert--Schmidt norm. The corresponding finite-dimensional
matrix result originates from \citet{EckartYoung1936}.
Consequently, we observe
\begin{equation*}
    \left\|
        K_H-K_H^{(r)}
    \right\|_{L^2([0,T]^2)}^2
    =
    \sum_{i>r}\lambda_i^2
    \longrightarrow0,
    \qquad r\to\infty.
\end{equation*}
Restricting the integral to \(D_T\) gives
\begin{equation}
    \left\|
        K_H-K_H^{(r)}
    \right\|_{L^2(D_T)}^2
    \leq
    \sum_{i>r}\lambda_i^2
    \longrightarrow0,
    \qquad
    r\to\infty.
    \label{eq:kernel-svd-triangle-bound}
\end{equation}
Moreover, we have
\begin{equation*}
    \left\|
        K_H-K_H^{(r)}
    \right\|_{L^2([0,T]^2)}
    =
    \inf_{\operatorname{rank}(G)\leq r}
    \left\|
        K_H-G
    \right\|_{L^2([0,T]^2)}.
\end{equation*}
Thus $K_H^{(r)}$ is the optimal rank-$r$ approximation of the
zero-extended kernel on the square $[0,T]^2$. Its restriction to the Volterra
triangle provides an $r$-factor separable approximation with the error
bound shown in Eq.~\eqref{eq:kernel-svd-triangle-bound}.

For each $i=1,\ldots,r$, we define the lifted state variable
\begin{equation*}
    U_t^{(r,i)}
    :=
    \frac{1}{s_L}
    \int_0^t
    \psi_i(v)\,\dd L_v^{\Pp},
    \qquad U_0^{(r,i)}=0,
\end{equation*}
so that
\begin{equation*}
    \dd U_t^{(r,i)}
    =
    \frac{1}{s_L}
    \psi_i(t)\,\dd L_t^{\Pp}.
\end{equation*}
The rank-$r$ approximation of the fractional L\'evy component is given by
    
\begin{align}
    \Upsilon_t^{H,\Pp,(r)}
    &:=
    \frac{1}{s_L}
    \int_0^t
    K_H^{(r)}(t,v)\,\dd L_v^{\Pp} =
    \sum_{i=1}^{r}
    \lambda_i\phi_i(t)U_t^{(r,i)}.
\end{align}
We write \(
    \mathbf U_t^{(r)}
    :=
    \bigl(U_t^{(r,1)},\ldots,U_t^{(r,r)}\bigr)^{\top}
\). 
{The reconstruction uses a measurable version of each \(\phi_i\). The vector \(\mathbf U_t^{(r)}\) is a time-inhomogeneous Markov additive process; equivalently, \((t,\mathbf U_t^{(r)})\) is Markov. The scalar \(\Upsilon^{H,\Pp,(r)}\) is a deterministic function of this state and itself need not be Markov. Eq.~\eqref{eq:svd-truncated-kernel} does not require \(\phi_i\in C^1\). The grid implementation uses sampled singular vectors and does not rely on a continuous \(C^1\) version.}
Thus, the vector process
is a finite-dimensional time-inhomogeneous Markov process under
$\mathbb P$.
{Moreover, we have 
\[
\operatorname{Cov}^{\Pp}\!\left(U_t^{(r,i)},U_t^{(r,j)}\right)
=\int_0^t\psi_i(v)\psi_j(v)\,\dd v.
\]
Thus, orthonormality gives
\(\operatorname{Cov}^{\Pp}(U_T^{(r,i)},U_T^{(r,j)})=\delta_{ij}\)
at the terminal horizon, but the lifted states need not be uncorrelated
for $t<T$.}
The corresponding approximating mixed driver is
\begin{equation*}
    M_t^{(r)}
    =
    p_1W_t^{\mathbb P}
    +
    p_2\Upsilon_t^{H,\Pp,(r)},\quad r = 1,2, \ldots
\end{equation*}
\noindent
If $\phi_i\in C^1([0,T])$ for $i=1,\ldots,r$, {an assumption not implied by the Hilbert--Schmidt SVD,} then
$\Upsilon^{H,\Pp,(r)}$ is a semimartingale and
\begin{align}
    \dd \Upsilon_t^{H,\Pp,(r)}
    &=
    \sum_{i=1}^{r}
    \lambda_i\phi_i'(t)U_{t-}^{(r,i)}\,\dd t
    +
    \frac{1}{s_L}
    \left(
        \sum_{i=1}^{r}
        \lambda_i\phi_i(t)\psi_i(t)
    \right)
    \dd L_t^{\Pp}.
\end{align}
\begin{remark}
\label{rem:svd-ou-special-case}
For a convolution kernel with
\[
    K^{(r)}(t-v)
    =
    \sum_{i=1}^{r}c_i e^{-x_i(t-v)},
\]
the corresponding factors
\[
    V_t^{(i)}
    :=
    \frac{1}{s_L}
    \int_0^t e^{-x_i(t-v)}\,\dd L_v^{\Pp}
\]
satisfy
\[
    \dd V_t^{(i)}
    =
    -x_iV_t^{(i)}\,\dd t
    +
    \frac{1}{s_L}
    \dd L_t^{\Pp}.
\]
Thus the usual multifactor OU lift is a special separable
representation. 
\end{remark}
Some of the convergence properties of the SVD approximation are {summarized} in \Cref{thm:svd-lifting-convergence}.
\begin{theorem}
\label{thm:svd-lifting-convergence}
Assume that $L^{\Pp}$ is a centered square-integrable L\'evy
process under $\mathbb P$ and
\[
    s_L^2
    =
        \operatorname{var}^{\mathbb P}(L_1^{\Pp})
    <\infty.
\]
The rank-$r$ truncated SVD approximation observes the following properties.
\begin{enumerate}
    \item For almost every $t\in[0,T]$,
    \begin{equation}
        \mathbb E^{\mathbb P}
        \left[
            \left|
                \Upsilon_t^{H,\Pp}
                -
                \Upsilon_t^{H,\Pp,(r)}
            \right|^2
        \right]
        =
        \int_0^t
        \left|
            K_H(t,v)-K_H^{(r)}(t,v)
        \right|^2\,\dd v.
        \label{eq:svd-pointwise-error}
    \end{equation}
    {The identity is valid for almost every \(t\in[0,T]\). It holds for every \(t\) only if the singular functions admit continuous versions. However, the Hilbert--Schmidt theory does not provide such condition.}

    \item The SVD approximation converges in integrated mean square:
    \begin{align}
        &\int_0^{T}
        \mathbb E^{\mathbb P}
        \left[
            \left|
                \Upsilon_t^{H,\Pp}
                -
                \Upsilon_t^{H,\Pp,(r)}
            \right|^2
        \right]\,\dd t
        \notag\\
        &\qquad \quad =
        \|K_H-K_H^{(r)}\|_{L^2(D_T)}^2
        \notag\\
        &\qquad \quad\leq
        \|K_H-K_H^{(r)}\|_{L^2([0,T]^2)}^2
        =
        \sum_{i>r}\lambda_i^2
        \longrightarrow0.
        \label{eq:svd-integrated-error}
    \end{align}
    {This gives convergence in \(L^2(\Omega\times[0,T])\). It does not imply mean-square convergence at every fixed \(t\), uniform-in-\(t\) convergence, nor pathwise/\(\sup_{t\le T}\) convergence.}

    \item The truncated SVD kernel $K_H^{(r)}$ gives the best rank-$r$
    approximation of the zero-extended kernel in the Hilbert--Schmidt
    norm, equivalently in $L^2([0,T]^2)$. Its restriction to $D_T$
    satisfies the convergence bound in
    Eq.~\eqref{eq:svd-integrated-error}.
\end{enumerate}
\end{theorem}
\noindent
Readers may refer to \ref{app:svd-continuous-proof} for the proof of Theorem~\ref{thm:svd-lifting-convergence}.

\section{Gompertz-Type mortality model with mixed fractional L\'evy shocks}
\label{sec:mortality model}
\noindent
Following \cite{Gompertz1825}, we model the deterministic aging component of the
mortality intensity through an exponential growth structure. 
While retaining the Gompertz-type drift, we add the 
stochastic mortality shock driven by a mixed {Brownian} motion and fractional L\'evy process.
Under the physical measure $\mathbb P$, we assume that the mortality intensity $\mu_t$ follows 
the dynamic equation:
\begin{equation}
 \dd\mu_t = a\mu_t\,\dd t+\sigma\,\dd M_t,
 \label{eq:sde}
\end{equation}
where $a>0$ is the aging coefficient, $M_t$ is the mixed 
driver as defined in Eq.~\eqref{eq:mixed_driver} and $\sigma>0$ is the common scale parameter, not to be interpreted as volatility. 
An important quantity in actuarial valuation is the conditional
survival probability as defined by
\begin{equation}
S^{\Pp}(t,T)
:=
\E^{\Pp}\left[
\left.
\exp\left(
-\int_t^T\mu_s\,\dd s
\right)
\right|
\mathcal F_t
\right].
\label{eq:survival}
\end{equation}
To demonstrate {analytical tractability} of our proposed
stochastic mortality model, we {present} the analytic
formula for the conditional survival probability under 
measure $\Pp$ in \Cref{prop:survival_prob_p}.
\begin{proposition}
\label{prop:survival_prob_p}
Assume that $\mu_t$ follows Eq.~\eqref{eq:sde} under 
the physical measure $\Pp$. 
The conditional
survival probability under the mortality
intensity $\mu_t$ is given by
\begin{equation}
 S^{\Pp}(t,T)
 =
 \exp\left(
 -\mu_t A(t,T)
 +B_{\mathrm{hist}}^{\Pp}(t,T)
 +B_{\mathrm{Brown}}^{\Pp}(t,T)
 +B_{\mathrm{\text{L\'evy}}}^{\Pp}(t,T)
 \right),
 \label{eq:affine_survival}
\end{equation}
where
\begin{align*}
 A(t,T)
 &=
 \frac{e^{a(T-t)}-1}{a},
 \\
 B_{\mathrm{hist}}^{\Pp}(t,T)
 &=
 -\frac{\sigma p_2}{s_L}
 \int_0^t
 (K_{t,T}^H f)(v)\,
 \dd L_v^{\Pp},
 \\
 B_{\mathrm{Brown}}^{\Pp}(t,T)
 &=
 \frac{\sigma^2p_1^2}{2a^2}
 \left[
 \frac{e^{2a(T-t)}-4e^{a(T-t)}+3}{2a}
 +(T-t)
 \right],
 \\
 B_{\mathrm{\text{L\'evy}}}^{\Pp}(t,T)
 &=
 \int_t^T
 \Psi_L^{\Pp}
 \left(
 -\frac{\sigma p_2}{s_L}
 (K_{t,T}^H f)(v)
 \right)
 \,\dd v,
\end{align*}
with
\begin{equation*}
 f(u,T)
 =
 \frac{e^{a(T-u)}-1}{a}.
\end{equation*}
At $t=0$, the historical term vanishes, and therefore
\begin{equation}
 S^{\Pp}(0,T)
 =
 \exp\left(
 -\mu_0A(0,T)
 +B_{\mathrm{Brown}}^{\Pp}(0,T)
 +B_{\mathrm{\text{L\'evy}}}^{\Pp}(0,T)
 \right).
 \label{eq:physical_survival_zero}
\end{equation}
\end{proposition}

{The above formula uses the stochastic Fubini property, which holds because \(f(\cdot,T)\in L^2([t,T])\) and \(K_{t,T}^Hf\in L^2([0,T])\), and the L\'evy--Khintchine factor is used on the interior of \(\operatorname{dom}\Psi_L^{\Pp}\).} 
\noindent
Readers may refer to \ref{app:proof-survival-prob-p} for the proof of Proposition~\ref{prop:survival_prob_p}.

\begin{corollary}
\label{cor:physical_compensated_cumulants}
The compensated cumulants under the physical measure $\Pp$ for various \llevy{} specifications are given by
\begin{subequations}
\begin{align}
 \Psi_P^{\Pp}(z)
 &=
 \lambda(e^z-1-z),
 &&\text{Poisson},
 \label{eq:pois_cum}\\
 \Psi_G^{\Pp}(z)
 &=
 -\alpha\log\left(1-\frac{z}{\beta}\right)-z\frac{\alpha}{\beta},
 \qquad z<\beta,
 &&\text{Gamma},
 \label{eq:gamma_cum}\\
 \Psi_{TS}^{\Pp}(z)
 &=
 c\Gamma(-\kappa)
 \left[
 (\lambda_{TS}-z)^\kappa-\lambda_{TS}^{\kappa}
 +z\kappa\lambda_{TS}^{\kappa-1}
 \right],
 \qquad z<\lambda_{TS},
 &&\text{TS}.
 \label{eq:ts_cum}
\end{align}
\end{subequations}
\end{corollary}
\noindent
Readers may refer to \ref{app:proof-physical-compensated-cumulants} for the proof of Corollary~\ref{cor:physical_compensated_cumulants}.

\begin{figure}[H]
    \centering

    \begin{minipage}[t]{0.48\textwidth}
        \centering
        \includegraphics[
            width=\linewidth
        ]{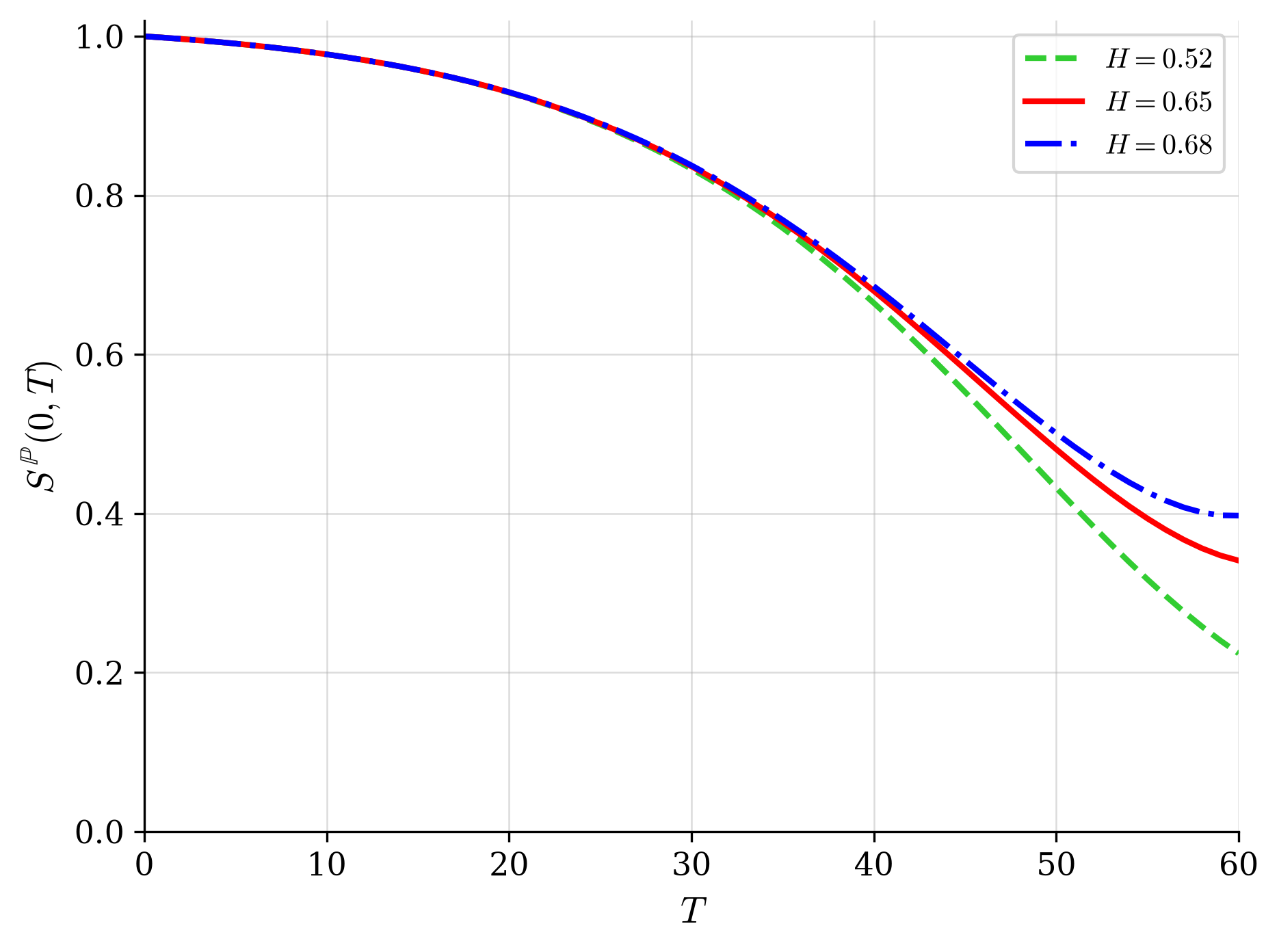}

        \smallskip
        (a) Different $H$.
    \end{minipage}
    \hfill
    \begin{minipage}[t]{0.48\textwidth}
        \centering
        \includegraphics[
            width=\linewidth
        ]{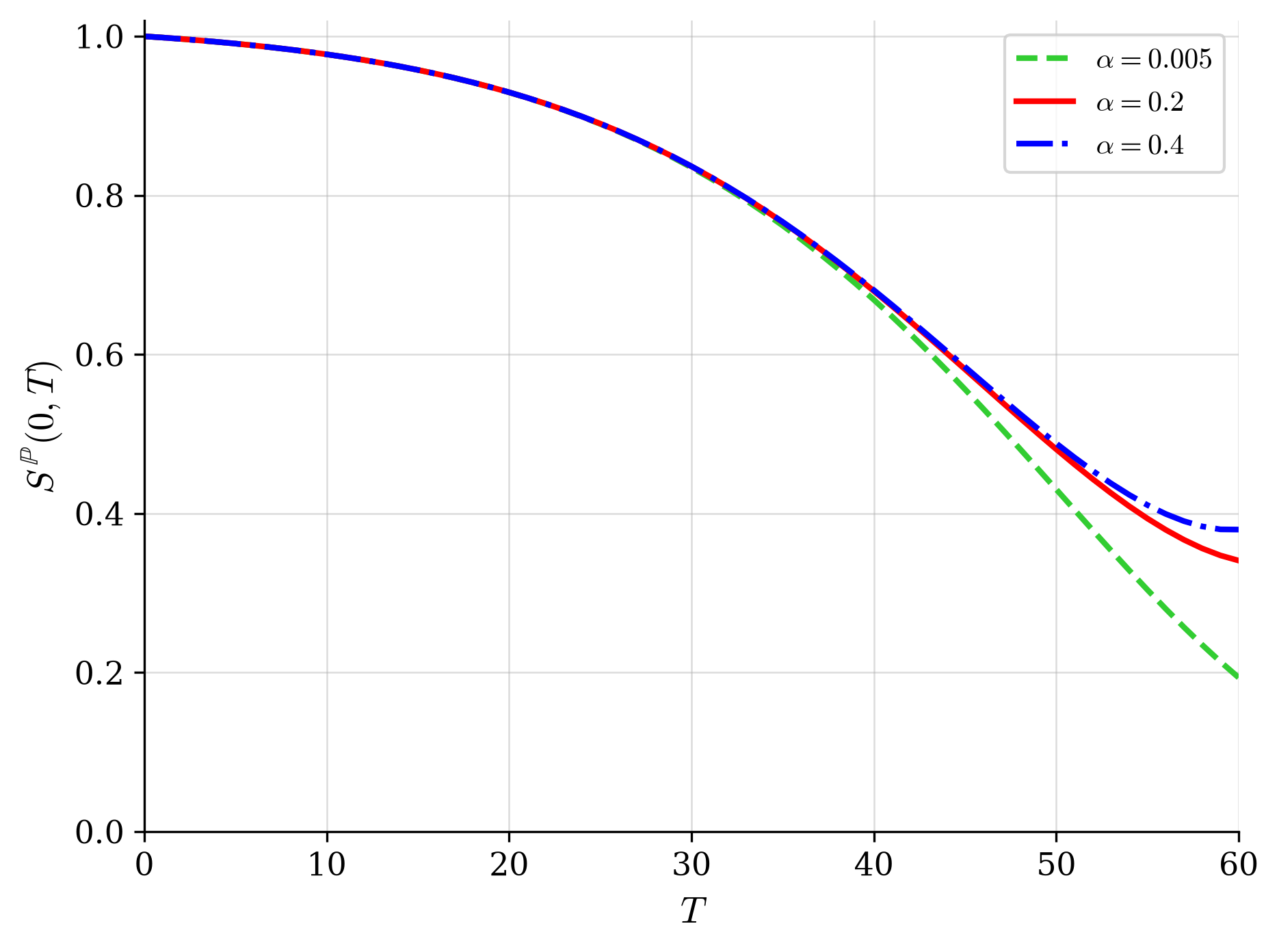}

        \smallskip
        (b) Different $\alpha$.
    \end{minipage}

    \vspace{0.8em}

    \begin{minipage}[t]{0.48\textwidth}
        \centering
        \includegraphics[
            width=\linewidth
        ]{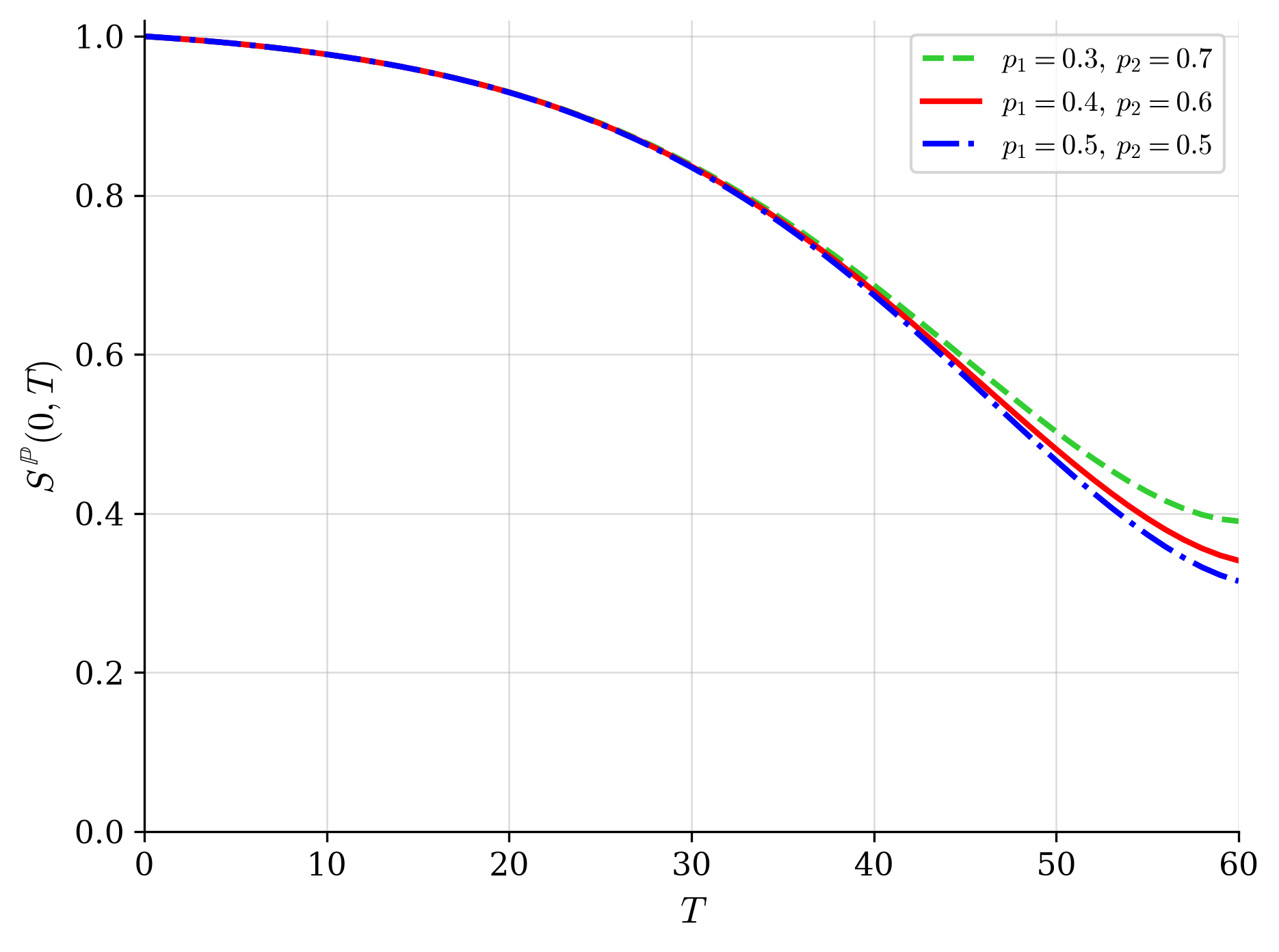}

        \smallskip
        (c) Different $p_1$ with $p_2=1-p_1$.
    \end{minipage}
    \hfill
    \begin{minipage}[t]{0.48\textwidth}
        \centering
        \includegraphics[
            width=\linewidth
        ]{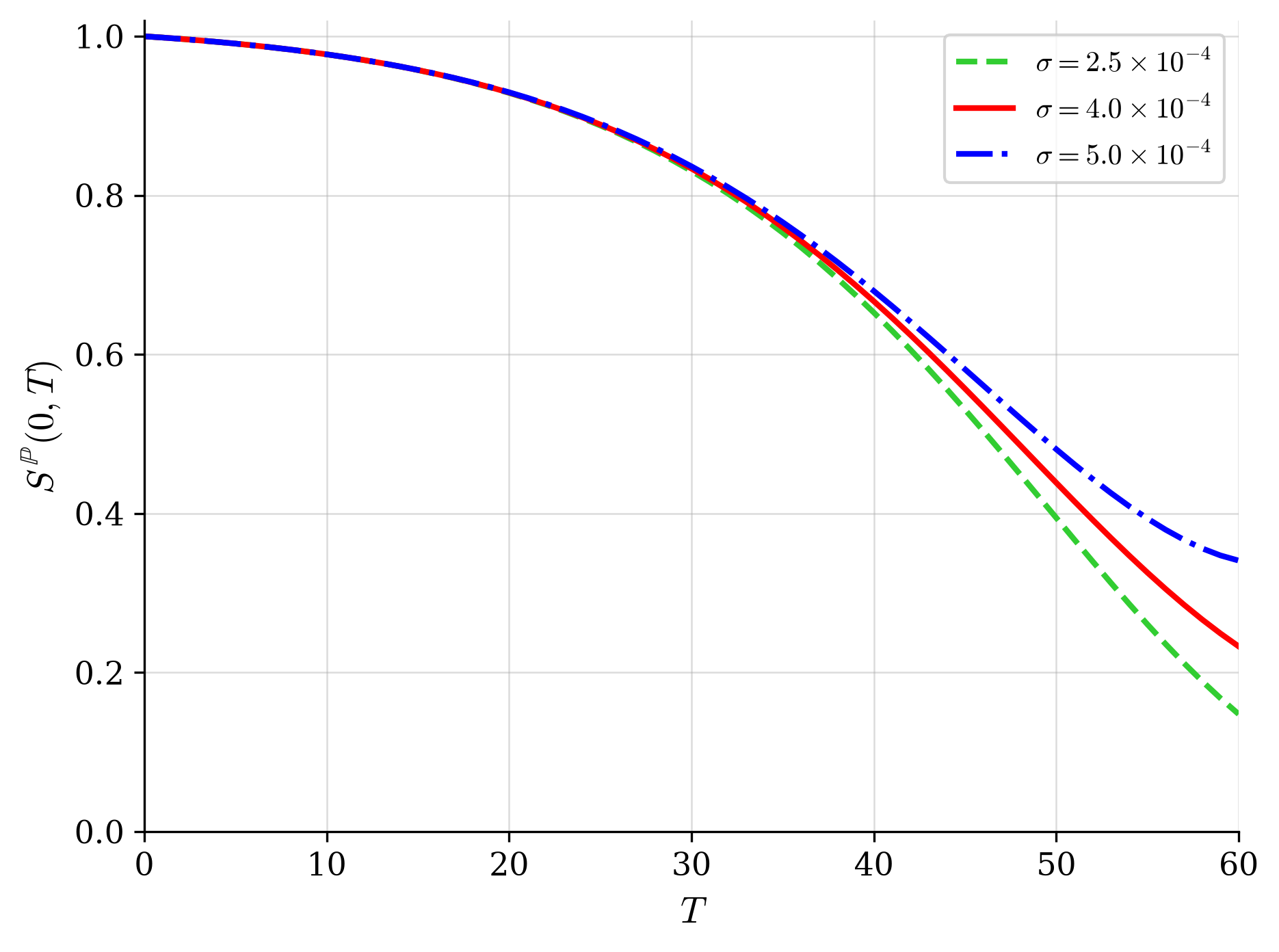}

        \smallskip
        (d) Different $\sigma$.
    \end{minipage}

    \caption{
        Sensitivity of the survival probability
        $S^{\mathbb P}(0,T)$ with respect to 
        different model parameters under the Gamma specification.
    }
    \label{fig:gamma-sensitivity}
\end{figure}

We analyze the sensitivity of the survival probability with respect to the model parameters. In \Cref{fig:gamma-sensitivity}, we illustrate the parameter
sensitivity of the model-implied cumulative survival probability
$S^{\mathbb P}(0,T)$ under the Gamma specification. The baseline parameter values are
$\mu_0=0.0015, a=0.08, H=0.65,
\sigma=5\times10^{-4}, p_1=0.40, p_2=0.60, \alpha=0.20, \beta=1.50$. In panels (a), (b), and (d), one parameter is varied while
the remaining parameters are held fixed. In panel (c), the normalized loading
$p_1$ is varied over $\{0.30,0.40,0.50\}$ and $p_2=1-p_1$ is adjusted accordingly.

As shown in Figure~\ref{fig:gamma-sensitivity}(a), a larger value of
$H$ produces a higher survival probability at medium and long maturities, while
the curves remain close to {each other under} short maturities. This reflects the cumulative
effect of the fractional dependence structure, which takes effect at medium and long maturities. 
Figure~
\ref{fig:gamma-sensitivity}(b) shows that increasing $\alpha$ 
also shifts the long-maturity survival curve upward
under the selected parameterization. 
Figure~\ref{fig:gamma-sensitivity}(c) shows the effect of reallocating the
normalized loading between the Brownian motion and fractional L\'evy process. Figure~\ref{fig:gamma-sensitivity}(d) reports the
sensitivity to the common scale parameter $\sigma$. The differences 
in $S^{\Pp}(0, T)$ become more
pronounced as the maturity increases. The
Poisson and TS specifications exhibit qualitatively
similar sensitivity patterns and the discussion on their
fractional {dependence} structure is omitted for brevity.

\section{Mortality dynamics under the risk-neutral measure}
\label{sec:risk-neutral}
\noindent 
The mortality dynamics introduced above are specified under the physical
measure \(\Pp\), whereas the valuation of longevity derivatives requires
an equivalent pricing measure $\Q$ that incorporates the market price
 of mortality risk. As
discussed in \citet{WangChiuWong2021}, an explicit link between the physical and
pricing measures is needed. The Esscher transform identifies a
pricing measure through an exponential change of measure, while Girsanov's
theorem provides the corresponding adjustment to the Brownian dynamics.

In the present mixed Brownian--fractional \llevy{} model, the two changes of
measure are applied to the corresponding sources of mortality risk. We apply a
Girsanov transform to the Brownian component and an Esscher transform to the
underlying uncompensated subordinator.

Since the Brownian motion and the \llevy{} driver are independent under
\(\Pp\), so their changes of measure can be specified {separately}. The
Radon--Nikodym density process can be decomposed as
\begin{equation*}
Z_u
=
\left.
\frac{\dd\Q}{\dd\Pp}
\right|_{\mathcal F_u}
=
Z_u^W Z_u^L,
\end{equation*}
where \(Z^W\) and \(Z^L\) denote the density processes associated with the
Brownian and \llevy{} components, respectively. The Brownian density ratio over
the interval \([t,T]\) is
\begin{equation}
\frac{Z_T^W}{Z_t^W}
=
\exp\left(
\int_t^T \gamma_s\,\dd W_s^{\Pp}
-
\frac{1}{2}\int_t^T\gamma_s^2\,\dd s
\right),
\label{eq:girsanov}
\end{equation}
where \(\gamma_s\) is a deterministic square-integrable function denoting the
market price of the Brownian risk.
As a result, we have
\(\dd W_s^{\Pp}=\dd W_s^\Q+\gamma_s\,\dd s\), where \(W^\Q\) is a standard Brownian motion
under \(\Q\).

For the \llevy{} component, let \(X^{\Pp}\) denote the uncompensated subordinator
with cumulant function \(\Psi_X^{\Pp}\). Its Esscher density ratio over
\([t,T]\) is
\begin{equation}
 \frac{Z_T^L}{Z_t^L}
 =
 \exp\left(
 \int_t^T \theta(v)\,\dd X_v^{\Pp}
 -
 \int_t^T \Psi_X^{\Pp}(\theta(v))\,\dd v
 \right),
 \label{eq:esscher_density}
\end{equation}
where \(\theta(v)\) is a deterministic function representing the market price
of mortality jump risk. The required exponential moments are assumed to be
finite. {The resulting measure \(\Q\) is a selected pricing measure in an incomplete market, not the unique equivalent martingale measure. If \(\theta(v)\) is not constant, \(X\) is in general only a time-inhomogeneous additive process under \(\Q\).} If
\(\theta(v)\equiv\theta\) is constant, \(X\) remains a subordinator under
\(\Q\), with cumulant function
\begin{equation}
 \Psi_X^\Q(z)
 =
 \Psi_X^{\Pp}(z+\theta)
 -
 \Psi_X^{\Pp}(\theta).
 \label{eq:q_uncompensated_cumulant}
\end{equation}
The independence of the Brownian and \llevy{} components is preserved under
\(\Q\) because the density factorizes into the product \(Z^WZ^L\).
We define the respective unit-time means of the subordinator by
\[
m_L^{\Pp}:=(\Psi_X^{\Pp})'(0),
\qquad
m_L^\Q:=(\Psi_X^\Q)'(0),
\qquad
\Delta m_L:=m_L^\Q-m_L^{\Pp}.
\]
The compensated L\'evy drivers under the two measures are
\begin{equation}
L_t^{\Pp}:=X_t^{\Pp}-tm_L^{\Pp},
\qquad
L_t^\Q:=X_t^\Q-tm_L^\Q.
\label{eq:p-q-compensated-drivers}
\end{equation}
Here, \(X^{\Pp}\) and \(X^\Q\) denote the same underlying subordinator
under the two equivalent measures. {Equivalently, one may use a single coordinate process \(X\) and write \(L_t^{\Pp}=X_t-tm_L^{\Pp}\), \(L_t^{\Q}=X_t-tm_L^{\Q}\). 
The realized path of \(X\) is unchanged, only its law changes.} Consequently,
\begin{equation}
\dd L_t^{\Pp}
=
\dd L_t^\Q+\Delta m_L\,\dd t.
\label{eq:p-q-levy-relation}
\end{equation}
The process \(L^\Q\) is centered under \(\Q\), and its cumulant is
\begin{equation}
 \Psi_L^\Q(z)
 =
 \Psi_X^\Q(z)-zm_L^\Q.
 \label{eq:q_levy_cumulant}
\end{equation}
To retain the physical normalization of the mortality model, we define
\[
\Upsilon_t^{H,\Q}
=
\frac{1}{s_L}\int_0^tK_H(t,v)\,\dd L_v^\Q,
\qquad
\kappa_H(t)=\int_0^tK_H(t,v)\,\dd v,
\qquad
b_H(t)=\frac{\dd}{\dd t}\kappa_H(t).
\]
{The martingale factor \(s_L\) is the standard deviation under measure \(\Pp\). In general \(v_L^{\Q}:=\operatorname{var}^{\Q}(L_1^{\Q})\ne s_L^2\), so \(\operatorname{cov}^{\Q}(\Upsilon_t^{H,\Q},\Upsilon_s^{H,\Q})=(v_L^{\Q}/s_L^2)R_H(t,s)\). The Molchan--Golosov kernel satisfies \(K_H(ct,cv)=c^{H-1/2}K_H(t,v)\), hence \(\kappa_H(t)=t^{H+1/2}\kappa_H(1)\) and \(b_H(t)=(H+1/2)\kappa_H(1)\,t^{H-1/2}\) almost everywhere.}
Eq.~\eqref{eq:p-q-levy-relation} then gives
\begin{equation}
\Upsilon_t^{H,\Pp}
=
\Upsilon_t^{H,\Q}
+
\frac{\Delta m_L}{s_L}\kappa_H(t),
\label{eq:p-q-fractional-relation}
\end{equation}
where \(\Upsilon^{H,\Pp}\) is the fractional
driver under measure \(\Pp\) defined in Eq.~\eqref{eq:fractional}
and \(\Upsilon^{H,\Q}\) is its counterpart under measure \(\Q\).

\begin{proposition}
\label{prop:q_dynamics}
Under the measure change defined by Eqs.~\eqref{eq:girsanov} and
\eqref{eq:esscher_density}, with constant Esscher parameter \(\theta\), the
mortality intensity satisfies
\begin{equation}
 \dd\mu_s
 =
 \left(
 a\mu_s+\sigma p_1\gamma_s
 +\frac{\sigma p_2\Delta m_L}{s_L}b_H(s)
 \right)\,\dd s
 +
 \sigma p_1\,\dd W_s^\Q
 +
 \sigma p_2\,\dd\Upsilon_s^{H,\Q},
 \label{eq:q_dynamics}
\end{equation}
where \(W^\Q\) is a standard Brownian motion under \(\Q\), and
\(\Upsilon^{H,\Q}\) is driven by the centered process \(L^\Q\).
\end{proposition}
\noindent
Readers may refer to \ref{app:proof-q-dynamics} for the proof of Proposition~\ref{prop:q_dynamics}.

In our model, the cumulative mortality over the
period \([t,T]\) is represented by
$\int_t^T \mu_s\,\dd s.$
The corresponding survival index is the exponential of the negative
cumulative mortality intensity. 
Its conditional
expectation under \(\Q\) gives the risk-neutral survival probability:
\begin{equation}
    \label{eq:survial_prob_q}
S^\Q(t,T)
=
\E^\Q\left[
\left.
\exp\left(
-\int_t^T \mu_s\,\dd s
\right)
\right|
\mathcal F_t
\right].
\end{equation}
The conditional mean of the survival index is sufficient for linear
payoffs. Nonlinear payoffs, such as survival
caplets and corridor contracts, require its conditional distribution.
To characterize this distribution, we define the log-survival index by
\begin{equation}
Y_{t,T}
:=
-\int_t^T\mu_s\,\dd s,
\label{log_survial_index}
\end{equation}
so that the survival index over \([t,T]\) is \(e^{Y_{t,T}}\).
Solving Eq.~\eqref{eq:q_dynamics} and integrating over $[t,T]$ gives
\begin{align*}
 Y_{t,T}
 =
 &-\mu_tA(t,T)
 -D_\gamma(t,T)
 -D_L(t,T)
 -\sigma p_1
 \int_t^T
 f(u,T)\,\dd W_u^\Q\\
 &-\frac{\sigma p_2}{s_L}
 \int_0^t
 (K_{t,T}^H f)(v)\,\dd L_v^\Q
 -\frac{\sigma p_2}{s_L}
 \int_t^T
 (K_{t,T}^H f)(v)\,\dd L_v^\Q,
\end{align*}
where
\[
 A(t,T)=\frac{e^{a(T-t)}-1}{a},
 \qquad
 f(u,T)=\frac{e^{a(T-u)}-1}{a},
\]
\[
 D_\gamma(t,T)=\int_t^T \sigma p_1\gamma_u f(u,T)\,\dd u,
 \qquad
 V_f(t,T)=\int_t^T f(u,T)^2\,\dd u,
\]
\[
 D_L(t,T)
 =
 \frac{\sigma p_2\Delta m_L}{s_L}
 \int_t^T f(u,T)b_H(u)\,\dd u.
\]
For real \(u\), the conditional characteristic function of \(Y_{t,T}\) is defined by
\[
 \phi_{t,T}^\Q(u)
 =
 \E^\Q\left[
 e^{iuY_{t,T}}
 \middle|\mathcal F_t
 \right].
\]
We use the same notation for its extension to complex \(u\) whenever the
required conditional exponential moment is finite.
\noindent
Together with Eqs.~\eqref{eq:survial_prob_q} and \eqref{log_survial_index}, we observe that 
$S^\Q(t,T)
    =
    \phi_{t,T}^\Q(-i).$

\begin{proposition}
\label{prop:yt_characteristic_function}
For every $u\in\mathbb C$ for which the required exponential moment is
finite, the conditional characteristic function of $Y_{t,T}$ under $\Q$ is
\begin{equation}
\begin{aligned}
 \phi_{t,T}^{\Q}(u)
 =
 \exp\Bigg(
 &-iu\mu_tA(t,T)
 -iuD_\gamma(t,T)-iuD_L(t,T)
 -iu\frac{\sigma p_2}{s_L}
 \int_0^t
 (K_{t,T}^H f)(v)\,
 \dd L_v^\Q\\
 &-\frac12u^2\sigma^2p_1^2V_f(t,T) +
 \int_t^T
 \Psi_L^\Q
 \left(
 -iu\frac{\sigma p_2}{s_L}
 (K_{t,T}^H f)(v)
 \right)
 \,\dd v
 \Bigg).
\end{aligned}
\label{eq:yt_characteristic_function}
\end{equation}
\end{proposition}
\noindent
Readers may refer to \ref{app:proof-yt-characteristic-function} for the proof of Proposition~\ref{prop:yt_characteristic_function}.



Accordingly, the survival probability under $\Q$ is given by
\begin{equation}
\begin{aligned}
 S^\Q(t,T)
 &=
 \phi_{t,T}^\Q(-i)\\
 &=
 \exp\Big(
 -\mu_tA(t,T)
 +B_{\mathrm{hist}}^\Q(t,T)
 +B_{\mathrm{drift}}^\Q(t,T)\\
 &\hspace{3.5cm}
 +B_{\mathrm{Brown}}^\Q(t,T)
 +B_{\mathrm{\text{L\'evy}}}^\Q(t,T)
 \Big),
\end{aligned}
 \label{eq:q_survival}
\end{equation}
where
\begin{align*}
 B_{\mathrm{hist}}^\Q(t,T)
 &=
 -\frac{\sigma p_2}{s_L}
 \int_0^t
 (K_{t,T}^H f)(v)\,
 \dd L_v^\Q,
 \\
 B_{\mathrm{drift}}^\Q(t,T)
 &=
 -D_L(t,T),
 \\
 B_{\mathrm{Brown}}^\Q(t,T)
 &=
 \frac12\sigma^2p_1^2V_f(t,T)
 -
 D_\gamma(t,T),
 \\
 B_{\mathrm{\text{L\'evy}}}^\Q(t,T)
 &=
 \int_t^T
 \Psi_L^\Q
 \left(
 -\frac{\sigma p_2}{s_L}
 (K_{t,T}^H f)(v)
 \right)
 \,\dd v.
\end{align*}
At $t=0$, the historical term vanishes, and therefore
\begin{equation*}
 S^\Q(0,T)
 =
 \exp\left(
 -\mu_0A(0,T)
 +B_{\mathrm{drift}}^\Q(0,T)
 +B_{\mathrm{Brown}}^\Q(0,T)
 +B_{\mathrm{\text{L\'evy}}}^\Q(0,T)
 \right).
\end{equation*}

We now examine whether the SVD approximation preserves the
risk-neutral survival index. Let \(K_{0,T}^{H,(r)}\) denote the
counterpart of \(K_{0,T}^{H}\) obtained by replacing the exact kernel
\(K_H\) with its rank-\(r\) SVD approximation \(K_H^{(r)}\). We define the
exact and rank-\(r\) SVD approximation transformed kernels by
\begin{align}
g_{0,T}(v)
&:=
\bigl(K_{0,T}^{H}f\bigr)(v),
\qquad 0\leq v\leq T,
\label{eq:svd-original-survival-kernel}\\
g_{0,T}^{(r)}(v)
&:=
\bigl(K_{0,T}^{H,(r)}f\bigr)(v),
\qquad 0\leq v\leq T.
\label{eq:svd-approximating-survival-kernel}
\end{align}
By the adjoint relation for \(K_{0,T}^H\), the full correction defined
above and its rank-$r$ counterpart can be written as
\begin{equation}
D_L(0,T)
=
\frac{\sigma p_2\Delta m_L}{s_L}
\int_0^Tg_{0,T}(v)\,\dd v,
\qquad
D_L^{(r)}(0,T)
:=
\frac{\sigma p_2\Delta m_L}{s_L}
\int_0^Tg_{0,T}^{(r)}(v)\,\dd v.
\label{eq:q-lifted-levy-drift-correction}
\end{equation}
We define the corresponding rank-\(r\) log-survival index under
\(\Q\) by
\begin{align}
Y_{0,T}^{(r)}
&=
-\mu_0A(0,T)
-D_\gamma(0,T)
-D_L^{(r)}(0,T)
-\sigma p_1
\int_0^T
f(u,T)\,\dd W_u^{\mathbb Q}
\notag\\
&\quad
-
\frac{\sigma p_2}{s_L}
\int_0^T
g_{0,T}^{(r)}(v)\,\dd L_v^\Q.
\label{eq:q-lifted-log-survival-index}
\end{align}
The corresponding exact and rank-$r$ SVD approximation survival indices are
    $S_T=
    e^{Y_{0,T}} \text{and }
    S_T^{(r)}=
    e^{Y_{0,T}^{(r)}},$ respectively.

\begin{theorem}
\label{thm:q-lifted-survival-index-convergence}
{The kernel bound in Eq.~\eqref{eq:kernel-svd-triangle-bound} and the Cauchy--Schwarz inequality give \(g_{0,T}^{(r)}\to g_{0,T}\) in \(L^2([0,T])\).}
Suppose that \(L^\Q\) has a finite second moment under \(\mathbb Q\), and
write
\[
    v_L^{\mathbb Q}
    :=
    \operatorname{var}^{\mathbb Q}(L_1^\Q)
    <\infty.
\]
We then have the convergence of the rank-\(r\) SVD approximation of the log-survival index, where
\begin{equation}
 Y_{0,T}^{(r)}
\longrightarrow
 Y_{0,T}
\quad\text{in }L^2(\mathbb Q).
\label{eq:y_convergence}
\end{equation}
More precisely, we obtain
\begin{equation}
\begin{aligned}
    \mathbb E^{\mathbb Q}
    \left[
        \left|
            Y_{0,T}^{(r)}-Y_{0,T}
        \right|^2
    \right]
    \leq
    \frac{\sigma^2p_2^2}{s_L^2}
    \left(
        v_L^{\mathbb Q}
        +
        T|\Delta m_L|^2
    \right)
    \left\|
        g_{0,T}^{(r)}-g_{0,T}
    \right\|_{L^2([0,T])}^2.
\end{aligned}
\label{eq:q-log-survival-convergence-bound}
\end{equation}
In addition, suppose there exists \(\eta>1\) such that
\begin{equation}
    \sup_{r\geq1}
    \mathbb E^{\mathbb Q}
    \left[
        e^{\eta Y_{0,T}^{(r)}}
    \right]
    <\infty,
    \qquad
    \mathbb E^{\mathbb Q}
    \left[
        e^{\eta Y_{0,T}}
    \right]
    <\infty,
    \label{eq:q-uniform-survival-moment}
\end{equation}
{which is an additional restriction on the interior of \(\operatorname{dom}\Psi_L^{\Q}\). For the Gamma and TS specifications, it requires the real cumulant arguments of \(g_{0,T}^{(r)}\) to remain uniformly inside the moment boundary. It is not implied by square integrability of \(L^{\Q}\) alone.}
We then have
\[
    S_T^{(r)}
    \longrightarrow
    S_T
    \quad\text{in }L^1(\mathbb Q).
\]
Consequently, we obtain
\[
    \mathbb E^{\mathbb Q}[S_T^{(r)}]
    \longrightarrow
    \mathbb E^{\mathbb Q}[S_T]
    =
    S^{\mathbb Q}(0,T).
\]
\end{theorem}
\noindent
Readers may refer to \ref{app:svd-q-survival-proof} for the proof of Theorem~\ref{thm:q-lifted-survival-index-convergence}.

\section{Valuation of longevity derivatives}
\label{sec:longevity_derivatives}
\noindent This section applies the risk-neutral survival probability derived above to valuation
of standard longevity contracts, including continuous life annuity,
longevity corridor, and longevity swap. 
Throughout this section, we assume that interest-rate risk and mortality
risk are conditionally independent under \(\Q\), given the information
available at the valuation time.

\subsection{Continuous life annuity with longevity risk}

\noindent
We first record the valuation formula for a continuous life annuity with longevity risk.
The contract pays one unit continuously while the insured individual
remains alive, up to maturity {\(T\)}. We condition on the insured individual being alive at
the valuation time $t$. The formula is written in terms of the
risk-neutral conditional survival probability $S^\Q(t,s)$ and the
zero-coupon bond price $P(t,s)$.

\begin{proposition}
\label{prop:life_annuity}
Assume that the insured individual is alive at time $t$. Let $P(t,s)$
denote the time-$t$ price of a default-free zero-coupon bond maturing
at time $s$. Using the risk-neutral survival probability in
Eq.~\eqref{eq:q_survival}, the time-$t$ value of a continuous life
annuity with maturity $T$ is
\begin{align}
\bar a(t,T)
&=
\int_t^T P(t,s)S^\Q(t,s)\,\dd s
\notag\\
&=
\int_t^T
P(t,s)
\exp\Big(
-\mu_t A(t,s)
+B_{\mathrm{hist}}^\Q(t,s)
+B_{\mathrm{drift}}^\Q(t,s)
\notag\\
&\qquad\qquad\qquad\qquad\, \,
+B_{\mathrm{Brown}}^\Q(t,s)
+B_{\mathrm{\text{L\'evy}}}^\Q(t,s)
\Big)\,\dd s.
\label{eq:life_annuity_value}
\end{align}
\end{proposition}
This price formula follows by discounting the payment made at time $s$ by
$P(t,s)$, weighting it by the conditional survival probability
$S^\Q(t,s)$, and integrating over $s\in[t,T]$.

\subsection{Pricing of survival caplet and longevity corridor contracts}
\noindent At time zero, we define the realized survival index at maturity \(T\) by
$ S_T
 =
 \exp\left(
 -\int_0^T\mu_s\,\dd s
 \right)
 =
 e^{Y_{0,T}}.$
Hence, we have 
\[
 \E^\Q[S_T]
 =
 S^\Q(0,T).
\]
For \(K>0\) and $k = \log K$, the undiscounted survival caplet value with strike $K$ is given by
\begin{equation*}
 \mathcal C^\Q(K,T)
 :=
 \E^\Q[(S_T-K)^+]
 =
 \E^\Q[(e^{Y_{0,T}}-e^k)^+].
\end{equation*}

\begin{proposition}
\label{prop:caplet_fourier}
Assume there exists a damping parameter \(\delta>0\) such that
\[
 \E^\Q\left[e^{(\delta+1)Y_{0,T}}\right]<\infty.
\]
The survival caplet value admits the Fourier inversion representation
\begin{equation}
 \mathcal C^\Q(K,T)
 =
 \frac{e^{-\delta k}}{\pi}
 \int_0^\infty
 \operatorname{Re}
 \left[
 e^{-ivk}
 \frac{
 \phi_{0,T}^\Q\left(v-i(\delta+1)\right)
 }{
 \delta^2+\delta-v^2+i(2\delta+1)v
 }
 \right]\,\dd v.
 \label{eq:caplet_fourier}
\end{equation}
{The Fourier inversion above is defined in the \(L^1\) sense when the damped Fourier inversion is integrable; otherwise it is understood as a generalized Fourier inversion.}
The time-zero caplet price is given by
\begin{equation}
 C_0(K,T)
 =
 NP(0,T)\mathcal C^\Q(K,T),
 \label{eq:caplet_price_zero}
\end{equation}
where \(N\) is the notional and \(P(0,T)\) is the discount factor.
\end{proposition}
\noindent
Readers may refer to \ref{app:proof-caplet-fourier} for the proof of Proposition~\ref{prop:caplet_fourier}.

\begin{corollary}
\label{cor:floorlet_parity}
Define the undiscounted survival floorlet value by
\begin{equation*}
 \mathcal P^\Q(K,T)
 :=
 \E^\Q[(K-S_T)^+].
\end{equation*}
The caplet and floorlet values satisfy the parity relation:
\begin{equation}
 \mathcal P^\Q(K,T)
 =
 \mathcal C^\Q(K,T)
 -
 \E^\Q[S_T]
 +
 K.
 \label{eq:floorlet_parity}
\end{equation}
The parity relation follows from
\((K-S_T)^+=(S_T-K)^+-S_T+K\) and taking expectations under \(\Q\).
Since \(\E^\Q[S_T]=S^\Q(0,T)\), the floorlet value can be obtained directly from
the caplet value and the risk-neutral survival probability. Similar 
to the caplet price, the time-zero floorlet
price is
\begin{equation}
 P_0(K,T)
 =
 NP(0,T)\mathcal P^\Q(K,T).
 \label{eq:floorlet_price_zero}
\end{equation}
\end{corollary}



Lastly, we consider a terminal-index longevity corridor written on
the realized survival index \(S_T\). 
Let \(S_l\) and \(S_u\) denote the lower and upper corridor levels,
respectively, where \(0<S_l<S_u\). If the realized survival index is
below \(S_l\), the contract makes no payment. If \(S_T\) lies between
\(S_l\) and \(S_u\), the payoff increases linearly with \(S_T\). Once
\(S_T\) exceeds \(S_u\), the payoff is capped at \(S_u-S_l\). Thus, the
contract compensates the holder for adverse longevity outcomes within
a prescribed range while limiting the maximum payment.

More precisely, the normalized payoff at maturity \(T\) is
\begin{equation*}
G_T
=
\begin{cases}
0, & S_T\leq S_l,\\
S_T-S_l, & S_l<S_T<S_u,\\
S_u-S_l, & S_T\geq S_u.
\end{cases}
\end{equation*}
For a corridor contract with notional \(N\), the monetary payoff is \(NG_T\).
Equivalently, we obtain
\begin{equation}
G_T
=
(S_T-S_l)^+-(S_T-S_u)^+ = \min \{S_u-S_l, \max(S_T-S_l,0)\}.
\label{eq:corridor_decomposition}
\end{equation}
Hence, the longevity corridor can be viewed as a long position in a
survival caplet with strike \(S_l\) combined with a short position in
a survival caplet with strike \(S_u\). This decomposition
allows the corridor value to be obtained
directly from the caplet pricing formula.

\begin{proposition}
\label{prop:corridor_fourier}
The undiscounted longevity corridor value is given by
\begin{equation*}
 \mathcal G(0,T)
 :=
 \E^\Q[G_T]
 =
 \mathcal C^\Q(S_l,T)-\mathcal C^\Q(S_u,T),
\end{equation*}
where \(\mathcal C^\Q(\cdot,T)\) is the survival caplet value. Equivalently, we have
\begin{align}
\mathcal G(0,T)
=&
\frac{e^{-\delta \log S_l}}{\pi}
\int_0^\infty
\operatorname{Re}
\left[
e^{-iv\log S_l}
\frac{
\phi_{0,T}^\Q\left(v-i(\delta+1)\right)
}{
\delta^2+\delta-v^2+i(2\delta+1)v
}
\right]\dd v
\nonumber\\
&-
\frac{e^{-\delta \log S_u}}{\pi}
\int_0^\infty
\operatorname{Re}
\left[
e^{-iv\log S_u}
\frac{
\phi_{0,T}^\Q\left(v-i(\delta+1)\right)
}{
\delta^2+\delta-v^2+i(2\delta+1)v
}
\right]\dd v.
\label{eq:corridor_fourier}
\end{align}
The time-zero corridor price is given by
\begin{equation}
 V_0^{\mathrm{corridor}}
 =
 NP(0,T)
 \left[
 \mathcal C^\Q(S_l,T)-\mathcal C^\Q(S_u,T)
 \right].
 \label{eq:corridor_price_zero}
\end{equation}
\end{proposition}
\noindent
These price formulas follow by taking expectations in
Eq.~\eqref{eq:corridor_decomposition} and applying
Eq.~\eqref{eq:caplet_fourier} at \(S_l\) and \(S_u\).

\subsection{Longevity swap}
\noindent
A longevity swap is based on the survival experience of a specified
reference population. At time \(t_k\), the realized survival index
\(S_{t_k}\) represents the proportion of the initial reference population
that is still alive. In practice, this index is calculated from published
mortality data. In our model, it is represented by
\[
S_{t_k}
=
\exp\left(
-\int_0^{t_k}\mu_s\,\dd s
\right).
\]
We consider the following tenor of payment dates 
\(
 0<t_1<t_2<\cdots<t_m.
\)
At time \(t_k\), the floating leg pays the realized survival index \(S_{t_k}\), while the
fixed leg pays the survival rate \(K_k\) agreed upon when the contract is
initiated. The net payoff to the receiver of the floating leg is
\[
 N(S_{t_k}-K_k).
\]

\begin{proposition}
\label{prop:swap_value}
The time-zero value of the longevity swap is
\begin{equation}
 V_0^{\mathrm{swap}}
 =
 N\sum_{k=1}^m
 P(0,t_k)
 \left[
 \E^\Q[S_{t_k}]-K_k
 \right].
 \label{eq:swap_value}
\end{equation}
In terms of the characteristic function $\phi_{0,t_k}^\Q$ of \(Y_{0,t_k}\), the swap value can be written as
\begin{equation}
 V_0^{\mathrm{swap}}
 =
 N\sum_{k=1}^m
 P(0,t_k)
 \left[
 \phi_{0,t_k}^\Q(-i)-K_k
 \right].
 \label{eq:swap_cf_value}
\end{equation}
Here, $K_k$ is the fixed leg payment at time \(t_k\).
Equivalently, we have
\begin{align}
V_0^{\mathrm{swap}}
=
N\sum_{k=1}^m
P(0,t_k)
\bigg[
&\exp\bigg(
-\mu_0A(0,t_k)
-D_\gamma(0,t_k)
-D_L(0,t_k)
+\frac12\sigma^2p_1^2V_f(0,t_k)
\nonumber\\
&\qquad
+\int_0^{t_k}
\Psi_L^\Q\left(
-\frac{\sigma p_2}{s_L}(K_{0,t_k}^Hf)(v)
\right)\dd v
\bigg)
-K_k
\bigg].
\label{eq:swap_closed_form}
\end{align}
\end{proposition}

To derive the above price formula, we use
discounting and summing the payments to give
Eq.~\eqref{eq:swap_value}. Next, by observing \(S_{t_k}=e^{Y_{0,t_k}}\), this gives
\(\E^\Q[S_{t_k}]=\phi_{0,t_k}^\Q(-i), k =1, 2,\cdots, m\).

\subsection{Convergence of rank-\(r\) approximate derivative prices}
\noindent
We examine the convergence of the longevity derivative prices under the rank-\(r\) SVD approximation to the exact derivative prices.
We apply Theorem~\ref{thm:q-lifted-survival-index-convergence}
to the longevity derivatives discussed in this section. Under the rank-\(r\) framework,
we define
\begin{align*}
\bar a^{(r)}(0,T)
&:=
\int_0^T P(0,s)\mathbb E^{\mathbb Q}[S_s^{(r)}]\,\dd s,\\
G_T^{(r)}
&:=
\min\left\{S_u-S_l,
\max\left(S_T^{(r)}-S_l,0\right)\right\},\\
V_0^{\mathrm{corridor},(r)}
&:=
NP(0,T)\mathbb E^{\mathbb Q}[G_T^{(r)}],\\
V_0^{\mathrm{swap},(r)}
&:=
N\sum_{k=1}^m P(0,t_k)
\left(
\mathbb E^{\mathbb Q}[S_{t_k}^{(r)}]-K_k
\right).
\end{align*}

\begin{corollary}
\label{cor:lifted-price-convergence}
For the life annuity, suppose that the assumptions of
Theorem~\ref{thm:q-lifted-survival-index-convergence} hold for every
\(s\in[0,T]\), and there exists \(b\in L^1([0,T])\) such that
\begin{equation}
P(0,s)\mathbb E^{\mathbb Q}
\left[
\left|S_s^{(r)}-S_s\right|
\right]
\leq b(s)
\label{eq:annuity-lifting-dominating-condition}
\end{equation}
for almost every \(s\in[0,T]\) and all \(r\). Then
\begin{equation*}
\bar a^{(r)}(0,T)
\longrightarrow
\bar a(0,T),
\qquad r\to\infty,
\end{equation*}
with
\begin{equation*}
\left|
\bar a^{(r)}(0,T)-\bar a(0,T)
\right|
\leq
\int_0^T
P(0,s)\mathbb E^{\mathbb Q}
\left[
\left|S_s^{(r)}-S_s\right|
\right]\,\dd s.
\end{equation*}

For the longevity corridor contract and longevity swap contract, suppose that the assumptions
of Theorem~\ref{thm:q-lifted-survival-index-convergence} hold at \(T\) and at
each payment date \(t_k\), respectively. Then their rank-\(r\) approximate
prices converge to the corresponding prices in
Eqs.~\eqref{eq:corridor_price_zero} and \eqref{eq:swap_value}:
\begin{equation}
V_0^{\mathrm{corridor},(r)}
\longrightarrow
V_0^{\mathrm{corridor}},
\qquad
V_0^{\mathrm{swap},(r)}
\longrightarrow
V_0^{\mathrm{swap}}.
\label{eq:lifted-swap-corridor-price-convergence}
\end{equation}
More precisely, we have 
\begin{align}
\left|
V_0^{\mathrm{corridor},(r)}-V_0^{\mathrm{corridor}}
\right|
&\leq
NP(0,T)
\mathbb E^{\mathbb Q}
\left[
\left|S_T^{(r)}-S_T\right|
\right],
\label{eq:corridor-lifting-price-bound}\\
\left|
V_0^{\mathrm{swap},(r)}-V_0^{\mathrm{swap}}
\right|
&\leq
N\sum_{k=1}^m P(0,t_k)
\mathbb E^{\mathbb Q}
\left[
\left|S_{t_k}^{(r)}-S_{t_k}\right|
\right].
\notag
\end{align}
\end{corollary}
\noindent
Readers may refer to \ref{app:proof-lifted-price-convergence} for the proof of Corollary~\ref{cor:lifted-price-convergence}.

\section{Numerical validation of the singular value decomposition approximation}
\label{sec:svd-lifting-validation}
\noindent
In this section, we examine the numerical performance of the SVD fitting procedure and its rank-\(r\) SVD approximation. The assessment of accuracy of calculating the prices of mortality derivatives are applied to the three products discussed in the previous sections. The rank-\(r\) approximation price of each product is compared with the corresponding benchmark price formula. Specifically, we use the integral formula \eqref{eq:life_annuity_value} for the life annuity, the Fourier inversion formula \eqref{eq:corridor_fourier} for the corridor contract and the integral formula \eqref{eq:swap_closed_form} for the longevity swap. We also derive a residual correction scheme to reduce the numerical error arising from the rank-\(r\) SVD truncation.

\subsection{Numerical performance of the rank-r SVD approximation}

\noindent
First, we consider the full SVD lifting without rank-$r$ approximation.
Let \(N\) denote the number of time intervals on \([0,T]\), and let
\(K_N\) denote the causal \(N\times N\) kernel matrix without SVD
truncation. Next, for a fixed \(N\), let \(K_N^{(r)}\) denote its rank-\(r\)
SVD approximation. In our numerical tests, the lifting is implemented on 
\(N=160\). 

Our numerical calculations compute three values for each derivative and
three different L\'evy specification: the rank-\(r\) approximation value 
$V_{160}^{(r)}$ on the \(N=160\)
grid, the full-kernel value $V_{160}^{\text{full}}$ on the \(N=160\) grid, and the benchmark
value $V^{\text{bench}}$. The errors are quantified by
\begin{equation}
E_{\mathrm{rank}}(r)
:=
\left|V_{160}^{(r)}-V_{160}^{\mathrm{full}}\right|,
\qquad
E_{\mathrm{total}}^{\mathrm{rel}}(r)
:=
\frac{\left|V_{160}^{(r)}-V^{\mathrm{bench}}\right|}{\left|V^{\mathrm{bench}}\right|} \times 100\%.
\label{eq:error_def}
\end{equation}
Here, \(E_{\mathrm{rank}}(r)\) measures the error incurred from the rank-\(r\) SVD approximation on the \(N=160\) grid. On the other hand, \(E_{\mathrm{total}}^{\mathrm{rel}}(r)\) measures the total errors arising from the discretization into \(N\) time steps in the SVD lifting and the SVD truncation relative to the benchmark price.

The numerical values of the model parameters used in the numerical
tests are listed below:
\[
H=0.65,\qquad T=5,\qquad
\mu_0=0.01,\qquad a=0.12,\qquad \sigma=0.03,
\]
with \(p_1=0.20\), \(p_2=0.80\). The Brownian risk premium is
\(\gamma=0.30\), and the Esscher parameter is \(\theta=0.10\). The risk-free
discount curve is \(P(0,t)=e^{-0.02t}\). For the corridor contract, we set
\(S_l=0.94\) and \(S_u=0.99\). For the longevity swap, the payment
dates are \(t_k=1,\ldots,5\) and the fixed rate is \(K_k=0.90\). For the
Poisson specification, \(\lambda=0.50\). For the
Gamma specification, \(\alpha=0.30\) and \(\beta=1.50\). For the TS
specification, \(c=0.20\), \(\kappa=0.40\), and
\(\lambda_{\mathrm{TS}}=1.20\).

For the numerical implementation, we divide \([0,T]\) into \(N\) intervals
of length \(\Delta t=T/N\), and set
\[
t_i=i\Delta t,
\qquad
v_j=\left(j-\frac12\right)\Delta t,
\qquad i,j=1,\ldots,N.
\]
Here \(t_i\) is a right-endpoint time and \(v_j\) is the midpoint
integration node in the \(j\)-th interval. The continuous kernel is
replaced by the matrix
\[
K_N=(K_H(t_i,v_j))_{i,j=1}^N.
\]
{More precisely, \((K_N)_{ij}=\mathbf 1_{\{j\le i\}}K_H(t_i,v_j)\). The kernel is undefined for \(v_j>t_i\) and those entries are set to zero.}
The matrix representing the integral operator under this quadrature rule
is \(A_N:=\Delta t K_N\). We compute its SVD via
\[
A_N
=U_N\operatorname{diag}
\bigl(\lambda_1^{(N)},\ldots,\lambda_N^{(N)}\bigr)V_N^\top,
\qquad
\lambda_1^{(N)}\geq\cdots\geq\lambda_N^{(N)}\geq0.
\]
Thus, \(\lambda_i^{(N)}\) is a singular value of \(A_N\). Retaining the
first \(r\) singular components gives
\[
A_N^{(r)}
:=
U_N^{(r)}\Sigma_r\bigl(V_N^{(r)}\bigr)^\top,
\qquad
K_N^{(r)}:=\frac{A_N^{(r)}}{\Delta t}.
\]
By the Eckart--Young theorem, \(A_N^{(r)}\) is the best rank-\(r\)
approximation to \(A_N\) in the Frobenius norm. This statement applies
before the causality constraint is imposed. Indeed, though \(K_N\) is
lower triangular, its rank-\(r\) SVD approximation \(K_N^{(r)}\) is
generally a full matrix and may contain nonzero entries with \(j>i\).
Such entries would make the value at \(t_i\) depend on future L\'evy
increments. We therefore use the causal projection
\[
\bigl(K_N^{(r),\triangle}\bigr)_{ij}
:=
\mathbf 1_{\{j\leq i\}}\bigl(K_N^{(r)}\bigr)_{ij}.
\]
The projection restores the Volterra structure, but
\(K_N^{(r),\triangle}\) need not have matrix rank \(r\). Nevertheless,
its action at time \(t_n\) can still be evaluated through the same \(r\)
cumulative states:
\[
\sum_{j=1}^{n}\bigl(K_N^{(r),\triangle}\bigr)_{nj}\Delta L_j
=
\frac{1}{\Delta t}\sum_{\ell=1}^{r}
\lambda_\ell^{(N)}\bigl(u_\ell^{(N)}\bigr)_n
\left[
\sum_{j=1}^{n}\bigl(v_\ell^{(N)}\bigr)_j\Delta L_j
\right].
\]
For each \(\ell\), the inner sum is updated recursively as \(n\)
increases. Hence, causal masking may change the matrix rank, but it does
not increase the number of states required by lifting.
Figures~\ref{fig:app-svd-factor-schema}--\ref{fig:app-svd-lift-schema}
in \ref{app:svd-discretization} illustrate the retained SVD
factors, the subsequent causal masking, and the assembly of the lifted
states from the L\'evy increments.
\begin{samepage}
For \(i=1,\ldots,N\), we define the cumulative SVD energy through
index \(i\) by
\[
E_i^{(N)}
:=
\frac{
\sum_{j=1}^{i}\left(\lambda_j^{(N)}\right)^2
}{
\sum_{j=1}^{N}\left(\lambda_j^{(N)}\right)^2
}.
\]
It measures the proportion of the total squared singular values
captured by the first \(i\) singular components.
\end{samepage}


Table~\ref{tab:svd-rank-diagnostics-new} reports the selected singular
values of \(A_{160}\) and the corresponding cumulative SVD energy
\(E_i^{(160)}\).
The spectrum of magnitudes decay from \(\lambda_1^{(160)}=3.943625\) to
\(\lambda_{16}^{(160)}=0.074745\). The first singular component accounts
for approximately \(87.86\%\) of the total energy, the first five
approximately \(98.37\%\), and the first eight approximately
\(99.11\%\).

\begin{table}[htbp]
\centering
\begin{tabular}{rcc}
\toprule
Index \(i\) & \(\lambda_i^{(160)}\) & \(E_i^{(160)}\)\\
\midrule
1  & 3.943625 & 0.878583 \\
2  & 1.103813 & 0.947414 \\
3  & 0.611609 & 0.968546 \\
5  & 0.310311 & 0.983688 \\
8  & 0.172213 & 0.991116 \\
16 & 0.074745 & 0.996378 \\
40 & 0.025827 & 0.998907 \\
80 & 0.012363 & 0.999604 \\
\bottomrule
\end{tabular}
\caption{Singular values and values of cumulative energy at \(N=160\)
for varying index values.}
\label{tab:svd-rank-diagnostics-new}
\end{table}


Next, we examine numerical accuracy of the rank-\(r\) SVD approximation
in calculating mortality derivative prices. The three
price formulas are evaluated under each L\'evy specification for
\(r=1,5,16,40,80\). Table~\ref{tab:svd-paired-rank-errors} lists
the values of $E_{\mathrm{rank}}(r)$ for different mortality derivatives
evaluated under various \llevy{} specifications and varying rank values.
This serves to compare each
rank-\(r\) approximation price with the corresponding full-kernel price on the same
\(N=160\) grid. 
At rank 80, the truncation errors $E_{\mathrm{rank}}(r)$ are at most
\(1.44\times10^{-5}\), \(4.18\times10^{-6}\), and
\(2.40\times10^{-5}\) for the annuity, corridor, and swap,
respectively. 

\begin{table}[H]
\centering
\fontsize{9.5}{11}\selectfont
\begin{tabular}{clrrr}
\toprule
\llevy{} Specification & Rank \(r\) & Annuity & Corridor & Swap \\
\midrule
\multirow{5}{*}{Poisson}
 & 1  & $1.739\times10^{-4}$ & $7.036\times10^{-4}$ & $5.566\times10^{-4}$ \\
 & 5  & $3.376\times10^{-4}$ & $7.151\times10^{-5}$ & $1.834\times10^{-4}$ \\
 & 16 & $9.487\times10^{-5}$ & $1.749\times10^{-5}$ & $5.642\times10^{-5}$ \\
 & 40 & $3.315\times10^{-5}$ & $5.867\times10^{-6}$ & $1.990\times10^{-5}$ \\
 & 80 & $1.313\times10^{-5}$ & $2.246\times10^{-6}$ & $7.993\times10^{-6}$ \\
\midrule
\multirow{5}{*}{Gamma}
 & 1  & $3.834\times10^{-3}$ & $9.131\times10^{-4}$ & $5.337\times10^{-3}$ \\
 & 5  & $4.247\times10^{-4}$ & $1.215\times10^{-4}$ & $7.003\times10^{-4}$ \\
 & 16 & $1.094\times10^{-4}$ & $3.102\times10^{-5}$ & $1.811\times10^{-4}$ \\
 & 40 & $3.720\times10^{-5}$ & $1.054\times10^{-5}$ & $6.185\times10^{-5}$ \\
 & 80 & $1.437\times10^{-5}$ & $4.089\times10^{-6}$ & $2.397\times10^{-5}$ \\
\midrule
\multirow{5}{*}{TS}
 & 1  & $3.839\times10^{-3}$ & $9.158\times10^{-4}$ & $5.348\times10^{-3}$ \\
 & 5  & $4.254\times10^{-4}$ & $1.236\times10^{-4}$ & $7.019\times10^{-4}$ \\
 & 16 & $1.095\times10^{-4}$ & $3.166\times10^{-5}$ & $1.814\times10^{-4}$ \\
 & 40 & $3.725\times10^{-5}$ & $1.077\times10^{-5}$ & $6.198\times10^{-5}$ \\
 & 80 & $1.439\times10^{-5}$ & $4.180\times10^{-6}$ & $2.402\times10^{-5}$ \\
\bottomrule
\end{tabular}
\caption{Values of $E_{\mathrm{rank}}(r)$ for different mortality 
derivatives evaluated under various L\'evy specifications and varying rank values.}
\label{tab:svd-paired-rank-errors}
\end{table}

Lastly, we examine the numerical accuracy of the rank-\(r\) SVD approximation in calculating prices of mortality derivatives when compared with the benchmark derivative values. 
In Table~\ref{tab:svd-total-errors}, we list the values of relative errors \(E_{\mathrm{total}}^{\mathrm{rel}}(r)\) for different mortality derivatives evaluated under various Lévy specifications and varying rank values. Good accuracy of the rank-\(r\) SVD approximation is revealed. Even with rank value as low as 5, the percentage errors are typically well below \(1\%\) already.


\begin{table}[H]
\centering
\small
\begin{tabular}{clrrr}
\toprule
Specification & Rank \(r\) & Annuity & Corridor & Swap \\
\midrule
\multirow{5}{*}{Poisson}
 & 1  & $0.0716\%$ & $3.9439\%$ & $0.5431\%$ \\
 & 5  & $0.0680\%$ & $0.5173\%$ & $0.2278\%$ \\
 & 16 & $0.0733\%$ & $0.2245\%$ & $0.2819\%$ \\
 & 40 & $0.0746\%$ & $0.1615\%$ & $0.2974\%$ \\
 & 80 & $0.0751\%$ & $0.1419\%$ & $0.3025\%$ \\
\midrule
\multirow{5}{*}{Gamma}
 & 1  & $0.1517\%$ & $4.2145\%$ & $2.2697\%$ \\
 & 5  & $0.0773\%$ & $0.6306\%$ & $0.4797\%$ \\
 & 16 & $0.0705\%$ & $0.2207\%$ & $0.2793\%$ \\
 & 40 & $0.0689\%$ & $0.1280\%$ & $0.2333\%$ \\
 & 80 & $0.0684\%$ & $0.0988\%$ & $0.2187\%$ \\
\midrule
\multirow{5}{*}{TS}
 & 1  & $0.1518\%$ & $4.3398\%$ & $2.2739\%$ \\
 & 5  & $0.0774\%$ & $0.6519\%$ & $0.4805\%$ \\
 & 16 & $0.0705\%$ & $0.2237\%$ & $0.2796\%$ \\
 & 40 & $0.0689\%$ & $0.1265\%$ & $0.2335\%$ \\
 & 80 & $0.0684\%$ & $0.0958\%$ & $0.2189\%$ \\
\bottomrule
\end{tabular}
\caption{Values of $E_{\mathrm{total}}^{\mathrm{rel}}$ for different
mortality derivatives evaluated under various L\'evy specifications
and varying rank values.}
\label{tab:svd-total-errors}
\end{table}


\subsection{Second-order residual correction for the rank-$r$ SVD approximation}
\label{sec:numerical-residual-correction}

\noindent
Next, we derive a residual correction scheme to reduce the numerical
error arising from the rank-$r$ SVD approximation. For this purpose, we let
\[
\delta g_{0,T}^{(r)}(v)
:=
g_{0,T}(v)-g_{0,T}^{(r)}(v)
\]
denote the truncation residual of the transformed kernel. For the
time-zero survival transform, we define the corresponding full and
rank-$r$ L\'evy arguments by
\[
z(v,T):=-\frac{\sigma p_2}{s_L}g_{0,T}(v),
\qquad
z^{(r)}(v,T):=-\frac{\sigma p_2}{s_L}g_{0,T}^{(r)}(v),
\]
respectively. Their difference is given by
\[
\delta z^{(r)}(v,T)
:=
z(v,T)-z^{(r)}(v,T)
=
-\frac{\sigma p_2}{s_L}\delta g_{0,T}^{(r)}(v).
\]
Although the truncation residual decreases as $r$ increases, the
price formulas apply a nonlinear L\'evy exponent to these arguments.
In our residual correction scheme, we attempt to approximate the change in the exponent up to the second
order. This residual correction preserves the rank-$r$ state
representation because it does not introduce an additional stochastic
factor.

Recall that the L\'evy
contribution to the time-zero log-survival transform is given by
\[
-D_L(0,T)
+\int_0^T
\Psi_L^\Q\bigl(z(v,T)\bigr)\,\dd v =
\int_0^T
\left[
\Psi_L^\Q\bigl(z(v,T)\bigr)
+\Delta m_L z(v,T)
\right]\dd v.
\]
by virtue of Eq.~\eqref{eq:q-lifted-levy-drift-correction}.
Hence, the centered cumulant and the deterministic measure-change term
enter into the pricing formula through the above sum.
This observation leads to the consideration of
\begin{equation}
\Lambda_L^\Q(z)
:=
\Psi_L^\Q(z)+\Delta m_L z
=\Psi_X^\Q(z)-m_L^{\Pp}z.
\label{eq:q-effective-levy-exponent}
\end{equation}
The second equality follows from
\(\Psi_L^\Q(z)=\Psi_X^\Q(z)-m_L^\Q z\) and
\(\Delta m_L=m_L^\Q-m_L^{\Pp}\). Equivalently, we have
\[
\Lambda_L^\Q(z)
=
\log\E^\Q\left[
\exp\left(z\bigl(X_1^\Q-m_L^{\Pp}\bigr)\right)
\right].
\]
Here, \(\Lambda_L^\Q\) is the cumulant under
\(\Q\) of the increment \(X_1^\Q\) centered by its \(\Pp\)-mean. 

For the numerical implementation, we apply the residual correction on
the midpoint grid. Using the nodes \(v_j\) defined above, we set
\begin{equation}
z_j(T):=z(v_j,T)=-\frac{\sigma p_2}{s_L}g_{0,T}(v_j),
\qquad
z_j^{(r)}(T):=z^{(r)}(v_j,T)
=-\frac{\sigma p_2}{s_L}g_{0,T}^{(r)}(v_j),
\label{eq:residual-cumulant-arguments}
\end{equation}
and
\[
\delta z_j^{(r)}(T)
:=\delta z^{(r)}(v_j,T)
=z_j(T)-z_j^{(r)}(T).
\]
Thus, \(\delta z_j^{(r)}(T)\) denotes the truncation residual in the
cumulant argument. Eq.~\eqref{eq:residual-cumulant-arguments} gives the
real arguments for the survival exponent. For the
Fourier inversion in the corridor contract, the same construction applies with
\[
z_j(u,T)=-iu\frac{\sigma p_2}{s_L}g_{0,T}(v_j),
\qquad
z_j^{(r)}(u,T)=-iu\frac{\sigma p_2}{s_L}g_{0,T}^{(r)}(v_j),
\]
where \(u=v-i(\delta+1)\). The correction derived below is then applied to
\(\Lambda_L^\Q\) at the complex argument \(z_j^{(r)}(u,T)\).
The full-grid and rank-$r$ L\'evy exponents are, respectively,
\begin{equation}
B_{\mathrm{\text{L\'evy}},N}^{\Q}(0,T)
:=\Delta t\sum_{j=1}^{N}\Lambda_L^{\Q}\bigl(z_j(T)\bigr),
\qquad
B_{\mathrm{\text{L\'evy}},N}^{\Q,(r)}(0,T)
:=\Delta t\sum_{j=1}^{N}
\Lambda_L^{\Q}\bigl(z_j^{(r)}(T)\bigr).
\label{eq:full-grid-levy-exponent}
\end{equation}
The Taylor expansion of \(\Lambda_L^\Q\) around \(z_j^{(r)}(T)\) up to
the second order term gives the
second-order residual-corrected exponent as follows:
\begin{align}
B_{\mathrm{\text{L\'evy}},N}^{\Q,(r,2)}(0,T)
:&=
\Delta t\sum_{j=1}^{N}
\Bigg[
\Lambda_L^{\Q}\bigl(z_j^{(r)}(T)\bigr)
+\delta z_j^{(r)}(T)(\Lambda_L^{\Q})'
\bigl(z_j^{(r)}(T)\bigr) \notag\\
& \qquad \qquad +\frac{(\delta z_j^{(r)}(T))^2}{2}
(\Lambda_L^{\Q})''\bigl(z_j^{(r)}(T)\bigr)
\Bigg].
\label{eq:second-order-levy-correction}
\end{align}
The first term is the usual rank-$r$ exponent, while the next two
terms are the first- and second-order adjustments associated
with the truncation residual. If \(\Lambda_L^{\Q}\) is differentiable up to
the third order on a convex set
containing the line segments between \(z_j^{(r)}(T)\) and \(z_j(T)\),
the Taylor remainder satisfies
\[
\bigl|B_{\mathrm{\text{L\'evy}},N}^{\Q}(0,T)-B_{\mathrm{\text{L\'evy}},N}^{\Q,(r,2)}(0,T)\bigr|
\leq
\frac{\Delta t}{6}
\sup_z\bigl|(\Lambda_L^{\Q})'''(z)\bigr|
\sum_{j=1}^{N}\bigl|\delta z_j^{(r)}(T)\bigr|^3,
\]
where the supremum is taken over these line segments. 
In particular, the above bound shows that
the remaining exponent error is of third order in
\(\delta z_j^{(r)}(T)\). In the numerical implementation, we replace the
usual rank-$r$ L\'evy exponent in each price formula by its
residual corrected counterpart.

We then examine whether this correction improves accuracy at the same
rank. For the uncorrected price $V_{160}^{(r)}$ and corrected prices $V_{160}^{(r,2)}$, respectively, we define
\begin{equation}
E_{\mathrm{rank}}^{\mathrm{rel}}(r)
:=
\frac{
\left|V_{160}^{(r)}-V_{160}^{\mathrm{full}}\right|
}{
\left|V_{160}^{\mathrm{full}}\right|
},
\qquad
E_{\mathrm{corr}}^{\mathrm{rel}}(r)
:=
\frac{
\left|V_{160}^{(r,2)}-V_{160}^{\mathrm{full}}\right|
}{
\left|V_{160}^{\mathrm{full}}\right|
}.
\label{eq:error_residual}
\end{equation}
Both errors use the full kernel on the same $N=160$ grid as the
benchmark and therefore isolate the SVD truncation error before and
after the residual correction. The comparison uses
$r=1,2,3,5,8,12,16$.

\begin{figure}[H]
\centering
\safeincludegraphics[width=\textwidth]{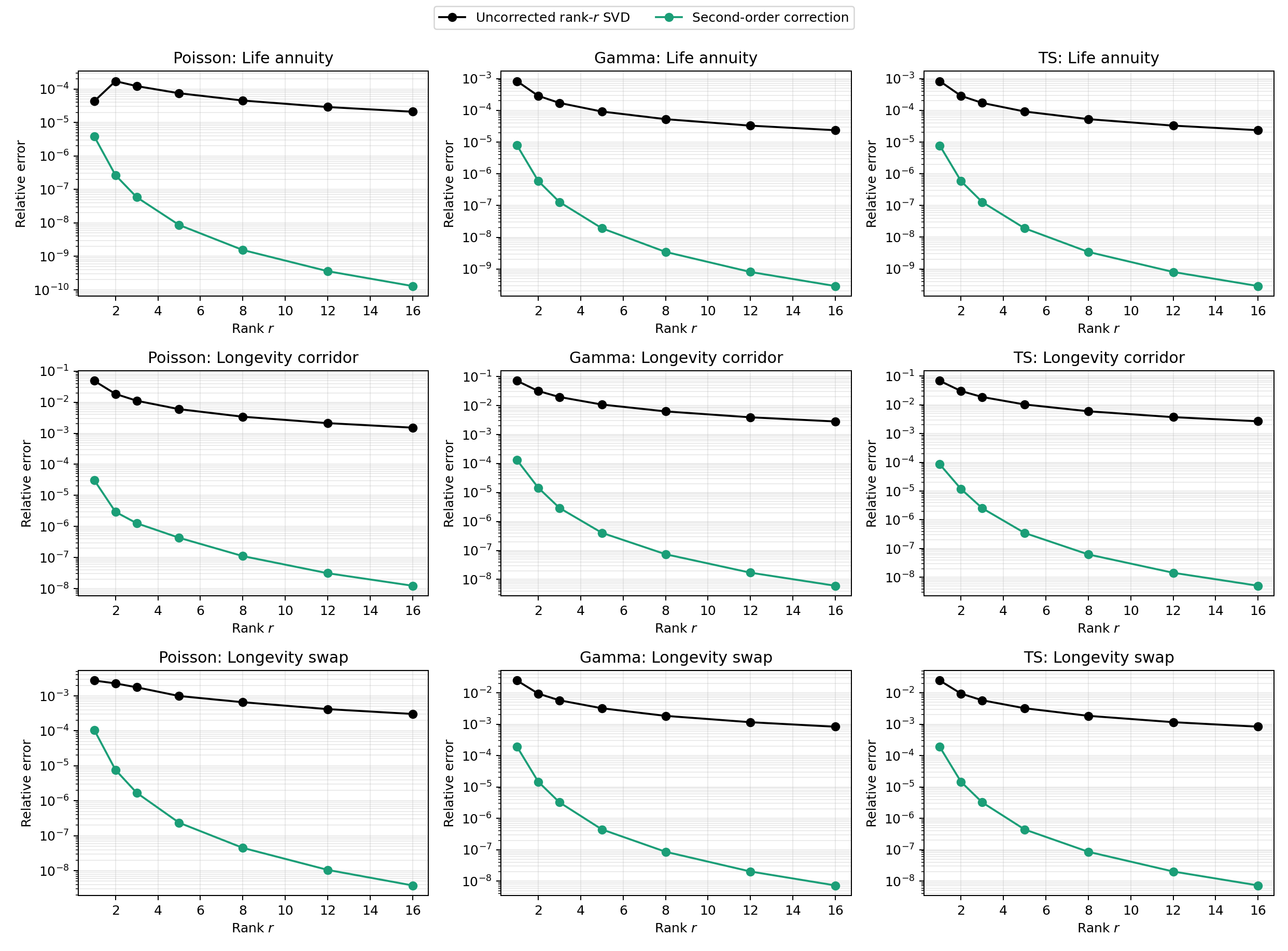}
\caption{Plots of relative pricing errors of the rank-\(r\) SVD (upper curves) and the second-order residual-corrected approximation (lower curves) for varying rank values under different Lévy specifications and mortality derivatives.}
\label{fig:svd-residual-correction-errors}
\end{figure}

Figure~\ref{fig:svd-residual-correction-errors} shows that the corrected
errors are below the usual SVD errors for every displayed rank,
derivative product, and L\'evy specification. The improvement is clearly visible
at even rank one and continues as the rank increases. 
A broader set of robustness experiment conducted by varying
the Hurst parameter, maturity, fractional L\'evy exposure, Esscher
parameter, L\'evy specification, and rank also give a smaller corrected
error in all $972$ comparisons. Thus, the residual correction improves
numerical accuracy for all three longevity derivatives in our numerical
experiments.

\section{Empirical calibration and stress tests on structural mortality shocks}
\label{sec:empirical_analysis}
\noindent
The COVID-19 is widely considered a structural shock in mortality. A structural shock is defined not only by its magnitude but by its ability to alter underlying patterns. COVID-19 achieved this through multiple channels. For example, it reversed a long-term trend of improving life expectancy and created a persistent elevation in death rates. It left a lasting impact on actuarial assumptions and public health system. Another major structural mortality shock is the death tolls in World War II. In this section, we performed empirical studies on these two structural mortality shocks under different specifications of the Lévy driver in our stochastic mortality model.

\subsection{Experimental design}

\noindent 
We use the mortality data from the \citet{HMD}
for France. The first is the 1885 birth cohort, which contains the mortality
shocks associated with World War II. The second is the 1935 birth
cohort, which is used to study the COVID-19 mortality shock.
We consider
three specifications of the L\'evy driver, namely the Poisson, Gamma, and
TS specifications. We compare the
finite-activity Poisson specification with the infinite-activity Gamma and
TS specifications under stress conditions to see which specifications
gives the lowest prediction error.

\vspace{\baselineskip}

\noindent \textit{Case 1: World War II stress test}
\mbox{}\\
The France 1885 birth cohort is examined under two designs. 
The mortality observation at age 50 in
1935 is used to determine the initial mortality intensity
\(
 \mu_0=-\log(1-q_{50}),
\)
which is not included as a separate survival-probability observation in the
calibration loss. 
In the first design, the model is calibrated using 20 cumulative survival
observations over \(T=1,\ldots,20\), corresponding to ages 51--70 and
calendar years 1936--1955. The fitted model is then extrapolated over
\(T=21,\ldots,40\), corresponding to ages 71--90 and calendar years
1956--1975.
In the second design, the model is calibrated using 30 cumulative survival
observations over \(T=1,\ldots,30\), corresponding to ages 51--80 and
calendar years 1936--1965. The out-of-sample extrapolation period is
\(T=31,\ldots,40\), corresponding to ages 81--90 and calendar years
1966--1975.
The two designs share
the same initial cohort observation and final evaluation age, but differ in
the length of the calibration period.

\vspace{\baselineskip}

\noindent \textit{Case 2: COVID-19 stress test}
\mbox{}\\
The France 1935 birth cohort is used to examine mortality dynamics around
the COVID-19 pandemic. The mortality observation at age 65 in 2000 is used
to determine the initial mortality intensity
\(
 \mu_0=-\log(1-q_{65}),
\)
and is not included as a separate cumulative survival observation in the
calibration loss. 
The models are calibrated using 14 cumulative survival observations over
\(T=1,\ldots,14\), corresponding to ages 66--79 and calendar years
2001--2014. The fitted models are then evaluated out of sample over
\(T=15,\ldots,23\), corresponding to ages 80--88 and calendar years
2015--2023. The out-of-sample period contains a pre-pandemic validation
phase from 2015 to 2019 and the COVID-19 mortality shock from 2020 onward.

\subsection{Calibration method}
\label{subsec:calibration-method}

\noindent The empirical analysis uses the annual male period life-table data for France
from \citet{HMD}. Specifically, we use the one-year age
by one-year calendar-year period life table for the French {male} population
series. The underlying mortality variable is the one-year probability of
death, denoted by \(q_{x,y}\), for an individual aged \(x\) in calendar
year \(y\).
Although the source file contains period life-table observations, cohort
mortality paths are constructed by extracting diagonal age--year sequences.
For a birth cohort \(b\), the observation at age \(x\) is selected from
calendar year
\(
 y=b+x.
\)
The observed cumulative survival probability at horizon \(T_i=i\) is
constructed as
\begin{equation*}
 S_i^{\mathrm{obs}}
 =
 \prod_{j=0}^{i-1}
 \left(
 1-q_{x_0+j,\,b+x_0+j}
 \right).
\end{equation*}
Thus, \(S_i^{\mathrm{obs}}\) is the cumulative survival probability from
the initial age, rather than a one-year conditional survival probability.

For each L\'evy specification, the model parameters are calibrated by
minimizing the mean squared error (MSE) between the observed cumulative survival
probabilities and the corresponding model-implied survival probabilities.
Let \(S_i^{\mathrm{obs}}\) denote the observed cumulative survival
probability at horizon \(T_i\), and let
\(S^{\mathrm{model}}(T_i;\Theta)\) denote the survival probability generated
by the model with parameter vector \(\Theta\). The calibration problem is
defined as
\begin{equation}
 \widehat{\Theta}
 =
 \arg\min_{\Theta\in\mathcal D}
 \frac{1}{n_{\mathrm{train}}}
 \sum_{i=1}^{n_{\mathrm{train}}}
 \left(
 S_i^{\mathrm{obs}}
 -
 S^{\mathrm{model}}(T_i;\Theta)
 \right)^2,
 \label{eq:calibration_def}
\end{equation}
where \(\mathcal D\) denotes the admissible parameter domain.


Appropriate parameter restrictions are imposed to preserve
the admissibility of the L\'evy cumulants and the long-range dependence
condition \(H>1/2\).
For the Gamma specification, the rate parameter is fixed at 1. For the
TS specification, the tempering parameter
\(\lambda_{\mathrm{TS}}\) is also fixed at 1. The remaining structural
and jump parameters are estimated subject to their respective admissible
bounds.
{The mixed specification is used as the general modelling framework rather than as a claim that both components must make comparable empirical contributions. The Brownian component represents regular Brownian fluctuations, whereas the fractional L\'evy component provides non-Brownian shocks and LRD. A calibration close to the L\'evy-dominated boundary is therefore admissible within the mixed model. Whenever \(p_2\ne0\), the fractional component remains non-Markovian and the SVD lifting is still required for finite-dimensional numerical pricing.}


\subsection{World War II stress test}
\begin{figure}[H]
    \centering
    \safeincludegraphics[width=0.9\textwidth]{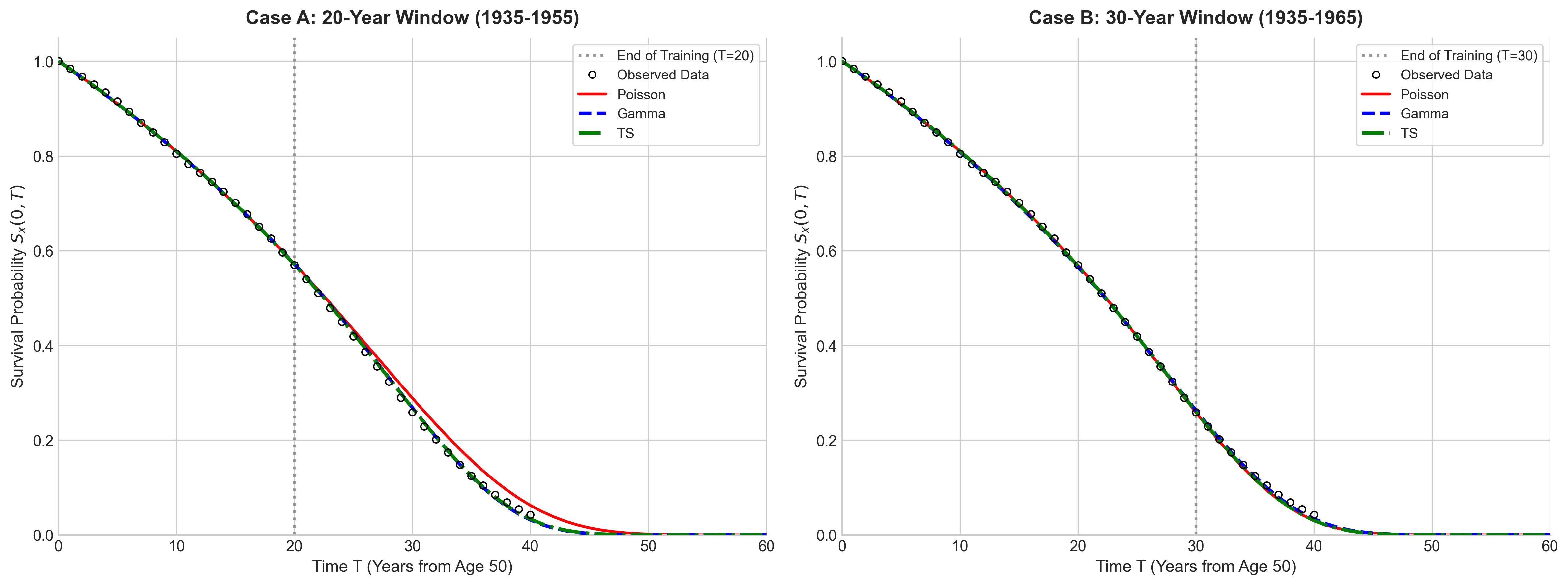}
    \caption{Empirical calibration results for the World War II stress test using two observation windows for the France 1885 cohort.}
    \label{fig:wwii_calibration}
\end{figure}

\noindent
Table~\ref{tab:wwii_calibration_results} and Figure~\ref{fig:wwii_calibration} report the
calibration results for the France 1885 cohort under two observation windows. 
In the 20-year window, all three specifications achieve similar in-sample fits, with mean
squared errors close to \(9\times 10^{-6}\). The difference is mainly observed in the
out-of-sample period. The Gamma and TS specifications obtain
out-of-sample MSEs of \(5.56\times 10^{-5}\) and \(6.41\times 10^{-5}\), respectively,
while the Poisson specification gives \(5.99\times 10^{-4}\). In this window, the
Gamma specification gives the lowest out-of-sample error, followed closely by the
TS specification.
This result is consistent with the different jump specifications. The Poisson specification
uses a finite-activity jump process, so mortality shocks are represented through
discrete jump arrivals. We observe that the estimated Poisson intensity is
small, which restricts the frequency of jump in the survival path.
The Gamma and TS specifications are infinite-activity and
can generate a larger number of small jump movements. One possible explanation 
is that the infinite-activity specifications allow a richer accumulation of 
small mortality disturbances. 
In the 30-year window, the performance gap becomes smaller. The Gamma and
TS specifications obtain out-of-sample MSEs of \(2.72\times10^{-5}\) and
\(6.90\times10^{-5}\), respectively, compared with \(8.23\times10^{-5}\) for the
Poisson specification. The Gamma specification remains the best performing specification,
while the TS specification still performs better than the Poisson benchmark.

The comparison between the two windows suggests that 
the infinite-activity specifications yield lower out-of-sample MSEs in the present calibration exercise.
 When the longer window includes the
post-war recovery phase, the fitted survival path is smoother and the gap between the
models narrows. Across both windows, the Gamma specification provides the most stable
out-of-sample results, while the TS specification also performs
better than the Poisson specification.

An additional feature of the calibration results concerns the estimate of
\(H\), where $H \in (\frac12, 1)$. For the Poisson specification, the constrained objective
function attains its minimum close to the lower boundary of $H={0.5}$, indicating that the Poisson
specification favors the weakest degree of fractional persistence allowed
by the admissible parameter region. By contrast, the calibrated values of
\(H\) for the Gamma and TS specifications are interior
solutions and do not reach the lower bound. This boundary behavior suggests
that, when mortality shocks are modeled through a finite-activity Poisson
driver, the data provide limited support for an additional strong
long-range dependence effect. 

\begin{table}[H]
\centering
\small
\setlength{\tabcolsep}{3pt}
\renewcommand{\arraystretch}{1.15}
\resizebox{\textwidth}{!}{%
\begin{tabular}{@{}lcccccccc@{}}
\toprule
\makecell{\textbf{Scenario}\\\textbf{Specification}}
& \(\mathbf{a}\) & \(\boldsymbol{\sigma}\) & \(\mathbf{H}\)
& \(\mathbf{p_1}\) & \(\mathbf{p_2}\)
& \makecell{\textbf{Jump}\\\textbf{Parameter}}
& \makecell{\textbf{In-sample}\\\textbf{MSE}}
& \makecell{\textbf{Out-of-sample}\\\textbf{MSE}} \\
\midrule
\multicolumn{9}{l}{\textit{Case A: 20-year window, ages 51--70, crisis dominated}} \\
Poisson & \(0.0929\) & \(0.0236\) & \(0.510\) & \(0.000041\) & \(0.999959\) & \(\lambda = 1.76 \times 10^{-3}\) & \(9.44 \times 10^{-6}\) & \(5.99 \times 10^{-4}\) \\
Gamma & \(0.126\) & \(0.0755\) & \(0.670\) & \(0.000206\) & \(0.999794\) & \(\alpha = 2.40 \times 10^{-4}\) & \(8.97 \times 10^{-6}\) & \(5.56 \times 10^{-5}\) \\
TS & \(0.128\) & \(0.131\) & \(0.718\) & \(0.000163\) & \(0.999837\) & \makecell{\(c = 4.78 \times 10^{-5}\) \\ \(\kappa = 1.34 \times 10^{-1}\)} & \(8.82 \times 10^{-6}\) & \(6.41 \times 10^{-5}\) \\
\midrule
\multicolumn{9}{l}{\textit{Case B: 30-year window, ages 51--80, long-term recovery}} \\
Poisson & \(0.114\) & \(0.0312\) & \(0.510\) & \(0.000278\) & \(0.999722\) & \(\lambda = 2.02 \times 10^{-3}\) & \(7.29 \times 10^{-6}\) & \(8.23 \times 10^{-5}\) \\
Gamma & \(0.118\) & \(0.0676\) & \(0.668\) & \(0.000036\) & \(0.999964\) & \(\alpha = 2.40 \times 10^{-4}\) & \(1.28 \times 10^{-5}\) & \(2.72 \times 10^{-5}\) \\
TS & \(0.121\) & \(0.102\) & \(0.664\) & \(0.000052\) & \(0.999948\) & \makecell{\(c = 8.43 \times 10^{-5}\) \\ \(\kappa = 1.76 \times 10^{-1}\)} & \(8.05 \times 10^{-6}\) & \(6.90 \times 10^{-5}\) \\
\bottomrule
\end{tabular}
}
\caption{Calibrated parameters and mean squared errors for the France 1885 cohort}
\label{tab:wwii_calibration_results}
\end{table}

To complement the MSE comparison for the World War II stress test, Table~\ref{tab:dm_test_summary} reports the Diebold--Mariano test results for the two
World War II windows. The null hypothesis
is equal predictive accuracy between the Poisson specification and an alternative
infinite-activity specification. The loss function is chosen to be the squared prediction error. We define
the loss differential as the squared error of the Poisson specification minus the squared
error of the alternative specification, so a positive DM statistic indicates lower average loss
for the alternative specification. Since the extrapolation samples are relatively small, we apply the
Harvey--Leybourne--Newbold small-sample correction to the DM statistic and compute
the \(p\)-value from a \(t\)-distribution with \(T-1\) degrees of freedom.
In Table~\ref{tab:dm_test_summary}, \(^{***}\) denotes statistical significance at the 0.1\% level.

For the 20-year window, the Poisson-versus-Gamma comparison gives a DM statistic of
\(6.3419\) with a \(p\)-value of \(4.38\times10^{-6}\). The Poisson-versus-tempered
stable comparison gives a DM statistic of \(6.2237\) with a \(p\)-value of
\(5.60\times10^{-6}\). Both tests reject equal predictive accuracy at the \(0.1\%\) significance level. This is consistent with the lower out-of-sample MSEs reported for
the Gamma and TS specifications in the 20-year window.

For the 30-year window, the Poisson-versus-Gamma comparison gives a DM statistic of
\(5.2584\) with a \(p\)-value of \(5.22\times10^{-4}\). The Poisson-versus-tempered
stable comparison gives a DM statistic of \(4.7923\) with a \(p\)-value of
\(9.84\times10^{-4}\). The tests again reject equal predictive accuracy at the \(0.1\%\) significance level.

\begin{table}[H]
\centering
\resizebox{\textwidth}{!}{%
\begin{tabular}{@{}lccccc@{}}
\toprule
 \makecell{\textbf{Window}\\\textbf{Comparison}}
&  \makecell{\textbf{Poisson}\\\textbf{MSE}}
&  \makecell{\textbf{Alternative}\\\textbf{MSE}}
& \textbf{DM Stat} 
& \textbf{\(p\)-value} 
& \textbf{Inference} \\
\midrule
\multicolumn{6}{l}{\textit{Window A: 20-year window, ages 51--70, crisis-dominated period}} \\
Poisson vs. Gamma 
& \(5.99 \times 10^{-4}\) 
& \(5.56 \times 10^{-5}\) 
& \(6.3419\) 
& \(4.38 \times 10^{-6}\) 
& Gamma lower loss*** \\
Poisson vs. TS
& \(5.99 \times 10^{-4}\) 
& \(6.41 \times 10^{-5}\) 
& \(6.2237\) 
& \(5.60 \times 10^{-6}\) 
& TS lower loss*** \\
\midrule
\multicolumn{6}{l}{\textit{Window B: 30-year window, ages 51--80, crisis and recovery period}} \\
Poisson vs. Gamma 
& \(8.23 \times 10^{-5}\) 
& \(2.72 \times 10^{-5}\) 
& \(5.2584\) 
& \(5.22 \times 10^{-4}\) 
& Gamma lower loss*** \\
Poisson vs. TS    
& \(8.23 \times 10^{-5}\) 
& \(6.90 \times 10^{-5}\) 
& \(4.7923\) 
& \(9.84 \times 10^{-4}\) 
& TS lower loss*** \\
\bottomrule
\end{tabular}%
}
\par\smallskip
\caption{Diebold--Mariano predictive accuracy test results for the World War II stress test}
\label{tab:dm_test_summary}
\end{table}

The DM test results provide an additional statistical comparison of the out-of-sample
errors. In both World War II windows, the null of equal predictive accuracy is rejected
when the Poisson specification is compared with either the Gamma or TS
specification. The evidence is strongest in the 20-year window and remains in the
30-year window. These results are consistent with the MSE comparison: the
infinite-activity specifications have lower out-of-sample squared prediction errors in
this exercise, with the Gamma specification giving the lowest MSE in both windows.

\subsection{COVID-19 stress test}
\noindent
The COVID-19 stress test focuses on a short and localized mortality shock at high ages. We use the France 1935 birth cohort. The models are calibrated over the stable 2001--2014 period and tested over 2015--2023. The out-of-sample window contains both a pre-pandemic phase and the COVID-19 mortality shock.

Table~\ref{tab:covid_calibration_parameters} shows that the Poisson specification fits the stable training period well but produces substantially larger prediction errors during the pandemic period. The calibrated Poisson intensity is small during the 2001--2014 baseline period, which suppresses jumps in the period. When the COVID-19 shock appears out of sample, the Poisson specification produces a large prediction error, with an out-of-sample MSE of \(1.03\times10^{-4}\) and a maximum absolute error of \(2.09\%\).

The Gamma and TS specifications give lower out-of-sample errors in the
same stress window. Their out-of-sample MSEs are \(2.03\times10^{-6}\) and
\(1.89\times10^{-6}\), respectively. Their maximum absolute errors are also smaller
than that of the Poisson specification. In this experiment, the TS specification gives the lowest out-of-sample MSE, while the Gamma specification gives a
similar level of predictive accuracy.

\begin{table}[H]
\resizebox{\textwidth}{!}{%
\begin{threeparttable}
\begin{tabular}{lcccccccc}
\toprule
 \makecell{\textbf{Model}\\\textbf{Framework}} & \(\mathbf{a}\) & \(\boldsymbol{\sigma}\) & \(\mathbf{H}\) & \(\mathbf{p_1}\) & \(\mathbf{p_2}\) &  \makecell{\textbf{Jump}\\\textbf{Parameter}} &  \makecell{\textbf{In-sample}\\\textbf{MSE}} &  \makecell{\textbf{Out-of-sample}\\\textbf{MSE}}\\
\midrule
\multicolumn{9}{l}{\textit{Baseline training, 2001--2014, and pandemic projection, 2015--2023}} \\
Poisson         & \(0.129\) & \(0.200\) & \(0.510\) & \(0.000192\) & \(0.999808\) & \(\lambda = 7.04 \times 10^{-5}\) & \(7.13 \times 10^{-8}\) & \(1.03 \times 10^{-4}\) \\
Gamma           & \(0.155\) & \(0.537\) & \(0.551\) & \(0.000020\) & \(0.999980\) & \(\alpha = 1.50 \times 10^{-4}\) & \(1.53 \times 10^{-7}\) & \(2.03 \times 10^{-6}\) \\
TS              & \(0.160\) & \(0.611\) & \(0.568\) & \(0.000437\) & \(0.999563\) & \makecell{\(c = 1.01 \times 10^{-5}\) \\ \(\kappa = 1.03 \times 10^{-1}\)} & \(2.70 \times 10^{-7}\) & \(1.89 \times 10^{-6}\) \\
\bottomrule
\end{tabular}
\end{threeparttable}%
}
\centering
\caption{Calibrated parameters and mean squared errors under the COVID-19 shock: France 1935 cohort}
\label{tab:covid_calibration_parameters}
\end{table}

\begin{figure}[H]
    \centering
    \safeincludegraphics[width=0.8\textwidth]{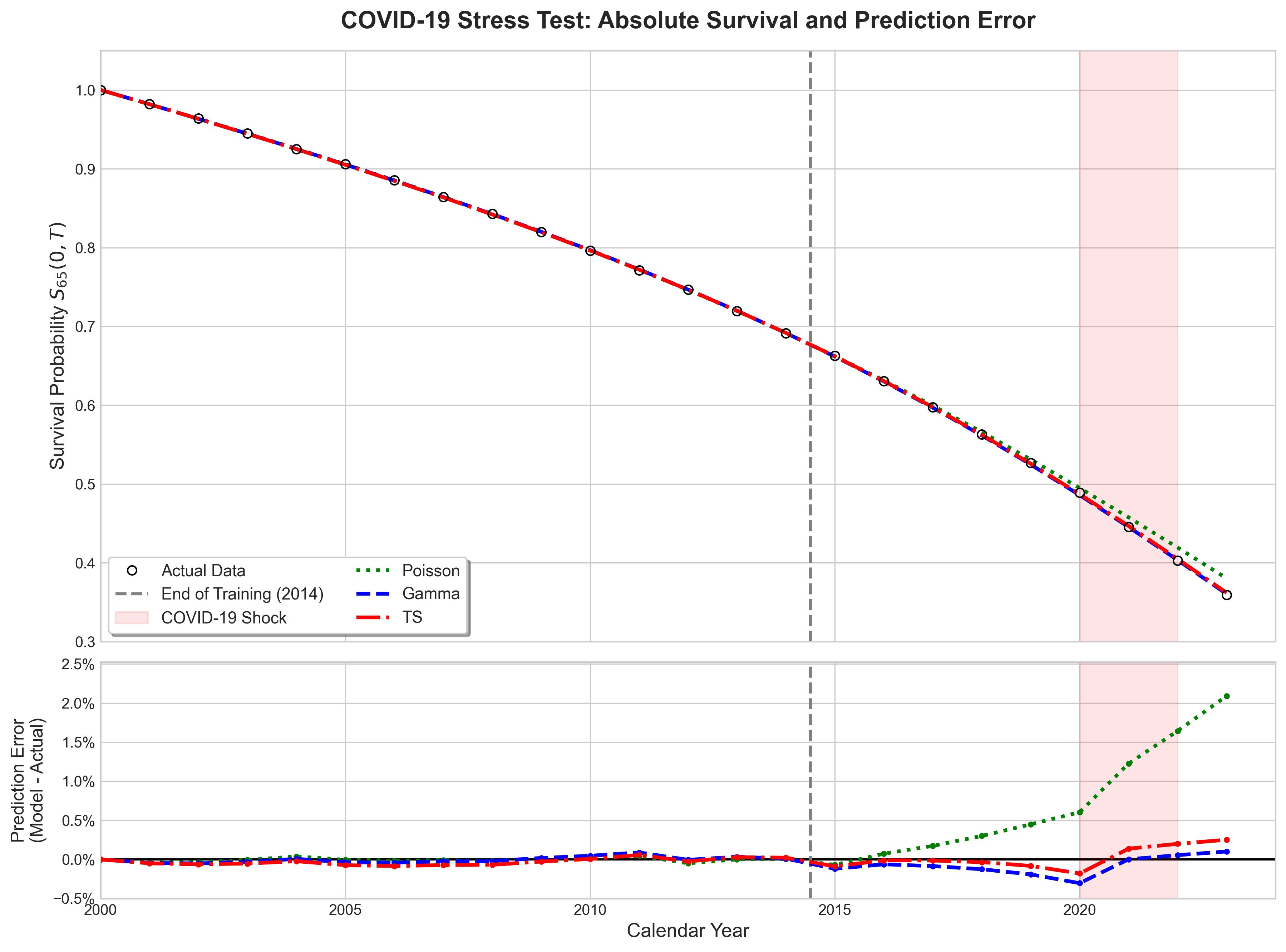}
    \caption{Prediction error under the COVID-19 shock.}
    \label{fig:covid_prediction_error}
\end{figure}

Since the COVID-19 out-of-sample period contains only a small number of annual
observations, the same Diebold--Mariano test used for the World War II windows is not
well supported in such test. We therefore report two other predictive metrics. First, the
out-of-sample coefficient of determination \(R^2_{oos}\) relative to the Poisson
baseline is defined as
\begin{equation}
    \label{eq:r_square}
     R_{\mathrm{oos},m}^2
 =
 1-
 \frac{\operatorname{MSE}_{\mathrm{oos},m}}
 {\operatorname{MSE}_{\mathrm{oos},\mathrm{Poisson}}},
\end{equation}
where \(m\) denotes either the Gamma or TS specification. A positive value
indicates that the alternative specification produces a lower
out-of-sample MSE than the Poisson benchmark. 
Second, the maximum absolute error is defined over the full out-of-sample period as
\begin{equation}
    \label{eq:maxae}
 \operatorname{MaxAE}_m
 =
 \max_{t\in \{2015,\ldots,2023\}}
 \left|
 S_t^{\mathrm{model},m}
 -
 S_t^{\mathrm{obs}}
 \right|.
\end{equation}
Relative
to the Poisson baseline, {the Gamma and TS specifications reduce the out-of-sample MSE by approximately
98\%, corresponding to out-of-sample $R^2$ values of approximately $98\%$.} These
results indicate that the infinite-activity specifications provide lower prediction errors
than the finite-activity Poisson benchmark in this COVID-19 stress test.

\begin{table}[H]
\resizebox{\textwidth}{!}{%
\begin{threeparttable}
\begin{tabular}{lcccc}
\toprule
\makecell{\textbf{Model}\\\textbf{Framework}} & \makecell{\textbf{In-sample}\\\textbf{MSE}} & \makecell{\textbf{Out-of-sample}\\\textbf{MSE}} & \makecell{\textbf{Out-of-sample}\\\textbf{MaxAE}} & \(\mathbf{R^2_{oos}}\) \textbf{(\%)} \\
\midrule
Poisson         & \(7.13 \times 10^{-8}\) & \(1.03 \times 10^{-4}\) & \(2.09 \times 10^{-2}\) & -- \\
Gamma           & \(1.53 \times 10^{-7}\) & \(2.03 \times 10^{-6}\) & \(3.03 \times 10^{-3}\) & 98.03\% \\
TS & \(2.70 \times 10^{-7}\) & \(1.89 \times 10^{-6}\) & \(2.52 \times 10^{-3}\) & 98.16\% \\
\bottomrule
\end{tabular}
\end{threeparttable}%
}
\centering
\caption{Predictive performance under the COVID-19 mortality shock: France 1935 cohort.
Baseline calibration : 2001-2014, out-of-sample calibration: 2015-2023.}
\label{tab:covid_stress_test}
\end{table}

\section{Conclusion}
\noindent
Long-range dependence in mortality risk represents a fundamental shift in actuarial science. The overwhelming empirical evidence across multiple countries, age groups, and time periods confirms that mortality rates possess memory properties that traditional Markovian models cannot capture. Our proposed stochastic mortality model includes a mixed driver of {a fractional L\'evy process} and Brownian motion. As revealed by the empirical calibration of our model to mortality shocks due to World War II and COVID-19, our modelling of stochastic mortality shows greater flexibility in the choice of {L\'evy} specification to capture LRD. In general, our mortality model {exhibits analytical tractability in deriving survival probabilities} and pricing annuity products. The non-Markovian nature of the {fractional L\'evy} process may pose some challenge in numerical pricing of exotic longevity-linked derivatives. We manage to design the singular value decomposition approximation under which the approximate {fractional L\'evy} kernel {admits a finite-dimensional time-inhomogeneous Markov state representation}. We provide convergence proof of the approximate solution under the singular value decomposition to the exact solution under the original {fractional L\'evy} kernel. 

Ignoring LRD in mortality modelling leads to systematic biases, like underestimation of life expectancy, overestimation of hedge effectiveness, and inadequate capital reserves. For insurers, pension funds, and regulators, incorporating LRD into mortality models is not merely an academic exercise but a practical necessity for sound financial management. The integration of LRD into standard actuarial practice will likely become essential for accurate pricing, effective risk management, and regulatory compliance. In summary, our stochastic mortality model exhibits empirical evidence to capture LRD. It also shows strong {analytical} and numerical tractability in actuarial valuations of survival probability and pricing of exotic longevity-linked derivatives.   

\section*{Acknowledgment}
\noindent The work of Yue Kuen Kwok was supported by Project 2023CX10X1 and the Guangzhou-HKUST (GZ) Joint Funding Program under Grant Number 2024A03J0630.
Yifan Ye was supported by the Guangdong and Hong Kong Universities ``1+1+1'' Joint Research Collaboration Scheme (Number R0800021-25) and the Beijing Normal-Hong Kong Baptist University Start Up Research Fund (Number UICR0700136-26).


\bibliography{references}

\appendix

\input{appendix_corrected}

\end{document}

%% file: appendix_corrected.tex

\setcounter{figure}{0}
\setcounter{table}{0}
\makeatletter
\@addtoreset{figure}{section}
\@addtoreset{table}{section}
\makeatother
\renewcommand{\theHfigure}{\thesection.\arabic{figure}}
\renewcommand{\theHtable}{\thesection.\arabic{table}}

\section{Proof of \Cref{prop:fractional_integral_existence}}
\label{app:fractional_integral_proof}
\noindent
{
The assumption \(f\mathbf 1_{[a,b]}\in\mathcal U_H^2([0,T])\) means precisely that
\(
K_T^H(f\mathbf 1_{[a,b]})\in L^2([0,T]).
\)
Since \(L^{\Pp}/s_L\) is a centered square-integrable L\'evy martingale, the following stochastic integral
\[
\frac{1}{s_L}
\int_0^T
K_T^H(f\mathbf 1_{[a,b]})(v)\,
{\dd L_v^{\Pp}}
\]
is well-defined in \(L^2(\Pp)\). This is the definition of
\(\int_a^b f(u)\,\dd\Upsilon_u^{H,\Pp}\). 
The integrand is deterministic, so the integral is centered. The isometry property
\[
\E^{\mathbb P}\left[
\left|
\int_a^b f(u)\,\dd\Upsilon_u^{H,\Pp}
\right|^2
\right]
=
\|K_T^H(f\mathbf 1_{[a,b]})\|_{L^2([0,T])}^2
\]
gives Eqs.~(\ref{eq:fractional_integral_mean}, \ref{eq:fractional_integral_variance}).
}

\section{Proof of \Cref{lem:fractional_levy_second_moment}}
\label{app:proof-fractional-levy-second-moment}
\noindent
By the definition of the fractional stochastic integral, we have
\[
\int_0^T f(u)\,\dd\Upsilon_u^{H,\Pp}
=
\frac{1}{{s_L}}
\int_0^T
\left(
 K_T^H f
\right)(v)
{\dd L_v^{\Pp}}.
\]
Since \(L^{\Pp}\) is a compensated square-integrable L\'evy process,
the deterministic integrals with
respect to \(L^{\Pp}\) satisfy the isometry
\[
\E^{\mathbb P}\left[
\left(
\frac{1}{{s_L}}
\int_0^T
a(v){\, \dd L_v^{\Pp}}
\right)
\left(
\frac{1}{{s_L}}
\int_0^T
b(v){\, \dd L_v^{\Pp}}
\right)
\right]
=
\int_0^T a(v)b(v)\,\dd v.
\]
Putting \(a=K_T^Hf\) and \(b=K_T^Hg\) into the above expression gives Eq.~\eqref{eq:kernel_cov_isometry}.
Equivalently, the covariance is the inner product
\[
\langle f,g\rangle_{\mathcal U_H^2}
:=
\langle K_T^H f,\,K_T^H g\rangle_{L^2([0,T])}.
\]

Next, we derive the fractional double-integral form. For \(H\in(1/2,1)\), we observe
\[
\frac12\left(t^{2H}+s^{2H}-|t-s|^{2H}\right)
=
H(2H-1)\int_0^s\int_0^t |y-z|^{2H-2}\,\dd z\,\dd y .
\]
Hence, for the indicator functions \(f=\mathbf 1_{(0,t]}\) and
\(g=\mathbf 1_{(0,s]}\), the covariance of the fractional L\'evy process can be written as
\[
\E^{\mathbb P}[\Upsilon_t^{H,\Pp}\Upsilon_s^{H,\Pp}]
=
\frac{1}{2}
\left(t^{2H}+s^{2H}-|t-s|^{2H}\right)
=
 H(2H-1)
\int_0^s\int_0^t |y-z|^{2H-2}\,\dd z\,\dd y .
\]

For the step functions \(f=\sum_i a_i\mathbf 1_{(t_{i-1},t_i]}\) and
\(g=\sum_j b_j\mathbf 1_{(s_{j-1},s_j]}\), the same formula follows by bilinearity of
covariance and linearity of the stochastic integral. In this class, the
double integral coincides with \(\langle f,g\rangle_{\mathcal U_H^2}\).
For ordinary functions \(f,g\in\mathcal U_H^2([0,T])\) for which
\(\int_0^T\int_0^T |f(y)g(z)||y-z|^{2H-2}\,\dd y\,\dd z<\infty\),
Eq.~\eqref{eq:fractional_cov_integral} is a Lebesgue integral. This is the setting of \Cref{lem:fractional_levy_second_moment}.

\section{Proof of \Cref{prop:mixed_driver_second_moments}}
\label{app:proof-mixed-driver-second-moments}
\noindent
Recall \(M_t=p_1W_t^{\mathbb P}+p_2\Upsilon_t^{H,\Pp}\), and since
\(\E^{\mathbb P}[W_t^{\mathbb P}]=0\) and \(\E^{\mathbb P}[\Upsilon_t^{H,\Pp}]=0\), we have
\[
\E^{\mathbb P}[M_t]
=
p_1\E^{\mathbb P}[W_t^{\mathbb P}]+p_2\E^{\mathbb P}[\Upsilon_t^{H,\Pp}]
=0.
\]
\noindent
The important assumption is that \(W^{\Pp}\) is independent of the
subordinator \(X^{\Pp}\), hence of \(L^{\Pp}\). Since
\(\Upsilon^{H,\Pp}\) is a (pathwise) Volterra functional of \(L^{\Pp}\), we observe
\[
\sigma(W_s^{\Pp}:s\le T)
\;\perp\;
\sigma(L_s^{\Pp}:s\le T)
\quad\Longrightarrow\quad
W^{\Pp}\;\perp\;\Upsilon^{H,\Pp}.
\]
For the covariance, using this independence property, we obtain
\begin{align*}
\operatorname{cov}(M_t,M_s)
&=
\operatorname{cov}\left(p_1W_t^{\mathbb P}+p_2\Upsilon_t^{H,\Pp},
p_1W_s^{\mathbb P}+p_2\Upsilon_s^{H,\Pp}\right)\\
&=
p_1^2\operatorname{cov}(W_t^{\mathbb P},W_s^{\mathbb P})
+p_2^2\operatorname{cov}(\Upsilon_t^{H,\Pp},\Upsilon_s^{H,\Pp})
+p_1p_2\operatorname{cov}(W_t^{\mathbb P},\Upsilon_s^{H,\Pp}) \\
&\quad +p_1p_2\operatorname{cov}(\Upsilon_t^{H,\Pp},W_s^{\mathbb P}).
\end{align*}
The last two covariance terms are zero by virtue of independence property. Since
\(
\operatorname{cov}(W_t^{\mathbb P},W_s^{\mathbb P})=t\wedge s,
\)
and by Eq.~\eqref{eq:fraclevy_properties}, we deduce
\[
\operatorname{cov}(\Upsilon_t^{H,\Pp},\Upsilon_s^{H,\Pp})
=
\frac{1}{2}
\left(t^{2H}+s^{2H}-|t-s|^{2H}\right).
\]
Combining the results, we finally obtain Eq.~\eqref{eq:mixed_cov}.

Next, to prove Eq.~\eqref{eq:mixed_integral_cov}, we consider
\begin{align*}
&\E^{\mathbb P}\left[
\left(\int_0^T f(u)\,\dd M_u\right)
\left(\int_0^T g(u)\,\dd M_u\right)
\right]\\
&\quad =
p_1^2
\E^{\mathbb P}\left[
\left(\int_0^T f(u)\,\dd W_u^{\mathbb P}\right)
\left(\int_0^T g(u)\,\dd W_u^{\mathbb P}\right)
\right]\\
&\quad+
p_2^2
\E^{\mathbb P}\left[
\left(\int_0^T f(u)\,\dd\Upsilon_u^{H,\Pp}\right)
\left(\int_0^T g(u)\,\dd\Upsilon_u^{H,\Pp}\right)
\right]\\
&\quad+
p_1p_2
\E^{\mathbb P}\left[
\left(\int_0^T f(u)\,\dd W_u^{\mathbb P}\right)
\left(\int_0^T g(u)\,\dd\Upsilon_u^{H,\Pp}\right)
\right] \\
&\quad+ 
p_1p_2
\E^{\mathbb P}\left[
\left(\int_0^T f(u)\,\dd\Upsilon_u^{H,\Pp}\right)
\left(\int_0^T g(u)\,\dd W_u^{\mathbb P}\right)
\right].
\end{align*}
The two cross terms vanish because \(W^{\mathbb P}\) and \(\Upsilon^{H,\Pp}\) are independent and both
integrals are centered. The Brownian isometry gives the first term in
Eq.~\eqref{eq:mixed_integral_cov}. {The second term is Eq.~\eqref{eq:fractional_cov_integral}, which requires the absolute-integrability assumptions stated in \Cref{lem:fractional_levy_second_moment}. Without that assumption, the fractional contribution is still given by the kernel isometry in Eq.~\eqref{eq:kernel_cov_isometry}.}

\section{Proof of \Cref{prop:long_range_dependence}}
\label{app:proof-long-range-dependence}
\noindent
For $n>m$, the Brownian increments over the two nonoverlapping time intervals $[n,n+m]$ and $[0,m]$ are independent. Hence only the fractional component contributes to the covariance. By Eq.~\eqref{eq:fraclevy_properties}, we have
\begin{equation*}
 \operatorname{cov}(\Delta_mM_n,\Delta_mM_0)
 =\frac{p_2^2}{2}\left[(n+m)^{2H}+(n-m)^{2H}-2n^{2H}\right].
\end{equation*}
Here, \(\Delta_mM_0=M_m-M_0\) is the change in the mixed driver over the
single interval \([0,m]\). The Brownian and fractional
L\'evy components are independent, with increment variances \(m\) and
\(m^{2H}\), respectively. Hence,
\begin{equation*}
 \operatorname{var}(\Delta_mM_0)=p_1^2m+p_2^2m^{2H}.
\end{equation*}
The Taylor expansion of $(n+m)^{2H}+(n-m)^{2H}-2n^{2H}$ gives
\begin{equation*}
 (n+m)^{2H}+(n-m)^{2H}-2n^{2H}
 \sim 2H(2H-1)m^2n^{2H-2},\qquad n\to\infty.
\end{equation*}
Therefore, we obtain
\begin{equation*}
 \mathrm{corr}(\Delta_mM_n,\Delta_mM_0)
 \sim C_m n^{2H-2},
\end{equation*}
where
\begin{equation*}
 C_m=\frac{p_2^2H(2H-1)m^2}{p_1^2m+p_2^2m^{2H}}.
\end{equation*}
The assumption \(p_2\ne0\) makes the numerator strictly positive and
the denominator strictly positive, so \(C_m>0\). In particular, the
correlations are eventually positive with increasing $n$. Long-range dependence is
understood in the sense of
\Cref{prop:long_range_dependence}: for the fixed increment length
\(m>0\), we have
\[
\sum_{n=1}^{\infty}
\bigl|\mathrm{corr}(\Delta_mM_n,\Delta_mM_0)\bigr|
=\infty.
\]
Since $2H-2\in(-1,0)$, the series $\sum_{n\geq1}n^{2H-2}$ diverges, and
the same holds for the infinite sum of correlations in the above. This proves Eq.~\eqref{eq:lrd_result_full}.

\section{Proof of \Cref{thm:conditional_exponential_formula}}
\label{app:proof-conditional-exponential-formula}
\noindent
By definition of the fractional integral, we obtain
\begin{equation*}
 \int_t^s f(u)\,\dd\Upsilon_u^{H,\Pp}
 =\frac{1}{s_L}\int_0^T  K_T^H(f1_{[t,s]})(v)\,{\dd L_v^{\Pp}}.
\end{equation*}
Since $K_T^H(f1_{[t,s]})(v)=0$ for $v>s$, the integral is split into
\begin{equation*}
 \int_t^s f(u)\,\dd\Upsilon_u^{H,\Pp}
 =\frac{1}{s_L}\int_0^t  K_T^H(f1_{[t,s]})(v)\,{\dd L_v^{\Pp}}
 +\frac{1}{s_L}\int_t^s  K_T^H(f1_{[t,s]})(v)\,{\dd L_v^{\Pp}}.
\end{equation*}
The first term is $\mathcal F_t$-measurable. The second term depends
only on future increments of $L^{\Pp}$ over $(t,s]$. Write
\(q(v):=K_T^H(f\mathbf 1_{[t,s]})(v)\). The L\'evy--Khintchine formula
for the deterministic integrands requires
\[
\frac{\eta}{s_L}q(v)\in\operatorname{dom}\Psi_L^{\Pp}
\quad\text{for a.e.\ }v\in[t,s],
\qquad
\int_t^s
\Bigl|\Psi_L^{\Pp}\Bigl(\frac{\eta}{s_L}q(v)\Bigr)\Bigr|\,\dd v
<\infty.
\]
These are the conditions imposed on \(\eta\) in
\Cref{thm:conditional_exponential_formula}. They are not implied by
square integrability of \(L^{\Pp}\) alone: for the Gamma specification,
the effective domain is \(\{z:z<\beta\}\); and for the TS specification, it is
\(\{z:z<\lambda_{\mathrm{TS}}\}\). If \(\eta\) is complex, the formula
uses the analytic continuation of \(\Psi_L^{\Pp}\) on a neighbourhood
of the range of \((\eta/s_L)q\), with real parts remaining in the
interior of the exponential-moment domain. Under these conditions,
independent increments give
\begin{equation*}
 \E^{\mathbb P}\left[\exp\left(\frac{\eta}{s_L}\int_t^s  K_T^H(f1_{[t,s]})(v)\,{\dd L_v^{\Pp}}\right)\middle|\mathcal F_t\right]
 =\exp\left(\int_t^s\Psi_L^{\mathbb P}\left(\frac{\eta}{s_L} K_T^H(f1_{[t,s]})(v)\right)\,\dd v\right).
\end{equation*}
Multiplying by the exponential of the first term proves Eq.~\eqref{eq:conditional_exp_full_history}.

\section{Proof of \Cref{thm:svd-lifting-convergence}}
\label{app:svd-continuous-proof}
\noindent
Throughout this proof, $\mathcal T>0$ is fixed and
$\mathcal D_{\mathcal T}=\{(t,v):0\leq v\leq t\leq\mathcal T\}$.
Since $L^{\Pp}/s_L$ is a centered square-integrable L\'evy
martingale under measure $\mathbb P$ with unit variance at time one, the deterministic-integrand
isometry gives
\begin{equation*}
    \mathbb E^{\mathbb P}
    \left[
        \left(
            \frac{1}{s_L}
            \int_0^{\mathcal T}q(v)
            {\, \dd L_v^{\Pp}}
        \right)^2
    \right]
    =
    \int_0^{\mathcal T}|q(v)|^2\, \dd v
\end{equation*}
for every $q\in L^2([0,\mathcal T])$.

For almost every fixed $t$, we apply the above equation to
$q(v)=\bigl(K_H(t,v)-K_H^{(r)}(t,v)\bigr)\mathbf 1_{[0,t]}(v)$.
This gives Eq.~\eqref{eq:svd-pointwise-error}. 
Integrating Eq.~\eqref{eq:svd-pointwise-error} over
\(t\in[0,\mathcal T]\) gives
\begin{align*}
    &\int_0^{\mathcal T}
    \E^{\Pp}
    \left[
        \left|
            \Upsilon_t^{H,\Pp}
            -
            \Upsilon_t^{H,\Pp,(r)}
        \right|^2
    \right]
    \,\mathrm{d}t\\
    &\quad \quad=
    \int_0^{\mathcal T}
    \int_0^t
    \left|
        K_H(t,v)-K_H^{(r)}(t,v)
    \right|^2
    \,\mathrm{d}v\,\mathrm{d}t =
    \left\|
        K_H-K_H^{(r)}
    \right\|_{L^2(\mathcal D_{\mathcal T})}^2.
\end{align*}
By Eq.~\eqref{eq:kernel-svd-triangle-bound}, we obtain
\[
    \left\|
        K_H-K_H^{(r)}
    \right\|_{L^2(\mathcal D_{\mathcal T})}^2
    \leq
    \sum_{i>r}\lambda_i^2
    \longrightarrow0.
\]
This proves the integrated mean-square convergence in
Eq.~\eqref{eq:svd-integrated-error}. The identity is an equality in
\(L^2(\Omega\times[0,\mathcal T])\). It does not by itself yield
mean-square convergence at every fixed \(t\), nor uniform-in-\(t\)
convergence, nor pathwise or \(\sup_{t\le\mathcal T}\) convergence.
Item~(i) of \Cref{thm:svd-lifting-convergence} is a slice identity for
almost every \(t\), not a statement that the process converges at each
\(t\).

It remains to verify that $K_H$ belongs to
$L^2(\mathcal D_{\mathcal T})$. For $0<v<t\leq\mathcal T$, we obtain
\[
    K_H(t,v)
    =
    c_H
    v^{\frac12-H}
    \int_v^t
    u^{H-\frac12}(u-v)^{H-\frac32}\, \dd u.
\]
Since $u\leq\mathcal T$, we have
\begin{align*}
    \int_v^t
    u^{H-\frac12}(u-v)^{H-\frac32}\, \dd u
    &\leq
    \mathcal T^{H-\frac12}
    \int_v^t(u-v)^{H-\frac32}\, \dd u =
    \frac{\mathcal T^{H-\frac12}}{H-\frac12}
    (t-v)^{H-\frac12}.
\end{align*}
Hence, for
\[
    C_{H,\mathcal T}
    :=
    \frac{|c_H|\mathcal T^{H-\frac12}}
    {H-\frac12},
\]
we have
\[
    |K_H(t,v)|
    \leq
    C_{H,\mathcal T}
    v^{\frac12-H}
    (t-v)^{H-\frac12}.
\]
Performing the integration, we obtain
\begin{align*}
    \int_0^{\mathcal T}\int_0^t|K_H(t,v)|^2\, \dd v\, \dd t
    &\leq
    C_{H,\mathcal T}^2
    \int_0^{\mathcal T}\int_0^t
    v^{1-2H}(t-v)^{2H-1}\, \dd v\, \dd t.
\end{align*}
Taking $v=ty$, the integral can be expressed in terms of the Beta Integral as follows:
\[
    \int_0^t
    v^{1-2H}(t-v)^{2H-1}\, \dd v
    =
    t\,\mathrm B(2-2H,2H),
\]
which is finite for $H\in(1/2,1)$. Integrating in \(t\) then gives
the explicit bound
\[
\int_0^{\mathcal T}\int_0^t|K_H(t,v)|^2\,\dd v\,\dd t
\leq
C_{H,\mathcal T}^2\,\mathrm B(2-2H,2H)\,
\frac{\mathcal T^2}{2}
<\infty.
\]
Thus, we conclude that
$K_H\in L^2(\mathcal D_{\mathcal T})$. The same calculation for fixed
$t$ shows that $K_H(t,\cdot)\in L^2([0,t])$ for every
$t\in[0,\mathcal T]$. After the zero extension to \([0,\mathcal T]^2\),
the Hilbert--Schmidt singular expansion of \(\mathcal K\) is taken on
the square. The inequality
\(\|K_H-K_H^{(r)}\|_{L^2(\mathcal D_{\mathcal T})}^2
\le\sum_{i>r}\lambda_i^2\) in Eq.~\eqref{eq:kernel-svd-triangle-bound}
is the restriction of that square-norm identity to the causal triangle,
not an Eckart--Young identity posed only on \(\mathcal D_{\mathcal T}\).
Finally, since Eq.~\eqref{eq:svd-truncated-kernel} is a finite sum, we obtain
\begin{align*}
   \frac{1}{s_L} \int_0^tK_H^{(r)}(t,v){\, \dd L_v^{\Pp}}
    &=
    \frac{1}{s_L}
    \sum_{i=1}^{r}
    \lambda_i\phi_i(t)
    \int_0^t\psi_i(v){\, \dd L_v^{\Pp}}
    =
    \sum_{i=1}^{r}
    \lambda_i\phi_i(t)U_t^{(r,i)},
\end{align*}
which is a finite-factor representation
\(\Upsilon_t^{H,\Pp,(r)}=F_r(t,U_t^{(r,1)},\ldots,U_t^{(r,r)})\)
with \(F_r(t,u)=\sum_{i=1}^r\lambda_i\phi_i(t)u_i\), after a measurable
version of each \(\phi_i\) is chosen. The Hilbert--Schmidt theory yields
only \(\phi_i,\psi_i\in L^2([0,T])\), as equivalence classes. It does
not yield continuity, absolute continuity, or finite variation of
\(\phi_i\). Consequently, we obtain the following results:
\begin{itemize}
\item The vector \((t,U_t^{(r,1)},\ldots,U_t^{(r,r)})\) is a
time-inhomogeneous Markov process, being driven by the additive
increments of \(L^{\Pp}\) with deterministic weights \(\psi_i\);
\item \(\Upsilon^{H,\Pp,(r)}\) is a deterministic function of that
augmented state, but need not itself be Markov;
\item \(\Upsilon^{H,\Pp,(r)}\) is a semimartingale only under an extra
regularity assumption on \(\phi_i\), such as the condition
\(\phi_i\in C^1([0,T])\) used for the differential formula in
\Cref{sec:svd-lifting}. That assumption is not part of the
Hilbert--Schmidt truncation.
\end{itemize}
The numerical scheme uses a finite-dimensional discrete state
representation on the implementation grid and does not rely on a
continuous-time semimartingale lift.


\section{Proof of \Cref{prop:survival_prob_p}}
\label{app:proof-survival-prob-p}
\noindent
For $s\geq t$, solving Eq.~\eqref{eq:sde} gives
\begin{equation*}
 \mu_s
 =
 \mu_t e^{a(s-t)}
 +
 \sigma
 \int_t^s
 e^{a(s-u)}\,\dd M_u.
\end{equation*}
Integrating over $[t,T]$ and applying stochastic Fubini's theorem yield
\begin{equation*}
 \int_t^T
 \mu_s\,\dd s
 =
 \mu_tA(t,T)
 +
 \sigma
 \int_t^T
 f(u,T)\,\dd M_u,
\end{equation*}
where
\[
 f(u,T)
 =
 \frac{e^{a(T-u)}-1}{a}
 \quad(a\ne0),
 \qquad
 f(u,T)=T-u
 \quad(a=0),
\]
and \(A(t,T)=f(t,T)\). The case \(a=0\) is the continuous extension of
the same formulae. The Brownian Fubini theorem applies because
\(f(\cdot,T)\in L^2([t,T])\). For the fractional L\'evy part, it is
enough that the transformed kernel lie in \(L^2\). Also, we have
\[
\int_t^T
\bigl|(K_{t,T}^H f)(v)\bigr|^2\,\dd v
<\infty,
\]
which holds because \(f(\cdot,T)\) is bounded on \([t,T]\) and
\(K_H\in L^2(D_T)\).
Using
\[
 M_u
 =
 p_1W_u^{\Pp}
 +
 p_2\Upsilon_u^{H,\Pp},
\]
we obtain
\begin{align*}
 \int_t^T
 \mu_s\,\dd s
 =
 &\,
 \mu_tA(t,T)
 +
 \sigma p_1
 \int_t^T
 f(u,T)\,\dd W_u^{\Pp} +
 \sigma p_2
 \int_t^T
 f(u,T)\,\dd\Upsilon_u^{H,\Pp}.
\end{align*}

By the definition of the fractional integral, we have 
\begin{align*}
 \int_t^T
 f(u,T)\,\dd\Upsilon_u^{H,\Pp}
 =
 &\,
 \frac{1}{s_L}
 \int_0^t
 (K_{t,T}^H f)(v) \, 
 {\dd L_v^{\Pp}} + 
 \frac{1}{s_L}
 \int_t^T
 (K_{t,T}^H f)(v) \, 
 {\dd L_v^{\Pp}}.
\end{align*}
The first term is $\mathcal F_t$-measurable, whereas the second term
depends only on the future increments of $L^{\Pp}$ over $(t,T]$.
The future Brownian increment over \((t,T]\) is independent of
\(\mathcal F_t\) and the future L\'evy increment since
\(W^{\Pp}\) and \(L^{\Pp}\) are independent under measure \(\Pp\). The two
conditional exponential factors may therefore be multiplied.

The Brownian integral is conditionally Gaussian with mean zero and
variance
\[
 \int_t^T f(u,T)^2\,\dd u.
\]
Hence its conditional exponential expectation gives
$B_{\mathrm{Brown}}^{\Pp}(t,T)$.

For the future L\'evy term, provided that
\[
-\frac{\sigma p_2}{s_L}(K_{t,T}^H f)(v)
\in\operatorname{dom}\Psi_L^{\Pp}
\quad\text{for a.e.\ }v\in[t,T]
\]
the resulting cumulant is integrable on \([t,T]\). Observing independent increments and applying the
L\'evy--Khintchine formula give
\begin{align*}
&\E^{\Pp}
\left[
 \exp\left(
 -\frac{\sigma p_2}{s_L}
 \int_t^T
 (K_{t,T}^H f)(v)\,
 \dd L_v^{\Pp}
 \right)
 \middle|
 \mathcal F_t
\right]
=
\exp\left[
 \int_t^T
 \Psi_L^{\Pp}
 \left(
 -\frac{\sigma p_2}{s_L}
 (K_{t,T}^H f)(v)
 \right)
 \,\dd v
\right].
\end{align*}
Combining the deterministic term, historical integral, Brownian
integral, and future L\'evy term gives
Eq.~\eqref{eq:affine_survival}. At $t=0$, the historical integral
vanishes, which gives Eq.~\eqref{eq:physical_survival_zero}.

\section{Proof of \Cref{cor:physical_compensated_cumulants}}
\label{app:proof-physical-compensated-cumulants}
\noindent
For a subordinator \(X^{\Pp}\) with cumulant \(\Psi_X^{\Pp}\), the compensated driver is
\[
 L_t^{\Pp}
 =
 X_t^{\Pp}-t(\Psi_X^{\Pp})'(0).
\]
Hence the cumulant of the compensated driver is
\[
 \Psi_L^{\Pp}(z)
 =
 \log \E^{\Pp}\left[e^{zL_1^{\Pp}}\right]
 =
 \Psi_X^{\Pp}(z)-z(\Psi_X^{\Pp})'(0).
\]
\noindent
For the Poisson subordinator, the uncompensated cumulant is
\[
 \Psi_{X,P}^{\Pp}(z)=\lambda(e^z-1),
 \qquad
 (\Psi_{X,P}^{\Pp})'(0)=\lambda.
\]
Therefore, we have
\[
 \Psi_P^{\Pp}(z)
 =
 \Psi_{X,P}^{\Pp}(z)-z(\Psi_{X,P}^{\Pp})'(0)
 =
 \lambda(e^z-1-z),
\]
which gives Eq.~\eqref{eq:pois_cum}.
\noindent
For the Gamma subordinator, the uncompensated cumulant is
\[
 \Psi_{X,G}^{\Pp}(z)
 =
 -\alpha\log\left(1-\frac{z}{\beta}\right),
 \qquad z<\beta,
\]
and
\[
 (\Psi_{X,G}^{\Pp})'(0)=\frac{\alpha}{\beta}.
\]
Therefore, we have
\[
 \Psi_G^{\Pp}(z)
 =
 \Psi_{X,G}^{\Pp}(z)-z(\Psi_{X,G}^{\Pp})'(0)
 =
 -\alpha\log\left(1-\frac{z}{\beta}\right)-z\frac{\alpha}{\beta},
\]
which gives Eq.~\eqref{eq:gamma_cum}.
\noindent
For the TS subordinator, with \(0<\kappa<1\) and
\[
 \Psi_{X,TS}^{\Pp}(z)
 =
 c\Gamma(-\kappa)
 \left[
 (\lambda_{TS}-z)^\kappa-\lambda_{TS}^{\kappa}
 \right],
 \qquad z<\lambda_{TS}.
\]
Evaluation of the derivative at zero gives
\[
 (\Psi_{X,TS}^{\Pp})'(0)
 =
 -c\kappa\Gamma(-\kappa)\lambda_{TS}^{\kappa-1}.
\]
Therefore, we have
\[
\begin{aligned}
 \Psi_{TS}^{\Pp}(z)
 &=
 \Psi_{X,TS}^{\Pp}(z)
 -
 z(\Psi_{X,TS}^{\Pp})'(0)\\
 &=
 c\Gamma(-\kappa)
 \left[
 (\lambda_{TS}-z)^\kappa-\lambda_{TS}^{\kappa}
 +z\kappa\lambda_{TS}^{\kappa-1}
 \right],
\end{aligned}
\]
which gives Eq.~\eqref{eq:ts_cum}.



\section{Proof of \Cref{prop:q_dynamics}}
\label{app:proof-q-dynamics}
\noindent
By Girsanov's theorem, the Brownian density in Eq.~\eqref{eq:girsanov}
implies that
\[
 W_s^\Q
 =
 W_s^{\Pp}-\int_0^s\gamma_u\,\dd u
\]
is a standard Brownian motion under \(\Q\). Equivalently,
\(\dd W_s^{\Pp}=\dd W_s^\Q+\gamma_s\,\dd s\). Substituting this relation into the
Brownian part under the physical measure gives
\[
 a\mu_s\,\dd s+\sigma p_1\,\dd W_s^{\Pp}
 =
 (a\mu_s+\sigma p_1\gamma_s)\,\dd s
 +
 \sigma p_1\,\dd W_s^\Q .
\]

For the jump part, the Esscher density in Eq.~\eqref{eq:esscher_density}
is applied to the same uncompensated subordinator \(X\), now reweighted
under \(\Q\). The notation \(X^{\Q}\) below refers to this process
under its \(\Q\)-law, not to a different pathwise integrator. The
parameter \(\theta\) is assumed to lie in the interior of
\(\operatorname{dom}\Psi_X^{\Pp}\), and every cumulant argument \(z\)
used below is required to satisfy
\(z+\theta\in\operatorname{dom}\Psi_X^{\Pp}\). For the Gamma specification, this means
\(\theta<\beta\) and \(z+\theta<\beta\); for the TS specification, we have
\(\theta<\lambda_{\mathrm{TS}}\) and \(z+\theta<\lambda_{\mathrm{TS}}\).
When \(\theta\) is constant, for \(s\ge t\), we obtain
\[
 \E^\Q\left[
 e^{z(X_s^\Q-X_t^\Q)}
 \middle|\mathcal F_t
 \right]
 =
 \exp\left(
 (s-t)\left[
 \Psi_X^{\Pp}(z+\theta)-\Psi_X^{\Pp}(\theta)
 \right]
 \right).
\]
Thus, under \(\Q\), the subordinator \(X\) has cumulant
\(\Psi_X^\Q\) given by Eq.~\eqref{eq:q_uncompensated_cumulant}.
By Eq.~\eqref{eq:p-q-compensated-drivers}, \(L^\Q\) is centered under
\(\Q\), with cumulant \(\Psi_L^\Q\) given by
Eq.~\eqref{eq:q_levy_cumulant}. Moreover,
Eq.~\eqref{eq:p-q-levy-relation} implies
\[
\Upsilon_s^{H,\Pp}
=
\frac{1}{s_L}\int_0^sK_H(s,v)\,\dd L_v^\Q
+
\frac{\Delta m_L}{s_L}\int_0^sK_H(s,v)\,\dd v
=
\Upsilon_s^{H,\Q}
+
\frac{\Delta m_L}{s_L}\kappa_H(s).
\]
The Molchan--Golosov kernel scales as
\(K_H(ct,cv)=c^{H-1/2}K_H(t,v)\) for \(c>0\). Hence, we obtain
\[
\kappa_H(t)
=
\int_0^t K_H(t,v)\,\dd v
=
t^{H+1/2}\kappa_H(1).
\]
In particular, \(\kappa_H\) is absolutely continuous on \([0,T]\), and observe
\[
b_H(t)
=
\kappa_H'(t)
=
\bigl(H+\tfrac12\bigr)t^{H-1/2}\kappa_H(1)
\]
almost everywhere. Since \(H>1/2\), the derivative is continuous on
\((0,T]\) and integrable near the origin. Taking differentials is
therefore justified in the Lebesgue sense, and
\[
\dd\Upsilon_s^{H,\Pp}
=
\dd\Upsilon_s^{H,\Q}
+
\frac{\Delta m_L}{s_L}b_H(s)\,\dd s.
\]
Combining this identity with the Brownian measure change gives
Eq.~\eqref{eq:q_dynamics}. The Girsanov factor \(Z^W\) depends only on
\(W^{\Pp}\) and the Esscher factor \(Z^L\) depends only on \(X\).
Independence of \(W^{\Pp}\) and \(X\) under \(\Pp\) is therefore
preserved under measure \(\Q\).

\section{Proof of \Cref{prop:yt_characteristic_function}}
\label{app:proof-yt-characteristic-function}
\noindent
The decomposition of $Y_{t,T}$ separates the
$\mathcal F_t$-measurable terms from the future Brownian and L\'evy
increments. The deterministic and historical terms contribute
\begin{equation*}
 \exp\left(
 -iu\mu_tA(t,T)
 -iuD_\gamma(t,T)
 -iuD_L(t,T)
 -iu\frac{\sigma p_2}{s_L}
 \int_0^t
 (K_{t,T}^H f)(v)\,
 \dd L_v^\Q
 \right).
\end{equation*}
\noindent
The Brownian integral is conditionally Gaussian with mean zero and
variance $V_f(t,T)$, and therefore contributes
\begin{equation*}
 \exp\left(
 -\frac12u^2\sigma^2p_1^2V_f(t,T)
 \right).
\end{equation*}
\noindent
The future Brownian and L\'evy increments remain conditionally
independent given \(\mathcal F_t\) under \(\Q\), because the density
factorizes as \(Z=Z^WZ^L\) and the two drivers are independent under
\(\Pp\). For the future L\'evy term, provided that
\(
-iu\frac{\sigma p_2}{s_L}(K_{t,T}^H f)(v)
\)
lies in the domain of the analytic continuation of \(\Psi_L^\Q\) for
almost every \(v\in[t,T]\) with integrable cumulant,
independent increments under
\(\Q\) and the L\'evy--Khintchine formula give
\begin{align*}
&\E^\Q
\left[
 \exp\left(
 -iu\frac{\sigma p_2}{s_L}
 \int_t^T
 (K_{t,T}^H f)(v)\,
 \dd L_v^\Q
 \right)
 \middle|
 \mathcal F_t
\right]
=
\exp\left(
 \int_t^T
 \Psi_L^\Q
 \left(
 -iu\frac{\sigma p_2}{s_L}
 (K_{t,T}^H f)(v)
 \right)
 \,\dd v
\right).
\end{align*}
Multiplying these terms together gives
Eq.~\eqref{eq:yt_characteristic_function}.

\section{Proof of \Cref{thm:q-lifted-survival-index-convergence}}
\label{app:svd-q-survival-proof}
\noindent
We define the rank-\(r\) transformed-kernel error by
\[
q_r(v)
:=
g_{0,T}^{(r)}(v)-g_{0,T}(v).
\]
By the adjoint relation used in
Eqs.~\eqref{eq:svd-original-survival-kernel}--\eqref{eq:svd-approximating-survival-kernel}, we have
\[
q_r(v)
=
\int_v^T
\bigl(K_H^{(r)}(u,v)-K_H(u,v)\bigr)f(u,T)\,\dd u.
\]
Applying the Cauchy--Schwarz inequality yields
\[
\bigl|q_r(v)\bigr|^2
\leq
\Biggl(
\int_v^T
\bigl|K_H^{(r)}(u,v)-K_H(u,v)\bigr|^2\,\dd u
\Biggr)
\Biggl(
\int_v^T
\bigl|f(u,T)\bigr|^2\,\dd u
\Biggr).
\]
Integrating in \(v\) then gives
\[
\|q_r\|_{L^2([0,T])}^2
\leq
\|f(\cdot,T)\|_{L^2([0,T])}^2
\,
\|K_H^{(r)}-K_H\|_{L^2(D_T)}^2.
\]
The second factor tends to zero by
Eq.~\eqref{eq:kernel-svd-triangle-bound}, and \(f(\cdot,T)\) is square
integrable on \([0,T]\). Hence, \(q_r\to0\) in \(L^2([0,T])\), which is
the assumption stated in
\Cref{thm:q-lifted-survival-index-convergence}.
By the representation of $Y_{0,T}$ above and
Eq.~\eqref{eq:q-lifted-log-survival-index},
\[
    Y_{0,T}^{(r)}-Y_{0,T}
    =
    -\frac{\sigma p_2}{s_L}
    \left[
    \int_0^T
    q_r(v)\,\dd L_v^\Q
    +
    \Delta m_L\int_0^Tq_r(v)\,\dd v
    \right].
\]
For a deterministic function \(q\in L^2([0,T])\), the independent
increments and centering of \(L^\Q\) under \(\mathbb Q\) give
\[
\begin{aligned}
    \mathbb E^{\mathbb Q}
    \left[
        \left|
            \int_0^Tq(v)\,\dd L_v^\Q
            +
            \Delta m_L\int_0^Tq(v)\,\dd v
        \right|^2
    \right]
    &=
    v_L^{\mathbb Q}
    \int_0^Tq(v)^2\,\dd v
    +
    |\Delta m_L|^2
    \left(
        \int_0^Tq(v)\,\dd v
    \right)^2.
\end{aligned}
\]
By the Cauchy--Schwarz inequality, we deduce
\[
    \left(
        \int_0^Tq(v)\,\dd v
    \right)^2
    \leq
    T
    \int_0^Tq(v)^2\,\dd v.
\]
Applying these identities with \(q=q_r\) gives
Eq.~\eqref{eq:q-log-survival-convergence-bound}. Combined with
\(q_r\to0\) in \(L^2([0,T])\), it follows that
\[
    Y_{0,T}^{(r)}
    \longrightarrow
    Y_{0,T}
    \quad\text{in }L^2(\mathbb Q),
\]
and hence also in probability.

By continuity of the exponential function, we obtain
\[
    S_T^{(r)}
    =
    e^{Y_{0,T}^{(r)}}
    \longrightarrow
    e^{Y_{0,T}}
    =
    S_T
\]
in probability. Convergence in probability of \(S_T^{(r)}\) does not
by itself yield \(L^1(\Q)\) convergence. Eq.~\eqref{eq:q-uniform-survival-moment}
is the uniform integrability assumption used for Vitali's theorem. A
sufficient condition is the existence of \(\varepsilon>0\) such that
\[
\sup_{r\ge1}
\E^{\Q}\bigl[(S_T^{(r)})^{1+\varepsilon}\bigr]
<\infty.
\]
For the Poisson specification the cumulant is entire, so the bound
holds as soon as \(g_{0,T}^{(r)}\) remains bounded in \(L^2\). For
the Gamma and TS specifications, the same moment requires that the real
cumulant arguments
\(-(1+\varepsilon)\sigma p_2 s_L^{-1}g_{0,T}^{(r)}\) remain in a
compact subset of the interior of \(\operatorname{dom}\Psi_L^{\Q}\).
This is an additional restriction on \((\sigma,p_2,\theta)\) and on the
range of the transformed kernel. It is not implied by square
integrability of \(L^{\Q}\) alone. Under
Eq.~\eqref{eq:q-uniform-survival-moment}, Vitali's convergence theorem
gives
\[
    S_T^{(r)}
    \longrightarrow
    S_T
    \quad\text{in }L^1(\mathbb Q).
\]
Taking expectations gives the final result.

\section{Proof of \Cref{prop:caplet_fourier}}
\label{app:proof-caplet-fourier}
\noindent
We introduce the damped value with the damping parameter \(\delta\):
\[
 \widetilde{\mathcal C}_\delta^\Q(k,T)
 :=
 e^{\delta k}\mathcal C^\Q(e^k,T),
 \qquad \delta>0.
\]
The assumption \(\E^\Q[e^{(\delta+1)Y_{0,T}}]<\infty\) in
\Cref{prop:caplet_fourier} is the damping condition. It makes
\(\widetilde{\mathcal C}_\delta^\Q(\cdot,T)\) integrable, justifies
Fubini below, and places the complex argument \(v-i(\delta+1)\) in the
domain of the analytic continuation of \(\phi_{0,T}^\Q\). For the Gamma and
TS specifications, this again requires that the corresponding real
cumulant arguments remain in the interior of
\(\operatorname{dom}\Psi_L^\Q\). Under that condition, the Fourier
transform of \(\widetilde{\mathcal C}_\delta^\Q\) is
\[
 \widehat{\mathcal C}_\delta^\Q(v,T)
 =
 \int_{-\infty}^{\infty}
 e^{ivk}\widetilde{\mathcal C}_\delta^\Q(k,T)\,\dd k.
\]
Using Fubini's theorem and the payoff identity above, we obtain
\[
\begin{aligned}
 \widehat{\mathcal C}_\delta^\Q(v,T)
 &=
 \E^\Q\left[
 \int_{-\infty}^{Y_{0,T}}
 e^{ivk}e^{\delta k}(e^{Y_{0,T}}-e^k)\,\dd k
 \right] \\
 &=
 \E^\Q\left[
 e^{(iv+\delta+1)Y_{0,T}}
 \left(
 \frac{1}{iv+\delta}
 -
 \frac{1}{iv+\delta+1}
 \right)
 \right] \\
 &=
 \frac{
 \phi_{0,T}^\Q\left(v-i(\delta+1)\right)
 }{
 \delta^2+\delta-v^2+i(2\delta+1)v
 }.
\end{aligned}
\]
Taking the Fourier inversion, which is legitimate in \(L^1\) because
the damped transform is integrable under the same exponential-moment
assumption, and knowing that \(\mathcal C^\Q(e^k,T)\) is real-valued,
\[
 \mathcal C^\Q(e^k,T)
 =
 \frac{e^{-\delta k}}{2\pi}
 \int_{-\infty}^{\infty}
 e^{-ivk}\widehat{\mathcal C}_\delta^\Q(v,T)\,\dd v \\
 =
 \frac{e^{-\delta k}}{\pi}
 \int_0^\infty
 \operatorname{Re}\left[
 e^{-ivk}\widehat{\mathcal C}_\delta^\Q(v,T)
 \right]\,\dd v.
\]
Substituting \(\widehat{\mathcal C}_\delta^\Q(v,T)\) gives
Eq.~\eqref{eq:caplet_fourier}. Multiplication by \(NP(0,T)\) gives
Eq.~\eqref{eq:caplet_price_zero}.

\section{Proof of \Cref{cor:lifted-price-convergence}}
\label{app:proof-lifted-price-convergence}
\noindent
For the life annuity, the definitions of the exact and rank-\(r\)
values, together with the triangle inequality, give
\[
\left|
\bar a^{(r)}(0,T)-\bar a(0,T)
\right|
 =
\left|
\int_0^T P(0,s)
\E^\Q\left[S_s^{(r)}-S_s\right]\dd s
\right|
\leq
\int_0^T
P(0,s)
\E^\Q\left[
\left|S_s^{(r)}-S_s\right|
\right]\dd s.
\]
The integrand converges to zero for each \(s\) by
\Cref{thm:q-lifted-survival-index-convergence} applied at maturity
\(s\). However, pointwise \(L^1(\Q)\) convergence does not by itself
justify passing the limit through the integral with respect to \(s\).
If the exact and approximating mortality intensities were nonnegative,
then \(S_s,S_s^{(r)}\in[0,1]\), which would provide the uniform bound
\(\left|S_s^{(r)}-S_s\right|\leq 1\).
In the present model, the Brownian and L\'evy components may cause the
intensities to take negative values, so this bound is not available.
We therefore impose the additional domination condition in
Eq.~\eqref{eq:annuity-lifting-dominating-condition}, which provides an
integrable dominating function independent of \(r\). Dominated
convergence then gives
\(\bar a^{(r)}(0,T)\to\bar a(0,T)\).

For the longevity corridor contract, the payoff function is given by
\[
h(x)=\min\{S_u-S_l,\max(x-S_l,0)\}
\]
which satisfies
\(\lvert h(x)-h(y)\rvert\leq\lvert x-y\rvert\). Therefore, we deduce
\[
\left|
V_0^{\mathrm{corridor},(r)}-V_0^{\mathrm{corridor}}
\right|
\leq
NP(0,T)
\E^\Q\left[
\left|S_T^{(r)}-S_T\right|
\right],
\]
which also converges to zero by the same theorem.

Finally, for the longevity swap, subtracting the two price formulas and
applying the triangle inequality gives
\[
\left|
V_0^{\mathrm{swap},(r)}-V_0^{\mathrm{swap}}
\right|
\leq
N\sum_{k=1}^m P(0,t_k)
\E^\Q\left[
\left|S_{t_k}^{(r)}-S_{t_k}\right|
\right].
\]
Each term on the right converges to zero by
Theorem~\ref{thm:q-lifted-survival-index-convergence}, and the sum
contains only finitely many payment dates.

\section{Discrete implementation of the SVD lifting}
\label{app:svd-discretization}
\noindent
The numerical implementation in \Cref{sec:svd-lifting-validation}
includes the following steps:
form the quadrature matrix \(A_N=\Delta t K_N\), perform the SVD decomposition
\(A_N=U_N\Sigma_N V_N^\top\), and retain the first \(r\) singular
components.  Dividing the reconstruction by \(\Delta t\) gives
the rank-\(r\) SVD approximation kernel
\[
K_N^{(r)}
=
\Delta t^{-1}U_N^{(r)}\Sigma_r(V_N^{(r)})^\top.
\]
Since this truncated reconstruction is generally full, the matrix
used in the Volterra sum is its causal projection
\[
\bigl(K_N^{(r),\triangle}\bigr)_{ij}
=
\mathbf 1_{\{j\le i\}}\bigl(K_N^{(r)}\bigr)_{ij}.
\]
The following two diagrams retain the operational details of these two
steps: which SVD factors are kept, and why the noncausal entries created
by truncation must be set back to zero.

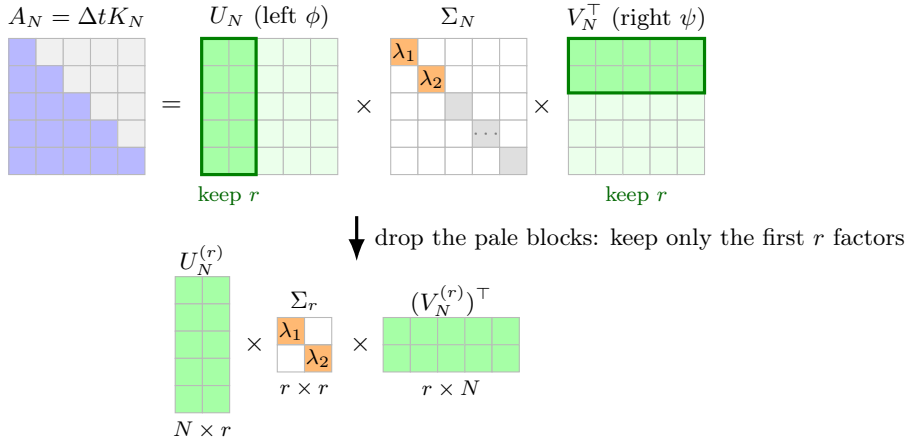
\begin{figure}[H]
\centering
{\normalcolor
\begin{tikzpicture}[>=Latex, font=\small]

\def\s{0.36}
\begin{scope}[shift={(0,0)}]
\foreach \i in {1,...,5} {
  \foreach \j in {1,...,5} {
    \pgfmathtruncatemacro{\ok}{\j<=\i ? 1 : 0}
    \ifnum\ok=1
      \fill[blue!28] ({(\j-1)*\s},{(5-\i)*\s}) rectangle ++(\s,\s);
    \else
      \fill[gray!12] ({(\j-1)*\s},{(5-\i)*\s}) rectangle ++(\s,\s);
    \fi
    \draw[gray!55] ({(\j-1)*\s},{(5-\i)*\s}) rectangle ++(\s,\s);
  }
}
\node[font=\footnotesize] at ({2.5*\s},{5*\s+0.28}) {\(A_N=\Delta t K_N\)};
\end{scope}
\node at (2.15,0.90) {\(=\)};
\begin{scope}[shift={(2.55,0)}]
\foreach \i in {1,...,5} {
  \foreach \j in {1,...,5} {
    \ifnum\j<3
      \fill[green!35] ({(\j-1)*\s},{(5-\i)*\s}) rectangle ++(\s,\s);
    \else
      \fill[green!8] ({(\j-1)*\s},{(5-\i)*\s}) rectangle ++(\s,\s);
    \fi
    \draw[gray!55] ({(\j-1)*\s},{(5-\i)*\s}) rectangle ++(\s,\s);
  }
}
\draw[very thick, green!50!black] (0,0) rectangle ({2*\s},{5*\s});
\node[font=\footnotesize] at ({2.5*\s},{5*\s+0.28}) {\(U_N\) (left \(\phi\))};
\node[font=\scriptsize, green!40!black] at ({1*\s},-0.28) {keep \(r\)};
\end{scope}
\node at (4.70,0.90) {\(\times\)};
\begin{scope}[shift={(5.05,0)}]
\foreach \i in {1,...,5} {
  \foreach \j in {1,...,5} {
    \ifnum\i=\j
      \ifnum\i<3
        \fill[orange!55] ({(\j-1)*\s},{(5-\i)*\s}) rectangle ++(\s,\s);
      \else
        \fill[gray!25] ({(\j-1)*\s},{(5-\i)*\s}) rectangle ++(\s,\s);
      \fi
    \else
      \fill[white] ({(\j-1)*\s},{(5-\i)*\s}) rectangle ++(\s,\s);
    \fi
    \draw[gray!55] ({(\j-1)*\s},{(5-\i)*\s}) rectangle ++(\s,\s);
  }
}
\node[font=\scriptsize] at ({0.5*\s},{4.5*\s}) {\(\lambda_1\)};
\node[font=\scriptsize] at ({1.5*\s},{3.5*\s}) {\(\lambda_2\)};
\node[font=\scriptsize, gray] at ({3.5*\s},{1.5*\s}) {\(\cdots\)};
\node[font=\footnotesize] at ({2.5*\s},{5*\s+0.28}) {\(\Sigma_N\)};
\end{scope}
\node at (7.05,0.90) {\(\times\)};
\begin{scope}[shift={(7.40,0)}]
\foreach \i in {1,...,5} {
  \foreach \j in {1,...,5} {
    \ifnum\i<3
      \fill[green!35] ({(\j-1)*\s},{(5-\i)*\s}) rectangle ++(\s,\s);
    \else
      \fill[green!8] ({(\j-1)*\s},{(5-\i)*\s}) rectangle ++(\s,\s);
    \fi
    \draw[gray!55] ({(\j-1)*\s},{(5-\i)*\s}) rectangle ++(\s,\s);
  }
}
\draw[very thick, green!50!black] (0,{3*\s}) rectangle ({5*\s},{5*\s});
\node[font=\footnotesize] at ({2.5*\s},{5*\s+0.28}) {\(V_N^\top\) (right \(\psi\))};
\node[font=\scriptsize, green!40!black] at ({2.5*\s},-0.28) {keep \(r\)};
\end{scope}
\draw[->, very thick] (4.6,-0.55) -- (4.6,-1.15);
\node[font=\footnotesize, right=3pt] at (4.6,-0.85) {drop the pale blocks: keep only the first \(r\) factors};
\begin{scope}[shift={(2.20,-3.15)}]
\foreach \i in {1,...,5} {
  \foreach \j in {1,...,2} {
    \fill[green!35] ({(\j-1)*\s},{(5-\i)*\s}) rectangle ++(\s,\s);
    \draw[gray!55] ({(\j-1)*\s},{(5-\i)*\s}) rectangle ++(\s,\s);
  }
}
\node[font=\footnotesize] at ({1*\s},{5*\s+0.22}) {\(U_N^{(r)}\)};
\node[font=\scriptsize] at ({1*\s},-0.22) {\(N\times r\)};
\end{scope}
\node at (3.25,-2.25) {\(\times\)};
\begin{scope}[shift={(3.55,-2.61)}]
\fill[orange!55] (0,\s) rectangle (\s,{2*\s});
\fill[white] (\s,\s) rectangle ({2*\s},{2*\s});
\fill[white] (0,0) rectangle (\s,\s);
\fill[orange!55] (\s,0) rectangle ({2*\s},\s);
\draw[gray!55] (0,0) rectangle (\s,\s);
\draw[gray!55] (\s,0) rectangle ({2*\s},\s);
\draw[gray!55] (0,\s) rectangle (\s,{2*\s});
\draw[gray!55] (\s,\s) rectangle ({2*\s},{2*\s});
\node[font=\scriptsize] at (0.5*\s,{1.5*\s}) {\(\lambda_1\)};
\node[font=\scriptsize] at (1.5*\s,{0.5*\s}) {\(\lambda_2\)};
\node[font=\footnotesize] at ({1*\s},{2*\s+0.22}) {\(\Sigma_r\)};
\node[font=\scriptsize] at ({1*\s},-0.22) {\(r\times r\)};
\end{scope}
\node at (4.65,-2.25) {\(\times\)};
\begin{scope}[shift={(4.95,-2.61)}]
\foreach \i in {1,...,2} {
  \foreach \j in {1,...,5} {
    \fill[green!35] ({(\j-1)*\s},{(2-\i)*\s}) rectangle ++(\s,\s);
    \draw[gray!55] ({(\j-1)*\s},{(2-\i)*\s}) rectangle ++(\s,\s);
  }
}
\node[font=\footnotesize] at ({2.5*\s},{2*\s+0.22}) {\((V_N^{(r)})^\top\)};
\node[font=\scriptsize] at ({2.5*\s},-0.22) {\(r\times N\)};
\end{scope}
\end{tikzpicture}
}
\caption{Matrix picture of the SVD. The operator matrix
\(A_N=\Delta t K_N\) factors as \(U_N\Sigma_N V_N^\top\). Columns of
\(U_N\) sample \(\phi_i\), columns of \(V_N\) sample \(\psi_i\), and
the diagonal of \(\Sigma_N\) holds the singular values. Rank \(r\)
keeps only the first \(r\) of these factors. 
In particular, the picture for 
\(N=5\) and \(r=2\) is illustrated. Dark green columns of \(U_N\) sample
\(\phi_1,\phi_2\), the dark green rows of \(V_N^\top\) sample
\(\psi_1,\psi_2\), the orange diagonal entries are the singular values,
and the pale blocks are truncated. The reconstructed kernel is
\(K_N^{(r)}=A_N^{(r)}/\Delta t\).}
\label{fig:app-svd-factor-schema}
\end{figure}

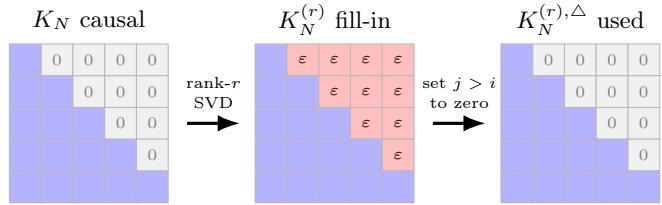
\begin{figure}[H]
\centering
{\normalcolor
\begin{tikzpicture}[>=Latex, font=\small]

\def\s{0.42}
\begin{scope}[shift={(0,0)}]
\foreach \i in {1,...,5} {
  \foreach \j in {1,...,5} {
    \pgfmathtruncatemacro{\ok}{\j<=\i ? 1 : 0}
    \ifnum\ok=1
      \fill[blue!30] ({(\j-1)*\s},{(5-\i)*\s}) rectangle ++(\s,\s);
    \else
      \fill[gray!12] ({(\j-1)*\s},{(5-\i)*\s}) rectangle ++(\s,\s);
      \node[font=\tiny, gray] at ({(\j-0.5)*\s},{(5-\i+0.5)*\s}) {\(0\)};
    \fi
    \draw[gray!55] ({(\j-1)*\s},{(5-\i)*\s}) rectangle ++(\s,\s);
  }
}
\node[font=\footnotesize] at ({2.5*\s},{5*\s+0.32}) {\(K_N\) causal};
\end{scope}
\draw[->, very thick] (2.35,1.05) -- (3.05,1.05);
\node[font=\tiny, align=center, above=2pt] at (2.70,1.05)
{rank-\(r\)\\SVD};
\begin{scope}[shift={(3.25,0)}]
\foreach \i in {1,...,5} {
  \foreach \j in {1,...,5} {
    \pgfmathtruncatemacro{\ok}{\j<=\i ? 1 : 0}
    \ifnum\ok=1
      \fill[blue!30] ({(\j-1)*\s},{(5-\i)*\s}) rectangle ++(\s,\s);
    \else
      \fill[red!25] ({(\j-1)*\s},{(5-\i)*\s}) rectangle ++(\s,\s);
      \node[font=\tiny] at ({(\j-0.5)*\s},{(5-\i+0.5)*\s}) {\(\varepsilon\)};
    \fi
    \draw[gray!55] ({(\j-1)*\s},{(5-\i)*\s}) rectangle ++(\s,\s);
  }
}
\node[font=\footnotesize] at ({2.5*\s},{5*\s+0.32}) {\(K_N^{(r)}\) fill-in};
\end{scope}
\draw[->, very thick] (5.60,1.05) -- (6.30,1.05);
\node[font=\tiny, align=center, above=2pt] at (5.95,1.05)
{set \(j>i\)\\to zero};
\begin{scope}[shift={(6.50,0)}]
\foreach \i in {1,...,5} {
  \foreach \j in {1,...,5} {
    \pgfmathtruncatemacro{\ok}{\j<=\i ? 1 : 0}
    \ifnum\ok=1
      \fill[blue!30] ({(\j-1)*\s},{(5-\i)*\s}) rectangle ++(\s,\s);
    \else
      \fill[gray!12] ({(\j-1)*\s},{(5-\i)*\s}) rectangle ++(\s,\s);
      \node[font=\tiny, gray] at ({(\j-0.5)*\s},{(5-\i+0.5)*\s}) {\(0\)};
    \fi
    \draw[gray!55] ({(\j-1)*\s},{(5-\i)*\s}) rectangle ++(\s,\s);
  }
}
\node[font=\footnotesize] at ({2.5*\s},{5*\s+0.32}) {\(K_N^{(r),\triangle}\) used};
\end{scope}
\end{tikzpicture}
}
\caption{Why the causal projection is needed. A lower-triangular kernel
does not in general remain triangular after rank-\(r\) SVD reconstruction,
which produces the full matrix \(K_N^{(r)}\). The red \(\varepsilon\)-cells
represent the resulting noncausal fill-in in the future triangle \(v>t\).
Since the discrete Volterra sum must not depend on future increments, these
entries are set back to zero before the driver is assembled. This recovers the
causal matrix \(K_N^{(r),\triangle}\), which is in general no longer of rank
\(r\).}
\label{fig:app-svd-fillin-schema}
\end{figure}

\paragraph*{Assembly of the lifted states}
Let \(\Delta L_j=L_{t_j}^{\Pp}-L_{t_{j-1}}^{\Pp}\).  In code, the
right singular vectors first accumulate these L\'evy increments into
\(r\) states,
\[
\widetilde U_n^{(i)}
=
\frac{1}{s_L}\sum_{j=1}^{n}
\bigl(v_i^{(N)}\bigr)_j\Delta L_j,
\qquad i=1,\ldots,r.
\]
The singular values and left singular vectors then reconstruct the
fractional driver.  Expanding the two finite sums gives the exact grid
identity
\begin{equation}
\Upsilon_{t_n}^{H,\Pp,(r)}
=
\sum_{i=1}^{r}
\frac{\lambda_i^{(N)}}{\Delta t}
\bigl(u_i^{(N)}\bigr)_n\widetilde U_n^{(i)}
=
\frac{1}{s_L}\sum_{j=1}^{n}
\bigl(K_N^{(r),\triangle}\bigr)_{nj}\Delta L_j.
\label{eq:app-svd-identity}
\end{equation}
Thus the state recursion and the causal matrix--vector product are two
implementations of the same rank-\(r\) approximation on the grid.

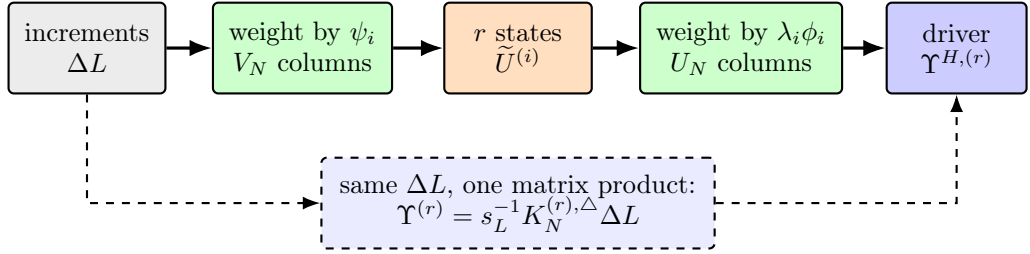
\begin{figure}[H]
\centering
{\normalcolor
\begin{tikzpicture}[>=Latex, font=\small,
  blk/.style={draw=black, thick, rounded corners=2pt, align=center,
  inner sep=6pt, minimum height=1.2cm, text=black}]

\node[blk, fill=gray!15, minimum width=1.95cm] (dL) at (0,0)
{increments\\[-1pt]\(\Delta L\)};
\node[blk, fill=green!20, minimum width=2.15cm] (psi) at (2.85,0)
{weight by \(\psi_i\)\\[-1pt]\(V_N\) columns};
\node[blk, fill=orange!25, minimum width=1.95cm] (Ust) at (5.70,0)
{\(r\) states\\[-1pt]\(\widetilde U^{(i)}\)};
\node[blk, fill=green!20, minimum width=2.25cm] (phi) at (8.65,0)
{weight by \(\lambda_i\phi_i\)\\[-1pt]\(U_N\) columns};
\node[blk, fill=blue!20, minimum width=1.85cm] (Ups) at (11.50,0)
{driver\\[-1pt]\(\Upsilon^{H,(r)}\)};
\draw[->, very thick, black] (dL) -- (psi);
\draw[->, very thick, black] (psi) -- (Ust);
\draw[->, very thick, black] (Ust) -- (phi);
\draw[->, very thick, black] (phi) -- (Ups);
\node[blk, fill=blue!8, dashed, minimum width=5.2cm] (alt) at (5.70,-2.05)
{same \(\Delta L\), one matrix product:\\[-1pt]
\(\Upsilon^{(r)}=s_L^{-1}K_N^{(r),\triangle}\Delta L\)};
\draw[->, dashed, thick, black] (dL.south) |- (alt.west);
\draw[->, dashed, thick, black] (alt.east) -| (Ups.south);
\end{tikzpicture}
}
\caption{Assembly of the discrete SVD lift. L\'evy increments are
accumulated against the right singular vectors \(\psi_i\) to form
\(r\) cumulative states. The left singular vectors \(\phi_i\) and
singular values \(\lambda_i\) then reconstruct the approximate
fractional driver. The dashed route is the same calculation written as
one causal matrix--vector product.}
\label{fig:app-svd-lift-schema}
\end{figure}

Figures~\ref{fig:app-svd-factor-schema}--\ref{fig:app-svd-lift-schema}
illustrate the three separate implementation steps: rank truncation, causal
masking, and assembly of the lifted states.
